\documentclass{emulateapj}
\usepackage{apjfonts}

\usepackage{amsmath}
\usepackage{color}

\lefthead{Ueda et al.}
\righthead{Standard Population Synthesis Model of the XRB}

\newcommand{\solarmass}{\mbox{${M_{\odot}}$}}

\newcommand{\solarlum}{\mbox{${L_{\odot}}$}}

\def\ltsima{$\; \buildrel < \over \sim \;$}
\def\simlt{\lower.5ex\hbox{\ltsima}}
\def\gtsima{$\; \buildrel > \over \sim \;$}
\def\simgt{\lower.5ex\hbox{\gtsima}}

\newcommand{\swift}{{\it Swift}}
\newcommand{\maxi}{{\it MAXI}}
\newcommand{\asca}{{\it ASCA}}
\newcommand{\suzaku}{{\it Suzaku}}
\newcommand{\rosat}{{\it ROSAT}}

\newcommand{\heao}{{\it HEAO1}}
\newcommand{\chandra}{{\it Chandra}}
\newcommand{\xmm}{{\it XMM-Newton}}
\newcommand{\subaru}{{\it Subaru}}

\newcommand{\sax}{{\it BeppoSAX}}
\newcommand{\integral}{{\it INTEGRAL}}

\newcommand{\logn}{log $N$ - log $S$ relation}

\newcommand{\erg}{erg s$^{-1}$}
\newcommand{\ergs}{erg cm$^{-2}$ s$^{-1}$}
\newcommand{\ergss}{erg cm$^{-2}$ s$^{-1}$ Str$^{-1}$}
\newcommand{\de}{deg$^2$}
\newcommand{\fct}{$f_{\rm CTK}$}
\newcommand{\rdisk}{$R_{\rm disk}$}
\newcommand{\nh}{$N_{\rm H}$}
\newcommand{\lx}{$L_{\rm X}$}
\newcommand{\lbol}{$L$} 

\newcommand{\fx}{$F_{\rm X}$}
\newcommand{\cx}{$C_{\rm X}$}
\newcommand{\IXRB}{$I_{\rm XRB,20-50}$}
\newcommand{\cosmo}{($H_0$, $\Omega_{\rm m}$, $\Omega_{\lambda}$)}

\begin{document}

\title{
Toward the Standard Population Synthesis Model of the X-Ray
Background: Evolution of X-Ray Luminosity and
Absorption Functions 
of Active Galactic Nuclei Including Compton-Thick
Populations
}

\author{Yoshihiro Ueda\altaffilmark{1},
Masayuki Akiyama\altaffilmark{2},
G\"unther Hasinger\altaffilmark{3},
Takamitsu Miyaji\altaffilmark{4,5},
Michael G. Watson\altaffilmark{6}
}

\altaffiltext{1}{Department of Astronomy, Kyoto University,
Kitashirakawa-Oiwake-cho, Sakyo-ku, Kyoto 606-8502, Japan}

\altaffiltext{2}{Astronomical Institute, Tohoku University, 6-3 Aramaki, Aoba-ku, Sendai 980-8578, Japan}

\altaffiltext{3}{Institute for Astronomy, 2680 Woodlawn Drive Honolulu,
HI 96822-1839, USA}

\altaffiltext{4}{Instituto de Astronom\'{i}a, Universidad Nacional
Aut\'{o}noma de M\'{e}xico, Ensenada, Baja California, Mexico; P.O.\ Box
439027, San Diego, CA 92143-9027, USA}

\altaffiltext{5}{University of California, San Diego, Center for
Astrophysics and Space Sciences, 9500 Gilman Drive, La Jolla, CA
92093-0424, USA}

\altaffiltext{6}{Department of Physics and Astronomy, University of Leicester, University Road, Leicester LE1 7RH, UK}

\begin{abstract}

We present the most up-to-date X-ray luminosity function (XLF) and
absorption function of Active Galactic Nuclei (AGNs) over the redshift
range from 0 to 5, utilizing the largest, highly complete sample
ever available obtained from surveys performed with \swift/BAT, \maxi,
\asca, \xmm, \chandra, and \rosat. The combined sample, including that
of the \subaru /\xmm\ Deep Survey, consists of 4039 detections in the
soft (0.5--2 keV) and/or hard ($>2$ keV) band. We utilize a maximum
likelihood method to reproduce the count-rate versus redshift
distribution for each survey, by taking into account the evolution of
the absorbed fraction, the contribution from Compton-thick (CTK) AGNs,
and broad band spectra of AGNs including reflection components from
tori based on the luminosity and redshift dependent unified scheme. 
We find that the shape of the XLF at $z \sim 1-3$ is significantly
different from that in the local universe, for which the luminosity
dependent density evolution model
gives much better description than the
luminosity and density evolution model. 
These results establish the
standard population synthesis model of the X-Ray Background (XRB), which
well reproduces the source counts, the observed fractions of CTK AGNs,
and the spectrum of the hard XRB. The number ratio of CTK AGNs to the
absorbed Compton-thin (CTN) AGNs is constrained to be $\approx$0.5--1.6
to produce the 20--50 keV XRB intensity within present uncertainties, by
assuming that they follow the same evolution as CTN AGNs. The growth
history of supermassive black holes is discussed based on the new AGN
bolometric luminosity function.

\end{abstract}

\keywords{diffuse radiation --- galaxies:active --- quasars:general ---
surveys --- X-rays:diffuse background}

\section{Introduction}
\label{sec-intro}

Understanding the cosmological evolution of supermassive black hole
(SMBHs) is a key issue in modern astrophysics. The good correlation of
the mass of a SMBH in a galactic center with that of the bulge in the
present-day universe \citep[e.g.,][]{mag98,fer00,geb00,mar03,har04,hop07,kor09,gul09} indicates that SMBHs and galaxies co-evolved in
the past. This idea is also supported by the similarity of the global
history between star formation and SMBH growth \citep[][]{mar04}. The
so-called down-sizing or anti-hierarchical evolution, the trend that
more massive systems formed in earlier cosmic time, has been revealed for
both SMBHs (\citealt{ued03}, U03; \citealt{has05}, H05) and galaxies
\citep[e.g.,][]{cow96,kod04,fon09}.

Active galactic nuclei (AGNs) are the phenomena where SMBHs gain their
masses from accreting gas by converting a part of their gravitational
energies into radiation. It is known that the majority of AGNs are
obscured by gas and dust surrounding the SMBHs, being classified as
``type-2'' AGNs. To elucidate the growth history of SMBHs, a complete
survey of AGNs including heavily obscured populations throughout the
history of the universe is necessary. X-ray observations, in particular
those at high energies above a few keV, provide one of the most powerful
approach for AGN detection thanks to the strong penetrating power
against absorption and little contamination from star lights in the host
galaxies. Furthermore, the deepest X-ray surveys currently available
achieve the highest sensitivity even for unobscured (``type-1'') AGNs
among those at any wavelengths \citep[see][and references
therein]{bra05}. The surface number density of the faintest X-ray AGNs
reaches $\sim 10^4$ deg$^{-2}$ \citep{xue11}.

The integration of emission from all accreting SMBHs in the
universe is observed as the X-ray background (XRB).
To quantitatively solve the XRB origin is equivalent to revealing the
cosmological evolution of AGNs that constitute the XRB. The XRB spans
over a wide energy range from $\sim$0.1 keV to $\sim$100 keV and then is
smoothly connected to the gamma-ray background at higher energies. Its
spectrum has a peak energy density around $\sim$20 keV. At energies
below $\sim$8 keV, now almost all of the XRB is resolved into discrete
sources, mainly AGNs. Enormous efforts have been made to identify AGNs
detected in X-ray surveys on the basis of multiwavelength observations,
and the redshifts (and hence luminosities) of a large fraction of these
sources are now estimated. These results make it possible to determine
the spatial number density of AGNs that constitute the XRB below $\sim$8
keV and its evolution. By contrast, in the hard X-ray band above
$\sim$10 keV, a significant fraction of the XRB is still left
unresolved. Therefore, at present, the whole origin of the XRB over the
wide range cannot be directly revealed by resolving individual sources.

It is very important to construct a ``population synthesis model'' of
the XRB where the evolutions of all X-ray emitting AGNs with various
types are formulated \citep[for previous works, see
e.g.][]{com95,gil99,ued03,bal06,gil07,tre09}. The model, in principle,
must explain all the observational constraints including source number
counts, redshift and luminosity distribution, the shape of the XRB. Once
established, it gives the basis to understand the accretion history of
the universe traced by X-rays, which is subject to least biases.

Two major elements in the population synthesis models are the X-ray
luminosity function (XLF) and absorption function (or \nh\ function,
\citealt{ued03}) of AGNs. The XLF represents the number density of AGN
per comoving space as a function of luminosity and redshift, one of the
most important statistical quantity that can be determined from unbiased
surveys.  Previously many studies have been made on the cosmological
evolutions of the XLF of AGNs in the soft band below 2 keV
\citep[e.g.,][]{mac91,boy93,jon97,pag97,miy00,has05} and the hard band
above 2 keV \citep[e.g.,][]{ued03,
laf05,bar05,sil08,ebr09,yen09,air10}. By using hard X-ray selected
samples that contain both type-1 (unabsorbed) and type-2 (absorbed)
AGNs, the absorption functions have been also investigated
\citep{ued03,laf05,bal06,tre06,has08}. Besides the well established
anti-correlation of absorption fraction with luminosity
\citep[e.g.,][]{ued03,ste03,sim05}, several works have reported that
the fraction of absorbed AGNs increased toward higher redshift from
$z=0$ to $z > 1$ \citep{laf05,bal06,tre06,has08}. More recent studies of
high redshift AGNs at $z>2$ in deep fields (\citealt{hir12};
\citealt{iwa12}) reveal larger absorption fractions of high luminosity
AGNs compared with the local universe. This redshift evolution is not
included in the population synthesis model by \citet{gil07}, one of the
most widely referred model available at present.

There still remain uncertainties on the evolution of AGNs, however.
The first issue is the shape of the XLF and its cosmological evolution. 
On the basis of hard X-ray surveys, \citet{ued03}, \citet{sil08},
\citet{ebr09}, and \citet{yen09} find that the XLF of AGN is best
described with the luminosity dependent density evolution (LDDE) model.
Later, \citet{air10} propose that the luminosity and density evolution
(LADE) where the shape of the XLF is constant over the whole redshift
range unlike the case of LDDE also gives a similarly good fit to their
data. While the down-sizing behavior is commonly seen in both models,
quite different number of AGNs are predicted particularly at high
redshifts of $z \geq 3$. The second one is the number density (or
fraction) of heavily obscured AGNs with \nh\ $>10^{24}$ cm$^{-2}$,
so-called ``Compton-thick'' AGNs (CTK AGNs). Even hard X-ray surveys
above 10 keV are subject to bias against detecting heavily CTK AGNs,
because the transmitted emission is significantly suppressed due to
repeated Compton scattering \citep[see e.g.,][]{wil99,ike09}. It is not
easy to identify individual CTK AGNs with limited photon statistics in
deep survey data, and to estimate their intrinsic number density by
correcting for such biases.

In this work, we present our latest results of AGN XLF over the
redshift range from 0 to 5, by utilizing one of the largest combined
sample ever available, obtained from surveys of various depth, width,
and energy bands performed with \swift /BAT, the Monitor of All-sky
X-ray Image (\maxi, \citealt{mat09}), \asca, \xmm, \chandra, and
\rosat. The sample consists of 4039 detections in the soft (0.5-2 keV)
and/or hard ($>2$ keV) bands. We utilize a maximum likelihood (ML)
method to reproduce the count-rate versus redshift distribution for
each survey, by taking into account the selection biases. In the
analysis, the contribution of CTK AGNs is considered, which is found
to be more important in harder bands and at fainter fluxes. This
enables us to determine the intrinsic XLF and the absorption function
of type-1 plus type-2 AGNs with an unprecedented accuracy, thus
establishing a standard population synthesis model of the XRB that is
consistent with most of observational constraints currently available.
 
The organization of the paper is as follows. Section~\ref{sec-sample}
explains the sample used in our analysis. In Section~\ref{sec-abs} the
absorption properties of AGNs are discussed and the absorption function
is formulated. Section~\ref{sec-template} introduces the ``template
model'' of broad band X-ray spectra of AGNs adopted in this work.  The
distribution of photon index is examined by using the \swift/BAT sample
in Section~\ref{sec-idx}. Section~\ref{sec-lf} describes the details of
main analysis using the whole sample. The best-fit results of the XLF
are presented there. The predictions from the population synthesis model
are given in Section~\ref{sec-model}. We also discuss the constraints on
the CTK AGN fraction and degeneracy with other
parameters. Section~\ref{sec-blf} represents an determination of the
bolometric luminosity function of AGNs based on our new XLF, and the
growth history of SMBHs. The conclusions of our work are summarized in
Section~\ref{sec-conclusions}. The cosmological parameters of \cosmo\ =
(70$h_{70}$ km s$^{-1}$ Mpc$^{-1}$, 0.3, 0.7) are adopted throughout the
paper. The ``log'' symbol represents the base-10 logarithm, while ``ln''
the natural logarithm.

\section{Sample}
\label{sec-sample}

In order to investigate the XLF and absorption function of AGNs
covering a wide range of luminosity and redshift, it is vital to
construct a sample combined from various surveys with different flux
limits and area. Also, high degrees of identification completeness in
terms of spectroscopic and/or photometric redshift determination are
required to minimize systematic uncertainties caused by sample
incompleteness. Basically, X-ray surveys at higher energies are more
suitable to detect obscured AGNs with less biases. Nevertheless, those
in the soft band ($\approx$0.5--2 keV) are also quite useful as long
as such biases are properly corrected. Generally, fainter flux limits
are achieved in softer energy band thanks to the larger collecting
area of X-ray telescopes. In particular, for high redshift AGNs, the
reduction of observed fluxes due to photo-electric absorptions becomes
less important thanks to the K-correction effect. Indeed, the soft
X-ray surveys are often utilized to search for high redshift AGNs even
including type-2 objects \citep[e.g.,][]{miy00,sil08}.

In our study, we collect the results from surveys performed with \swift
/BAT, \maxi, \asca, \xmm, \chandra, and \rosat, by utilizing the
heritage of X-ray astronomy accumulated up to present. Only those with
high identification completeness ($\simgt$90\%) are included. Our sample
is composed of those from the \swift /BAT 9-month survey (BAT9,
\citealt{tue08}), \maxi\ 7-month survey (MAXI7, \citealt{hir11, ued11}),
\asca\ Medium Sensitivity Survey (AMSS, \citealt{ued01,aki03}) and Large
Sky Survey (ALSS, \citealt{ued99,aki00}), \subaru /\xmm\ Deep Survey
(SXDS, \citealt{ued08,aki14}), \xmm\ survey of the Lockman Hole (LH/XMM,
\citealt{has01,bru08}), HELLAS2XMM survey (H2X, \citealt{fio03}), Hard
Bright Serendipitous Sample in the \xmm\ Bright Survey (HBSS,
\citealt{del04,del08}), \chandra\ Large Area Synoptic X-ray Survey
(CLASXS, \citealt{yan04,ste04,tro08}), \chandra\ Lockman Area North
Survey (CLANS, \citealt{tro08,tro09}), \chandra\ Deep Survey North
(CDFN, \citealt{ale03,bar03,tro08}) and South (CDFS, \citealt{xue11}),
and various \rosat\ surveys (see \citealt{miy00} and H05 and references
therein).

This paper presents the first work that makes use of the large X-ray
sample in the SXDS \citep{fur08,ued08}, one of the wide and deep
multiwavelength survey projects with a comparable survey area and depth
as the Cosmic Evolution Survey (COSMOS; \citealt{sco07}), in order to
constrain the XLF of AGNs with the best statistical accuracy. Also, new
hard X-ray all-sky surveys with \swift/BAT and \maxi\ are utilized
instead of \heao\ AGN samples that were usually employed in previous
studies. The other soft-band and hard-band samples adopted here are also
analyzed in H05 and \citet{has08}, respectively, although in some cases
different selection criteria are applied in our analysis to increase the
completeness (see below). The major difference in the soft-band sample
from that used by H05 is that we include both type-1 and type-2 AGNs
because we aim to investigate the evolution of the whole AGN
population. We do not use the sample from the Serendipitous
Extragalactic X-ray Source Identification (SEXSI) program \citep{eck06}
adopted by \citet{has08}, whose redshift identification completeness is
slightly worse ($\sim$84\%) than our threshold. The sample of the \xmm\
Medium-sensitivity Survey (XMS, \citealt{bar07}) used in \citet{has08}
are discarded because it partially overlaps with the SXDS sample. The
detailed description of each survey is given below.

\subsection{\swift/BAT}

We utilize the \swift/BAT 9-month catalog \citep{tue08}, which
originally contains 137 non-blazar AGNs with a flux limit of $2\times
10^{-11}$ erg cm$^{-2}$ s$^{-1}$ in the 14--195 keV band with the
signal-to-noise ratio (SNR) $> 4.0$. The \swift/BAT survey, performed
at energies above 10 keV, provides us with the least biased AGN sample
in the local universe against obscuration except for heavily CTK sources
absorbed with \nh\ $\gg 10^{24}$ cm$^{-2}$ along with those of
\integral\ \citep[e.g.,][]{bec09}. The X-ray spectra below 10 keV of all
the AGNs in the sample have been obtained from extensive follow-up
observations by other missions such as \swift/XRT, \xmm, and \suzaku\
\citep[e.g.,][]{win09a,egu09,win09b}. Thus, we can discuss their
statistical properties without incompleteness problems. Here we refer to
the absorption column densities summarized in Table~1 of \citet{ich12},
where new results obtained from broad-band X-ray spectra of \suzaku\ are
included. To define a statistical sample that is consistent with the
survey area curve given in Figure 1 of \citet{tue08}, we further impose
criteria that (1) the Galactic latitude is larger than 15 degrees ($|b|
> 15^\circ$) and (2) SNR $> 4.8$. We regard the pair of the interacting
galaxies NGC 6921 and MCG +04-48-002 unresolved with \swift/BAT as one
source, adopting the spectral information below 10 keV of the
latter. The sample consists of 87 non-blazar AGNs with identification
completeness of 100\%.

\subsection{\maxi}

We include the local AGN sample from the \maxi\ extragalactic survey
performed in the 4--10 keV band \citep{ued11} in our analysis. While the
sample significantly overlaps with the \swift/BAT 9 month sample, it is
useful to directly constrain the XLF in the 2--10 keV band. The sample
is collected from the first \maxi/GSC catalog at high Galactic latitudes
($|b|>10^\circ$) with a flux limit is $1.5\times10^{-11}$ \ergs\ (4--10
keV) based on the first 7 month data \citep{hir11}. It consists of 37
non-blazar AGNs by excluding Cen A and ESO~509--066, and is used by
\citet{ued11} to calculate the XLF and absorption function of AGNs in
the local universe. A merit is its high completeness of identification
(99.3\%), and similarly to the \swift/BAT 9 month catalog, the
information of the X-ray spectra below 10 keV are available for all the
objects as summarized in Table~1 of \citet{ued11}. Unlike in U03, we do
not use the local AGN samples by the \heao\ mission \citep{pic82},
considering the lower completeness than in the \maxi\ sample and
systematic uncertainties of the measured fluxes close to the sensitivity
limit that must be corrected for \citep[see][]{shi06}. Note that only 17
objects out of the total \maxi/GSC AGNs are listed in the \citet{pic82}
sample, suggesting significant long term variabilities of nearby AGNs
over 30 years.

\subsection{\asca}

\asca, the fourth Japanese X-ray astronomical satellite, performed first
imaging surveys in the energy band above 2 keV and provided statistical
X-ray samples at $>100$ times deeper levels than that of \heao\ A2
survey. These are still very useful to bridge the flux range between
all-sky surveys and deep surveys with \chandra\ and \xmm. In our paper we
utilize the two major samples, the \asca\ Large Sky Survey (ALSS) and
\asca\ Medium Sensitivity Survey (AMSS), both are firstly utilized by
U03 to calculate the hard XLF. The ALSS covers a continuous area of 5.5
\de\ near the North Galactic Pole and a flux limit of $\approx
1\times10^{-13}$ \ergs\ (2--10 keV) is achieved \citep{ued98,
ued99}. Thirty AGNs detected with the SIS instrument are optically
identified by \citet{aki00} with 100\% completeness by ignoring the one
unidentified source in \citet{aki00} as fake detection (see
Section~2.2.2 of U03). The AMSS is based on a serendipitous X-ray survey
with the GIS instrument, whose catalogs are published in \citet{ued01}
and \citet{ued05}. We use the ``AMSSn'' \citep{aki03} and ``AMSSs''
samples for which systematic identification programs were carried out in
the northern (DEC$>20^{\circ}$) and southern sky (DEC$< -20^{\circ}$),
respectively. The AMSSn and AMSSs samples contains 74 and 20
optically-identified non-blazar AGNs, respectively, with a detection
significance larger than 5.5 $\sigma$ at the flux range between
$5\times10^{-12}$ \ergs\ and $3\times10^{-13}$ in the 2--10 keV
band. Three X-ray sources are left unidentified, and thus the
completeness is 97\%. The total survey area covered by the AMSSn and
AMSSs is 90.8 \de.  We utilize the hardness ratio between the 2--10
(or 2--7) keV and 0.7--2 keV bands to estimate the absorption or photon
index of each source in the ALSS and AMSS, while results from follow-up
observations with \asca\ or \xmm\ are available for six and two sources
in the ALSS and AMSS, respectively (see Section~2.2 of U03).

\subsection{\xmm}

\subsubsection{SXDS}

The Subaru/\xmm\ Deep Survey (SXDS) is one of the largest
multi-wavelength surveys covering from radio to X-rays with an
unprecedented combination of depth and sky area. The SXDS field is a
contiguous region of $>$1 \de\ centered at R.A. = 02$^{h}$18$^{m}$ and
Dec. = $-05^{d}00^{m}$ (J2000). The large survey area makes it possible
to establish the statistical properties of extragalactic populations
without being affected by cosmic variance, and is also very useful to
construct a sample of sources with small surface densities, like
high-luminosity AGNs (i.e., QSOs) at high redshifts. The optical
photometric catalogs in the $B$, $V$, $R$, $i^\prime$, and $z^\prime$
bands obtained with Subaru/Suprime-Cam are presented in \citet{fur08},
achieving the $3\sigma$ sensitivities of $B$=28.4, $V$=27.8, $R_c$=27.7,
$i^\prime$=27.7, and $z^\prime$=26.6 (for the 2 arcsec aperture in the AB
magnitudes), respectively. The deep imaging data are particularly
important to reliably identify high-redshift AGNs by photometric
redshifts.

The X-ray catalog of the SXDS \citep{ued08} is based on the EPIC data in
the 0.3--10 keV band obtained from seven pointings performed with \xmm,
which covers a total of 1.14 \de. It contains 866, 1114, 645, and 136
sources with sensitivity limits of $6\times10^{-16}$, $8\times10^{-16}$,
$3\times10^{-15}$, and $5\times10^{-15}$ erg cm$^{-2}$ s$^{-1}$ in the
0.5--2, 0.5--4.5, 2--10, and 4.5--10 keV bands, respectively, with a
detection likelihood $\ge$ 7. The deep optical images were taken by five
pointing with Suprime-Cam with a slightly different shape from that of
the combined \xmm\ image. By limiting to the area of 1.02 \de\ where
these optical imaging data are available, we define two independent
parent samples consisting of 584 and 781 sources detected in the 2--10
keV and 0.5--2 keV bands, respectively.

The results from multiwavelength identification of the X-ray sources in
the SXDS is presented in \citet{aki14}. Out of the 584 (2--10 keV) and
781 (0.5--2 keV) X-ray source samples, 576 and 733 objects are left as
AGN candidates, respectively, after excluding Galactic stars, stellar
objects close to bright galaxies, and clusters of galaxies. Among them,
397 and 514 targets have spectroscopic redshift based on near-infrared
and/optical data. For the rest, \citet{aki14} determine the photometric
redshifts except for seven (2--10 keV) and eight (0.5--2 keV) objects
for which the fluxes in less than 4 bands are available. The redshifts
are estimated on the basis of HyperZ photometric redshift code with the
Spectral Energy Distribution (SED) templates of galaxies and QSOs. In
addition, constraints from the morphology and absolute magnitude limits
for the galaxy and QSO templates are applied (see \citep{hir12} and
\citep{aki14} for details). The accuracy of the photometric redshifts is
found to be $\Delta z/(1+z_{\rm spec} = 0.06$ as a median
value. Finally, 569 and 725 AGNs detected in the 2--10 keV and 0.5--2
keV bands have either spectroscopic or photometric redshifts, among
which 412 are common sources. The identification completeness is rather
high, 99\% both for the 2--10 keV and 0.5--2 keV samples. The hardness
ratio between the 2--10 keV and 0.5--2 keV is used to estimate the
absorption or photon index of each source.

\subsubsection{Lockman Hole}

We use both the hard-band (2--4.5 keV) and soft-band (0.5--2 keV)
selected samples from the \xmm\ deep survey of the Lockman Hole
\citep{has01,bru08}. 
They are essentially the same as those used in the
analysis by \citet{has08} and H05, respectively, except for small
differences that (1) type-2 AGNs detected in the soft-band are also
included in our analysis, unlike the case of H05 who only used type-1
AGNs, and that (2) some unpublished photometric redshifts for a few
hard-band selected sources mentioned in Section~2.8 of \citet{has08} are
not utilized and instead they are treated as unidentified objects in our
analysis. The effects by the latter difference are negligible. Before
the AO-2 phase 17 observations of this field were performed with
\xmm. The X-ray source catalog is provided by \citet{bru08} from 637 ks
(hard band) and the 770 ks (soft band) datasets. To maximize the
completeness of redshift identification, sub-samples are defined from
two off-axis intervals with different flux limits as detailed in
\citet{has08} and H05. The survey area after this selection becomes
0.183 \de\ and 0.126 \de\ with the fainter flux limits of
$6.1\times10^{-15}$ \ergs\ (2--10 keV) and $1.3\times10^{-15}$ \ergs\
(0.5--2 keV) for the hard- and soft-band samples, respectively. The soft
(hard) band sample consists of 57 (58) AGNs having either spectroscopic
or photometric redshifts 
\citep{fot12}
with identification completeness of 91\%
(88\%).
We refer to \citet{mat05} for the results of X-ray spectral
analysis, with some updates for sources whose redshifts are revised
after the publication of \citet{mat05} (Streblyanska 2006, private
comm.\ ).

\subsubsection{HELLAS2XMM}

The HELLAS2XMM \citep{bal02} is a serendipitous survey performed with
\xmm\ in the 2--10 keV band. We basically refer to the original sample
complied by \citet{fio03} selected from five \xmm\ fields, covering a
total area of 0.9 degree$^2$. Among the 122 hard X-ray selected sources,
\citet{fio03} present the spectroscopic redshifts for 97
AGNs. Additional information on the redshifts of the unidentified
sources in the original catalog of \citet{fio03} is reported in
\citet{mig04} and \citet{mai06}. To ensure a high completeness, we apply
a flux limit of $1.5\times10^{-14}$ \ergs\ (2--10 keV) band, which
leaves 95 sources after excluding one Galactic star and one extended
object. Among them, 87 AGNs have redshifts and 8 objects are left
unidentified. The completeness is thus 92\%. The results of X-ray
spectral analysis are presented in \citet{per04}. Unlike in
\citet{has08}, we do not include the new catalog of HELLAS2XMM by
\citet{coc07} in our analysis, considering the lower rate of redshift
identification (59 out of 110 sources).

\subsubsection{Hard Bright Survey}

The Hard Bright Serendipitous Sample (HBSS, \citealt{del04}) is a
subsample detected in the 4.5--7.5 keV band from those of the \xmm\
Bright Survey (XBS) aiming at relatively bright X-ray sources. From an
area of 25 degree$^2$ at $|b| > 20^\circ$, \citet{del08} define a
complete flux-limited sample consisting of 67 sources with the MOS2
count rates larger than 0.002 count s$^{-1}$ (4.5--7.5 keV),
corresponding to $2.2\times10^{-13}$ \ergs\ in the 2--10 keV band for a
photon index of 1.6. The sample is optically identified except for 2
sources, and in total 62 AGNs are cataloged in \citet{del08}. The
completeness of redshift identification is 97\%. The results from the
spectral analysis of the \xmm\ data are also summarized in \citet{del08}.

\subsection{\chandra}

\subsubsection{CLASXS}

The \chandra\ Large Area Synoptic X-ray Survey (CLASXS) is an
intermediate-depth survey covering a continuous area of $\approx$0.4
\de\ by 9 pointings in the Lockman Hole-Northwest field. \citet{yan04}
present the X-ray source catalog containing 525 sources. Following the
work by \citet{ste04}, \citet{tro08} report spectroscopic redshifts for
260 sources that are not stars and photometric redshifts for 134 sources
out of the 245 spectroscopically unidentified objects. We first select
sources detected in the 2--8 keV band at small off-set angles that are
used to calculate the \logn\ in their work (see their Figure 8).  The
survey area is thus reduced to be 0.28 \de. Then, to maximize the
completeness, we further impose a threshold to the hard-band count rate
above $6.6\times10^{-4}$ cts s$^{-1}$, corresponding to
$1.9\times10^{-14}$ \ergs\ in the 2--10 keV band for a photon index of
1.2. This leaves a total of 50 identified AGNs and 2 unidentified
sources after excluding 1 Galactic star. The redshift completeness is
96.2\%. We use the hardness ratio between the 2--8 keV and 0.4--2 keV
bands to estimate the absorption or photon index of each source in
Section~\ref{sec-sample-luminosity}.

\subsubsection{CLANS}

The \chandra\ Lockman Area North Survey (CLANS) is another
intermediate-depth survey in the Lockman Hole region, consisting of nine
separated \chandra\ pointings with an exposure of $\approx$70 ksec for
each. The field is a part of the of the Spitzer Wide-Area Infrared
Extragalactic Survey (SWIRE), and covers a total area of 0.6 \de . A
total of 761 X-ray sources are cataloged in \citet{tro08} with flux
limits of $3.5\times10^{-15}$ \ergs\ and $7\times10^{-16}$ \ergs\ in the
2--8 keV and 0.5--2 keV bands, respectively. \citet{tro08} also provide
327 spectroscopic redshifts (except for stars) and 234 photometric
redshifts out of the 425 spectroscopically unidentified objects. Some of
the redshifts are updated in \citet{tro09}. For our analysis, we only
select sources that are detected at offset angles from the pointing
center smaller than $8'$ and have the signal-to-noise ratios larger than
3 in the hard (2--8 keV) or soft (0.5--2 keV) band. Further, to achieve
high completeness of redshift identification (95\%), we impose flux
limits of $1.0\times10^{-14}$ \ergs\ (2--10 keV) and $2.3\times10^{-15}$
\ergs\ (0.5--2 keV) by assuming a photon index of 1.2 and 1.4 for the
hard-band and soft-band samples, which leave 159 and 191 identified AGNs
with 8 and 10 unidentified objects other than 0 and 3 Galactic stars,
respectively. Above these flux limits, the survey area can be regarded
to be constant to be 0.490 \de\ (see their Figure~5).  The number of
common AGNs in the two samples are 122.

\subsubsection{CDFN}

The \chandra\ Deep Survey North (CDFN) is the currently second deepest
survey ever performed in X-rays after the \chandra\ Deep Survey South
(CDFS). The 2-Ms catalog containing 503 sources from a survey area of
0.124 \de\ is presented by \citet{ale03}. To define the hard band (2--8
keV) and soft band (0.5--2 keV) selected samples presented in
\citet{ale03}, we apply a threshold of $\geq 3$ to the signal-to-noise
ratio defined as the count rate divided by its negative error in each
band. The sensitivities are $2.8\times10^{-17}$ \ergs\ and
$2.1\times10^{-16}$ \ergs\ in the 0.5--2 keV and 2--10 keV bands,
converted from the count-rate limits by assuming a photon index of 1.4
and 1.0, respectively. We basically refer to the redshift catalog
provided by \citet{tro08}, which contains the results from previous
works including \citet{bar03}, \citet{swi04}, \citet{cha05}, and
\citet{red06}. In total \citet{tro08} report 307 spectroscopic redshifts
(except for stars) and 107 photometric redshifts out of the 182
spectroscopically unidentified objects. To even increase the
completeness of redshift identification, we also utilize the compilation
of both spectroscopic and photometric redshifts by Hasinger (2008,
private comm.\ ) for the remaining unidentified objects, which is used in
\citet{has08}.
The hard and soft band samples consist of 286 and 375 AGN candidates
(including galaxies) that have redshift identification (with 179 and 252
spectroscopic redshifts), respectively, leaving no unidentified
objects. The number of common sources in the two samples is 195. The
completeness is thus 100\% for both samples. The hardness ratio between
the 2--8 keV and 0.5--2 keV count rates is used to estimate the
absorption or photon index of each source in
Section~\ref{sec-sample-luminosity}. The hard-band sample is the same as
used by \citet{has08} except that we do not apply the off-axis cut in
the sample selection to increase the sample size. Unlike in H05, we 
do not limit the soft-band sample to only type-1 AGNs.

\subsubsection{CDFS}

The CDFS provides the deepest X-ray survey dataset ever performed to
date obtained from a total of 4-Ms exposure of \chandra. We use the
source catalog complied by \citet{xue11}. It lists 740 X-ray sources
from an area of 464.5 arcmin$^2$. We define the hard band (2--8 keV) and
soft band (0.5--2 keV) selected samples that satisfy the
binomial-probability source-selection criterion of $P<0.004$ in each
survey band. They consist of 375 and 626 sources with flux limits of
$1.1\times10^{-16}$ \ergs\ (2--10 keV) and $1.1\times10^{-17}$ \ergs\
(0.5--2 keV), converted from the count-rate limits by assuming a photon
index of 1.0 and 1.4, respectively.  We basically refer to the redshift
identification presented in \citet{xue11} but adopt the new result
reported by \citet{vit13} for five $z>3$ AGNs (four spectroscopic and
one photometric redshifts). Extended sources and those identified as
``Star'' are excluded.  We keep ``AGN'' and ``Galaxy'' types in both
hard and soft samples before filtering them by the count rate and redshift
according to the procedure described in
Section~\ref{sec-sample-summary}, because the latter ones may well be
actually AGNs in some cases \citep[see][]{xue11}. The hard and soft band
samples consists of 358 and 583 redshift identified (228 and 378
spectroscopically identified) AGN candidates with 17 and 43 objects
without redshift information, thus achieving the completeness of 96\%
and 93\%, respectively. The number of common sources in both samples is
313. The hardness ratio between the two bands is utilized to estimate
the absorption or photon index in Section~\ref{sec-sample-luminosity}.

\subsection{\rosat}

We utilize a large, soft X-ray selected AGN sample obtained from
various \rosat\ surveys with different depth and area to cover the
brighter flux range than those of \chandra\ and \xmm . Essentially the
same sample was utilized by \citet{miy00} and H05 to construct
the soft XLF.  It is selected from the \rosat\ Bright Survey (RBS,
\citealt{sch00}), the RASS Selected-Area Survey North (SA-N,
\citealt{app98}), the \rosat\ North Ecliptic Pole Survey
(NEPS,\citealt{gio03,mul04}), the \rosat\ International X-ray/Optical
Survey (RIXOS, \citealt{mas00}), and the \rosat\ Medium Survey (RMS). We
refer the reader to \citet{miy00} and H05 for detailed
description of each survey. For our analysis we impose a conservative
flux cut of $S \geq 3.5\times10^{-14}$ \ergs, below which a sufficiently
large number of sources are available from the \chandra\ and \xmm\
surveys. Unlike in H05, we include both type-1 and type-2 AGNs
in our analysis as done in \citet{miy00}. In total 722 AGNs are
sampled. Information of the X-ray spectra covering the 2--10 keV band is
unavailable for most of the sources. In our main analysis
(Section~\ref{sec-lf}), we find that the source counts of the \rosat\
AGNs are better reproduced by any models when we adopt slightly
($\approx$10--20\%) higher fluxes than the original ones. We consider
that this is probably due to the cross-calibration error in the absolute
flux between different missions \citep[see e.g.,][]{ued99}.
To deal with this issue, we adopt $15\%$ higher fluxes than those
reported in the original tables for all \rosat\ sources. The uncertainty
does not affect the determination of the parameters and the XLF and
absorption function within the errors, however.

\subsection{Sample Summary}
\label{sec-sample-summary}

Table~\ref{tab-sample} summarizes the energy band, sensitivity limits,
survey area, number of sources with measured redshifts, and
identification completeness for each survey. Here the completeness is
defined as the fraction of all sources with redshift identification in
the total ones (including non AGNs) selected with the same detection
criteria\footnote{In the case of SXDS, CDFN, and CDFS, it refers to the
fraction of AGNs and galaxies with spectroscopic or photometric
redshifts in the total AGN candidates}. In total, we have 87 detections
in the very hard band (14--195 keV), 1791 in the hard band (within 2--10
keV), and 2654 in the soft band (0.5--2 keV). Although common sources
are contained in multiple samples obtained in different energy bands of
the same field, we basically treat them as independent detections in our
main analysis (Section~\ref{sec-lf}). The flux limits in units of \ergs\
are converted from the vignetting corrected count-rate limit in each
survey band by assuming a power law photon index given in the
parenthesis of the third column.
In the following analysis, we correct for the incompleteness of each
survey by multiplying the area by the completeness fraction
independently of flux. This procedure implicitly assumes that
unidentified sources follow the same redshift and luminosity
distribution as identified ones. Although it is a simplified assumption,
possible uncertainties in the correction little affect the overall
determination of the XLF, thanks to the high completeness of our sample.

\begin{figure}[h]
\epsscale{1.0}
\begin{center}
\includegraphics[angle=0,scale=0.45]{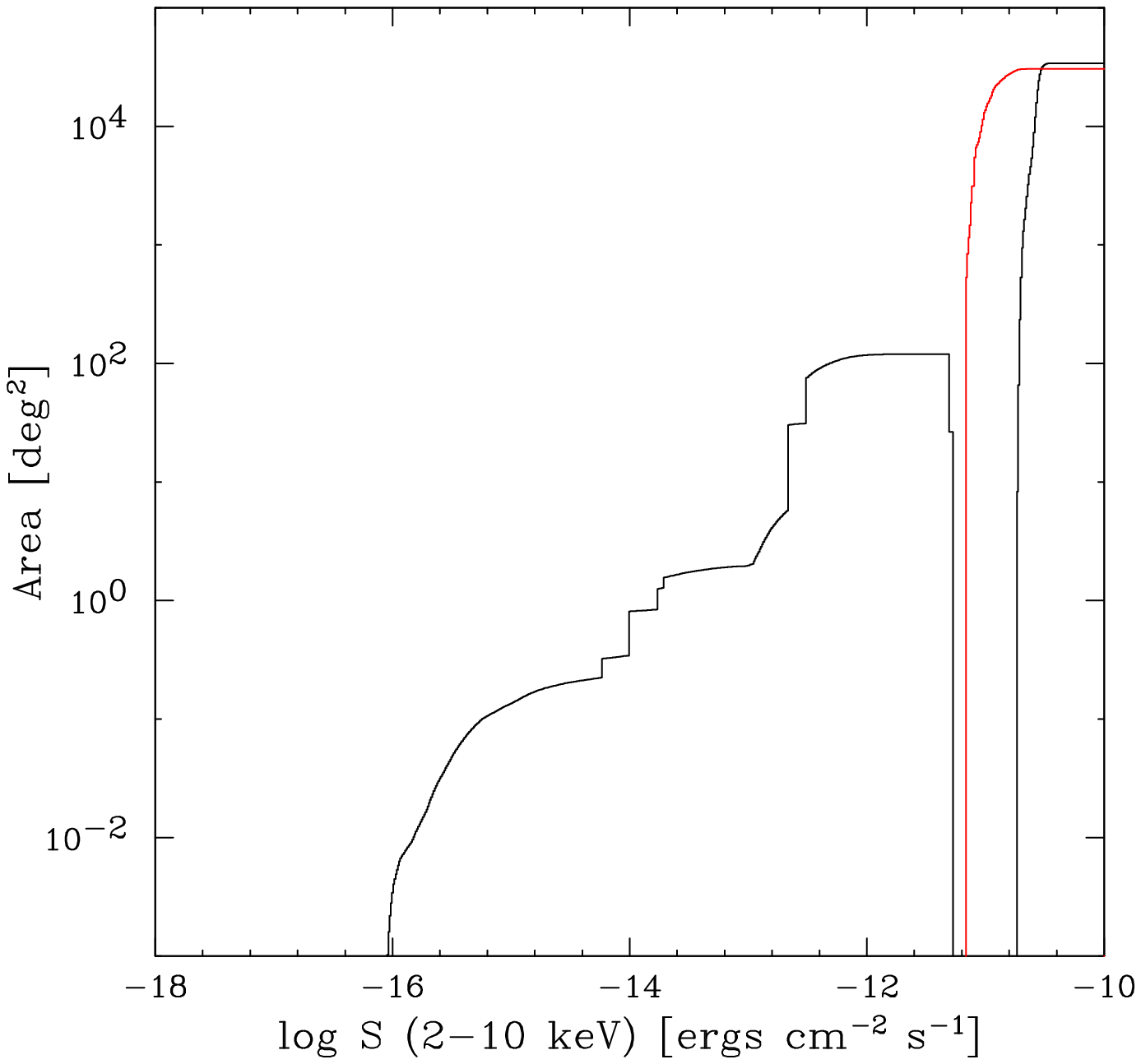}
\includegraphics[angle=0,scale=0.45]{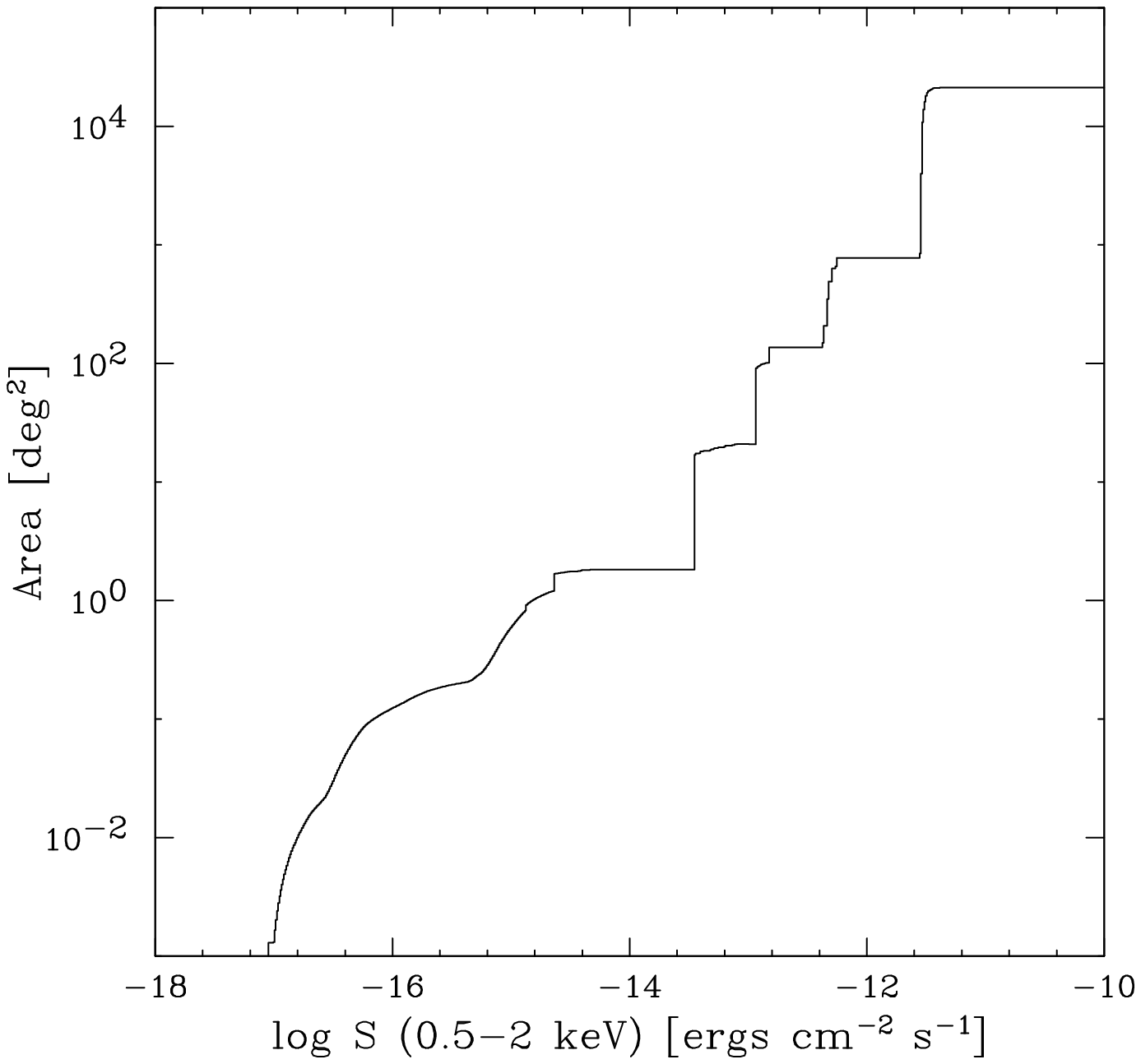}
\end{center}
\caption{
Survey area plotted against limiting flux in the hard (left)
and soft (right) bands. The red curve in the left panel 
corresponds to that of the \swift/BAT survey assuming $\gamma=1.6$.
}
\label{fig-area}
\end{figure}

\begin{figure}
\epsscale{1.0}
\begin{center}
\includegraphics[angle=0,scale=0.45]{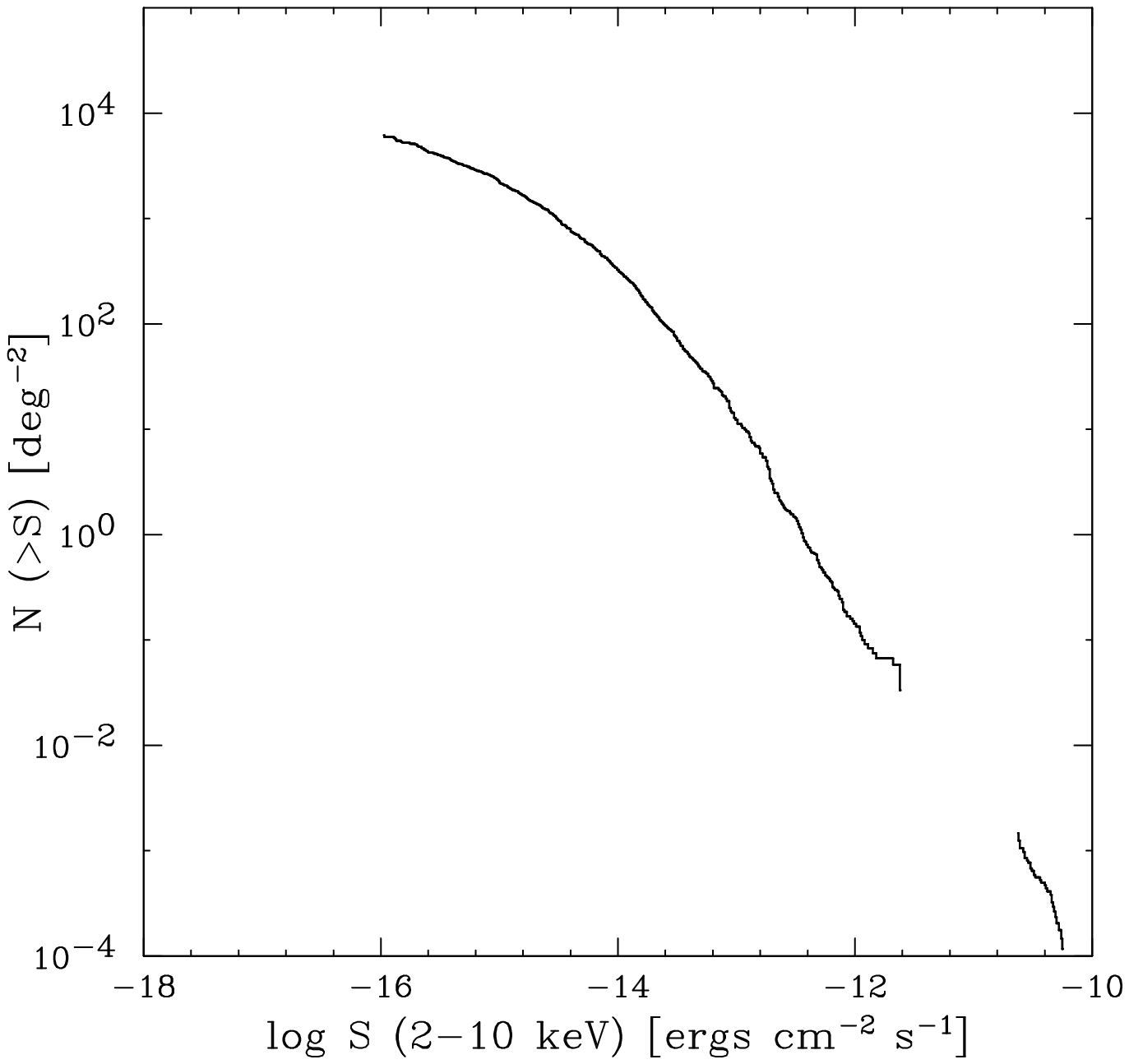}
\includegraphics[angle=0,scale=0.45]{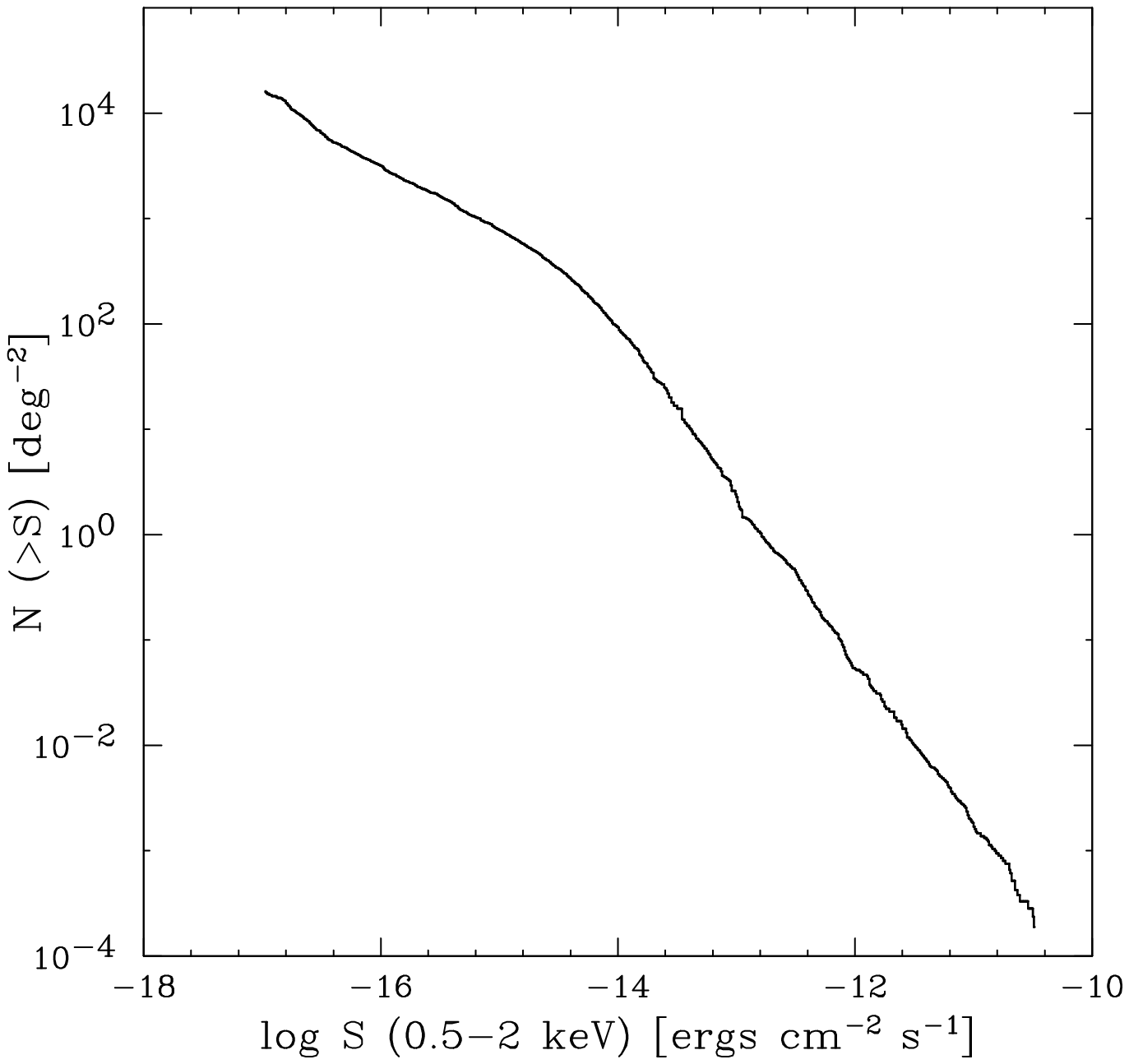}
\end{center}
\caption{
Observed source counts in the hard (left) and soft (right) bands.
}
\label{fig-logn}
\end{figure}

Figure~\ref{fig-area} plots the survey area against flux in the 2--10
keV and 0.5--2 keV bands. That of the \swift /BAT survey is overlaid in
Figure~\ref{fig-area} (left) by the red curve.  The \logn s in the
integral form are shown in Figure~\ref{fig-logn} (left) for the hard
band and Figure~\ref{fig-logn} (right) for the soft band. In the former,
the \swift /BAT sample is not included. The data at
$S=2.4\times10^{-11}-2.4\times10^{-12}$ \ergs\ are not shown because of
the survey flux gap in the hard band. The photon indices listed in
Table~\ref{tab-sample} for each survey are assumed for the count rate to
flux conversion in Figures~\ref{fig-area} and \ref{fig-logn}.

For the analysis of XLF presented in Section~\ref{sec-lf}, we limit the
redshift range to $z=0.002-5.0$. Although this lower limit is smaller
than $z=0.015$ adopted by previous works (e.g., \citealt{miy00}, U03,
H05), we confirm that excluding nearby AGNs at $z<0.015$ from the
analysis does not change our results over the statistical errors and
hence possible effects of the local over-density can be ignored. In very
deep surveys like CDFN and CDFS, we find non negligible contamination
from starburst galaxies in our sample at the lowest luminosity range,
which would affect our analysis of the XLF. To exclude such sources by
not relying their type identifications in the catalog that could be
quite uncertain \citep{xue11}, we impose a lower limit of the count rate
as a function of redshift in each survey, $c_{\rm lim}(z)$, that
corresponds to \lx\ = $10^{41}$ \erg\ or \lx\ = $10^{42}$ \erg\ for the
the hard band or soft band sample, respectively. We adopt a lower value
for the \lx\ threshold for the hard band sample because emission from
starburst galaxies is 
generally softer than that from AGNs, although X-ray binaries could
significantly contribute to the hard X-ray luminosity 
in galaxies with very high star formation rates and/or stellar masses
\citep[e.g.,][]{per02,leh10}.
We confirm 
that increasing the lower limit to \lx\ = $10^{41.5}$ \erg\
in the hard-band does not change the XLF parameters over the errors.
To be conservative, $c_{\rm lim}(z)$ is calculated by assuming
$\Gamma=1.9$ and no absorption. 
Applying
these cuts in addition to the redshift limit ($z=0.002-5.0$) slightly
reduces the number of sources in each sample, which are listed in the
fifth column of Table~\ref{tab-sample}. The numbers of detections used
in our main analysis (Section~\ref{sec-lf}) thus become 85 in the very
hard band (above 10 keV), 1770 in the hard band (within 2--10 keV), 2184
in the soft band (below 2 keV), and hence 4039 in total.

\subsection{Estimate of Luminosity}
\label{sec-sample-luminosity}

For convenience, here we calculate an intrinsic (de-absorbed) luminosity
in the rest-frame 2--10 keV band, \lx, for each object, following the
same procedure as described in Section~3.2 of U03. It can be calculated
as
\begin{equation}
        L_{\rm X} = 4\pi d_{\rm L}(z)^2 F_{\rm X},
\end{equation}
where $d_{\rm L}(z)$ is the luminosity distance at the source redshift $z$,
and \fx\ is the de-absorbed flux in the observer's frame 
$2/(1+z) - 10/(1+z)$ keV band. 
To convert the count rate into \fx, we need to consider the
spectrum of each source by taking into account with the energy response
of the instrument used in the survey.
We refer to the results of individual spectral analysis in terms of
absorption and photon index whenever such information is available. As
for the \swift/BAT sample, which contains identified CTK AGNs,
we assume the ``template spectra'' described in
Section~\ref{sec-template} with the \nh\ values available in Table~1 of
\citet{ich12} and a photon index of 1.94 for X-ray type-1 AGNs (with log
\nh\ $<$ 22) or 1.84 for X-ray type-2 AGNs (with log \nh\ $\geq$22).
For the rest, we utilize the hardness ratio between the hard and soft
band count rates to estimate the absorption or photon index. In this
case, we assume a cut-off power law model in the form of $E^{-\Gamma}
{\rm exp} (E/E_{\rm c})$ where $E_{\rm c}=$300 keV plus its
reflection component from cold matter calculated with the {\bf pexrav} code
\citep{mag95} for the intrinsic spectrum.  In the pexrav model, the solar
abundances, an inclination angle of $\theta_{\rm inc} = 60^\circ$, and a
reflection strength of $R_{\rm tot}\equiv\Omega/2\pi=1.0$ are adopted.
If the hardness ratio is found to be larger than that expected from
$\Gamma=1.9$, then we determine the intrinsic absorption \nh\ by
assuming $\Gamma=1.9$. Otherwise, we calculate a corresponding photon
index by assuming no absorption. The base spectrum with $\Gamma=1.9$,
$\theta_{\rm inc} = 60^\circ$, and $R_{\rm tot}=1.0$ is the same as
adopted in U03, which roughly corresponds to the averaged one of type-1
and type-2 AGNs in the template spectra
(Section~\ref{sec-template}). Note that with this procedure we can only
obtain absorptions of log \nh\ $ \leq24$. This means we ignore the
possibility that an object is a CTK AGN, whose spectrum is
quite complex and is not necessarily harder than CTN AGNs in
the energy band below 10 keV due to the relatively large contribution
from other softer components than the transmitted one (see
Section~\ref{sec-template}). Finally, for the case of the \rosat\
surveys where such hardness ratio information is not available, we
simply assume $\Gamma=1.9$ and no absorption in the above continuum.

\begin{figure}
\epsscale{1.0}
\begin{center}
\includegraphics[angle=0,scale=0.45]{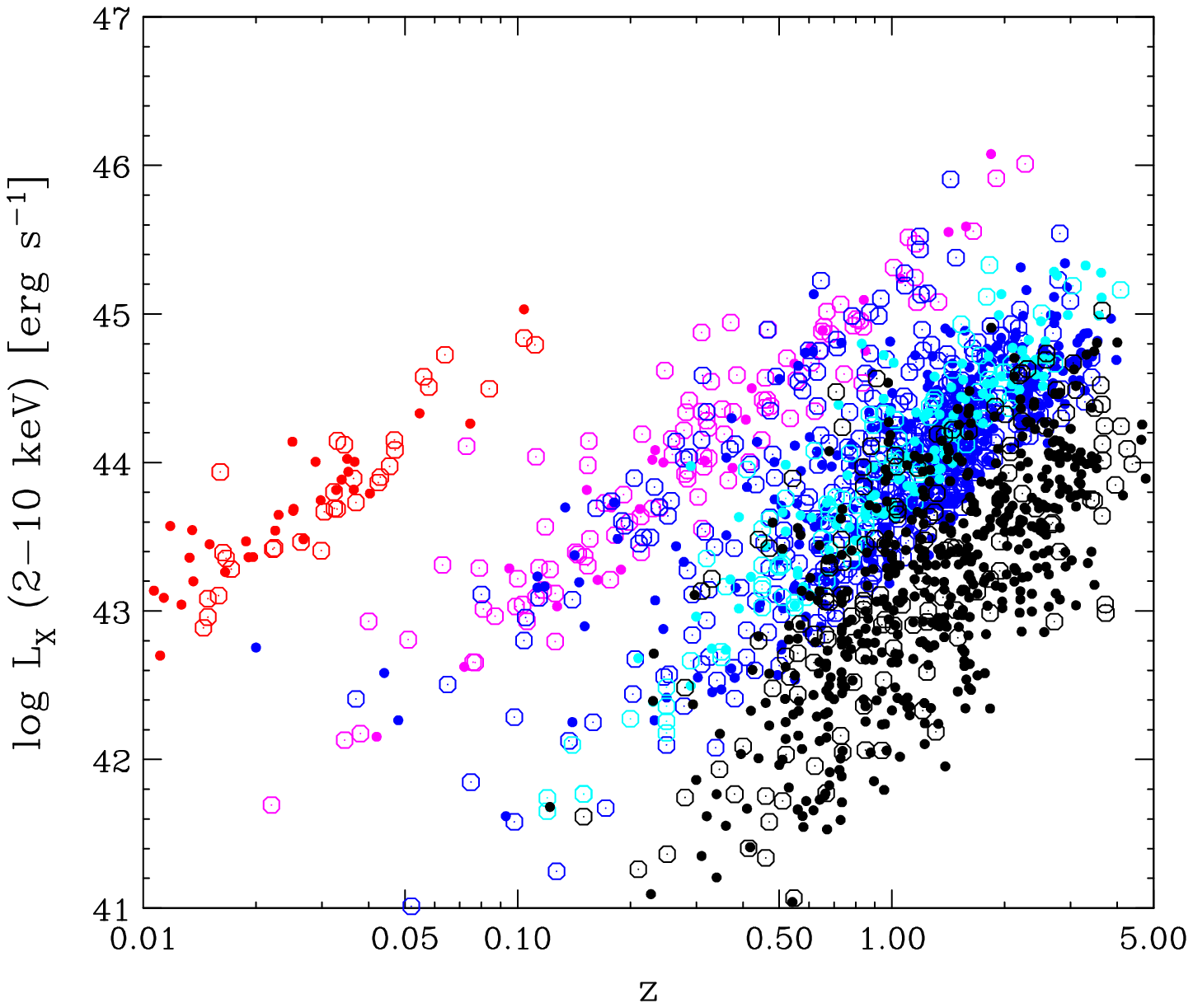}
\includegraphics[angle=0,scale=0.45]{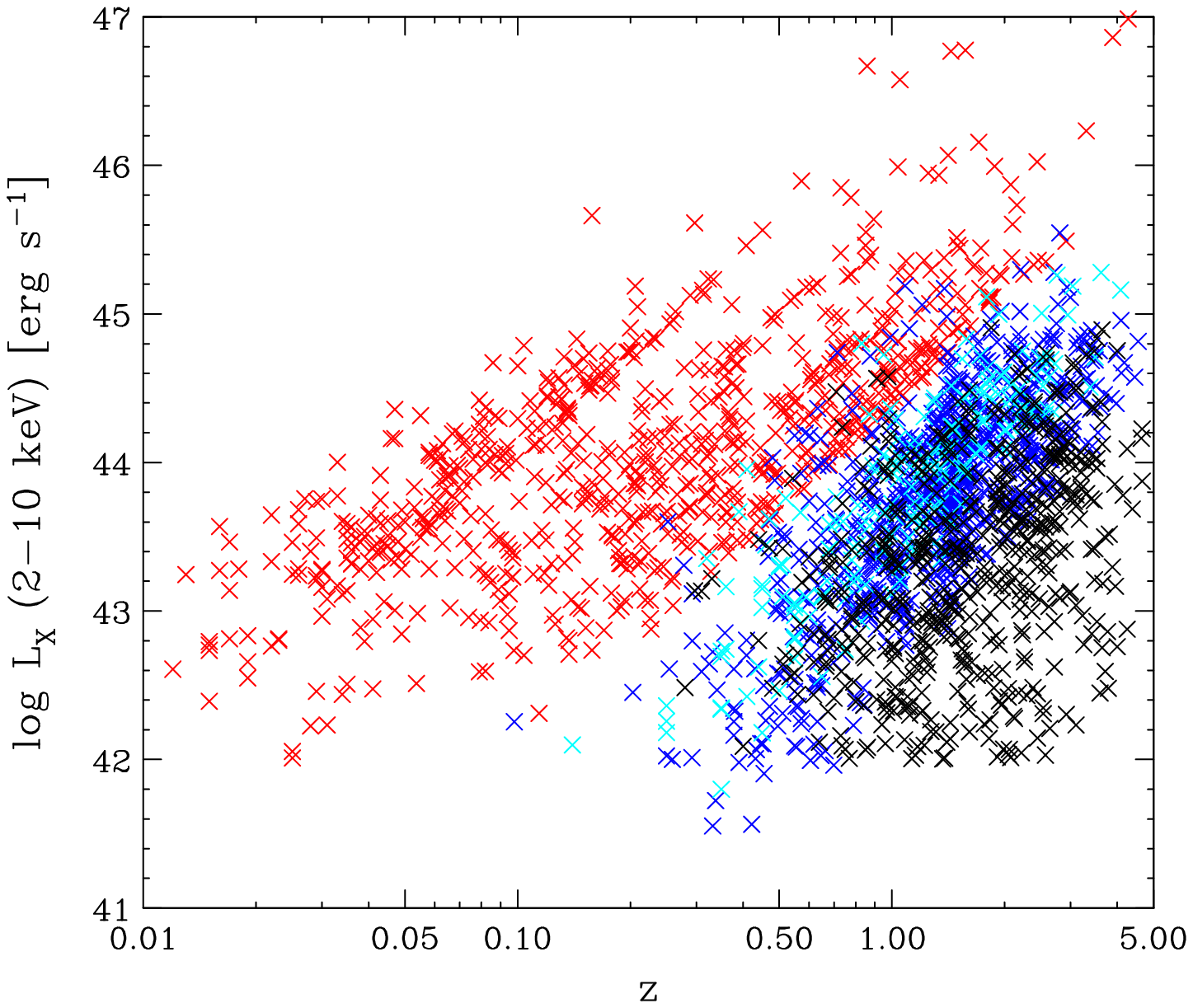}
\end{center}
\caption{
{\it Left}: redshift-luminosity plot of our sample detected in the hard ($> $2
 keV) band (red: \swift/BAT sample, magenta: \asca\ sample, blue: \xmm\
 sample, cyan: CLASX and CLANS sample, black: CDFN and CDFS sample). The
 open and filled circles represent X-ray type-1 AGNs (with log \nh\ $<$
 22) and X-ray type-2 AGNs (with log \nh\ $\geq$22), respectively. 
{\it Right}: redshift-luminosity plot of our sample detected in the soft (0.5--2
 keV) band (red: \rosat\ sample, blue: \xmm\ sample, cyan: CLANS sample,
 black: CDFN and CDFS sample).
}
\label{fig-plot-zl}
\end{figure}

Figure~\ref{fig-plot-zl} displays the \lx\ versus redshift
distribution for the hard band (left) and soft band (right) samples
after the above count-rate selection is applied. The different colors
correspond to different surveys. Here, the \maxi\ sample is not included
to avoid the overlap with the \swift /BAT sample. In
Figure~\ref{fig-plot-zl} (left), AGNs that are found to be absorbed with
\nh\ $>10^{22}$ cm$^{-2}$ (X-ray type-2 AGNs) are marked by filled
circles, while less absorbed ones (X-ray type-1 AGNs) are by open
circles. We do not distinguish the two classes for the soft-band sample
in Figure~\ref{fig-plot-zl} (right), however, which includes many \rosat\
sources.  Note that in our main analysis in Section~\ref{sec-lf}, we
perform ML fitting directly to the list of the count rate (not
luminosity) and redshift by consider a luminosity-dependent reflection
component and different photon index for type-1 and type-2 AGNs. The
contribution of CTK AGNs is also taken into account
there. Thus, the values of \lx\ calculated here should be regarded as
tentative ones and they will be mainly used for plotting purposes.

\section{Absorption Distribution}
\label{sec-abs}

The absorption properties of AGNs are important to understand the
circumnuclear environments such as dusty ``tori'' surrounding the SMBH.
In this section, we quantitatively formulate the absorption function (or
\nh\ function) and its evolution that must be taken into account in the
analysis of the XLF presented in next sections. The result is finally
included in the population synthesis model of the XRB. We first
determine the absorption function in the local universe by an analysis
of the \nh\ distribution of the \swift/BAT 9-month sample.  Then, its
redshift evolution is examined by using the \swift/BAT, AMSS, and SXDS
hard-band samples.

\subsection{Formulation}
\label{sec-abs-form}

Following U03 and later works, we introduce the absorption function or
\nh\ function, $f(L_{\rm x}, z; N_{\rm H})$, the probability
distribution function of absorption column density in the X-ray spectrum
of an AGN at a given luminosity and a redshift, in units of $({\rm log}
N_{\rm H})^{-1}$. We assume that it has no dependence on the photon index.
By adopting the same definition of U03, the absorption function is normalized
to unity in the ``Compton-thin'' region of log \nh\ $\leq$ 24 so that
\begin{equation}
\int_{20}^{24} f (L_{\rm X}, z; N_{\rm H}) d{\rm log} N_{\rm H} = 1.
\end{equation}
The lower limit of log \nh\ = 20 is a dummy value 
introduced for convenience, and we assign log \nh\ = 20 for unabsorbed
AGNs with log \nh\ $<$ 20. Note that $f(L_{\rm x}, z; N_{\rm H})$ is
defined also for CTK AGNs with log \nh\ $>$ 24.
The reason why we normalize
the absorption function within 20 $\leq$ log \nh\ $\leq$ 24 is that the XLF
can be most accurately determined for CTN AGNs from the survey
data. As done in the previous population synthesis models, we represent
the number density of CTK AGNs in terms of the ratio to that
of CTN ones at the same \lx\ and $z$.

The same formulation as presented in \citet{ued11} is adopted for the
shape of the absorption function.  First we introduce the $\psi (L_{{\rm
X}}, z)$ parameter, which represents the fraction of absorbed
CTN AGNs (i.e., log \nh\ = 22-24) in total CTN AGNs
(i.e., log \nh\ = 20-24) as a function of \lx\ and $z$. It is expressed
by a linear function of log \lx\ within a range between $\psi_{\rm min}$
and $\psi_{\rm max}$;
\begin{equation}
\psi(L_{{\rm X}}, z) = {\rm min} [\psi_{\rm max}, \; {\rm max} [\psi_{43.75}(z) - \beta ({\rm log} L_{\rm X} - 43.75), \; \psi_{\rm min}] ], 
\end{equation}
where $\psi_{43.75}(z)$ gives the absorption fraction of
CTN AGNs with log \lx\ = 43.75\footnote{
Here we change the  reference luminosity from log \lx\ = 44 
adopted in U03 and \citet{ued11} to log \lx\ = 
43.75 to match with the formulation by \citet{has08}.
} located at $z$. 
In this work we adopt $\psi_{\rm min} = 0.2$, the fraction of absorbed
AGNs at highest luminosity range found in the \swift/BAT survey \citep{bur11},
and $\psi_{\rm max} = 0.84$, the upper limit from the assumption on the
form of the absorption function explained below.
On the basis of the results by \citet{has08}, 
we take into account the redshift dependence as 
\begin{equation}
\label{eq-a1}
\psi_{43.75}(z) = \left\{
\begin{array}{ll}
 \psi_{43.75}^0 (1+z)^{a1} & [z < 2.0]\\
 \psi_{43.75}^0 (1+2)^{a1} & [z \geq 2.0]\\
\end{array} \right.
\end{equation}
Here $\psi_{43.75}^0 \equiv \psi_{43.75}(z=0)$ is the local value, 
a free parameter to be determined from the analysis 
in the next subsection. In our paper, we adopt $\beta
= 0.24$, the best-fit value obtained by \citet{ued11}, which also agree
with those in various redshift ranges obtained by \citet{has08}.

We define the absorption function separately for two ranges of the $\psi
(L_{{\rm X}},z)$ value;
\begin{eqnarray}
\label{eq-former}
({\rm for}\;\; \psi (L_{\rm X}, z) <&& \frac{1+\epsilon}{3+\epsilon}) \nonumber \\
f(L_{{\rm X}}, z; N_{{\rm H}}) &&= \left\{
\begin{array}{ll}
 1-\frac{2+\epsilon}{1+\epsilon}\psi (L_{{\rm X}}) & [20 \leq {\rm log} N_{\rm H} < 21]\\
\frac{1}{1+\epsilon} \psi (L_{{\rm X}}) & [21 \leq {\rm log} N_{\rm H} < 22]\\
\frac{1}{1+\epsilon} \psi (L_{{\rm X}}) & [22 \leq {\rm log} N_{\rm H} < 23]\\
\frac{\epsilon}{1+\epsilon} \psi (L_{{\rm X}}) & [23 \leq {\rm log} N_{\rm H} < 24]\\
\frac{f_{\rm CTK}}{2} \psi (L_{\rm X}) & [24 \leq {\rm log} N_{\rm H} < 26]\\
\end{array} \right.
\end{eqnarray}
and
\begin{eqnarray}
\label{eq-latter}
({\rm for}\;\; \psi (L_{\rm X}, z) \geq && \frac{1+\epsilon}{3+\epsilon}) \nonumber \\
f(L_{{\rm X}}, z; N_{{\rm H}}) &&= \left\{ 
\begin{array}{ll}
\frac{2}{3}-\frac{3+2\epsilon}{3+3\epsilon}\psi (L_{\rm X}) & [20 \leq {\rm log} N_{\rm H} < 21]\\
\frac{1}{3}-\frac{\epsilon}{3+3\epsilon}\psi (L_{\rm X}) & [21 \leq {\rm log} N_{\rm H} < 22]\\
\frac{1}{1+\epsilon} \psi (L_{\rm X}) & [22 \leq {\rm log} N_{\rm H} < 23]\\
\frac{\epsilon}{1+\epsilon} \psi (L_{\rm X}) & [23 \leq {\rm log} N_{\rm H} < 24]\\
\frac{f_{\rm CTK}}{2} \psi (L_{\rm X}) & [24 \leq {\rm log} N_{\rm H} < 26]\\
\end{array} \right.
\end{eqnarray}
The absorption function is flat above log \nh\ = 21 in the former case
(Equation~\ref{eq-former}), while the value in log \nh\ = 21--22 is set
as the mean of those at log \nh\ = 20--21 and log \nh\ = 22--23 in the
latter (Equation~\ref{eq-latter}).  The fraction of CTK AGNs
to the absorbed CTN AGNs (with log \nh\ = 22--24) is given by
the \fct\ parameter, and its absorption function is assumed to be flat over
the range of log \nh\ = 24--26. The maximum absorption fraction
corresponds to the case of $f(N_{{\rm H}})$ = 0 at log \nh\ = 20--21 and
thus $\psi_{max} = \frac{1+\epsilon}{3+\epsilon}$. The $\epsilon$
parameter represents the ratio of the absorption function in log \nh\ =
23--24 to that in log \nh\ = 22--23, which is fixed at $\epsilon = 1.7$
in our paper. This is the same value adopted by U03 and \citet{gil07},
based on the \nh\ distribution of [O-III] selected AGNs in the local
universe \citep{ris99}. Although \citet{ued11} adopt $\epsilon = 1.3$ on
the basis of the original \nh\ distribution of \swift/BAT 9-month sample
reported in \citet{tue08}, we find that $\epsilon = 1.7$ better fit the
revised \nh\ distribution that utilizes new \nh\ measurements as
reported in \citet{ich12}. Recent work by \citet{vas13} based on a
deeper survey of \swift/BAT also favors a larger value than that of 
\citet{tue08}.

\subsection{Absorption Function in the Local Universe}
\label{sec-abs-local}

\begin{figure}
\epsscale{1.0}
\begin{center}
\includegraphics[angle=0,scale=0.6]{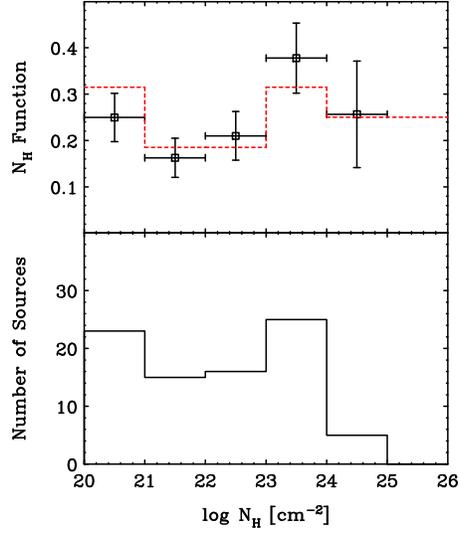}
\end{center}
\caption{
Observed histogram of $N_{\rm H}$ in the \swift/BAT sample (lower) and
estimated $N_{\rm H}$ function corrected for detection biases
(upper). The red lines represent an example of $N_{\rm H}$ function we
adopt.
}
\label{fig-bat-nh}
\end{figure}

To determine the absorption function in the local universe, we use the
revised \nh\ distribution of the \swift /BAT 9 month sample based on
Table~1 of \citet{ich12}. In this analysis, we limit the luminosity
range to log \lx\ = 42--46, which leaves 84 AGNs\footnote{NGC 4395, NGC
4051, and NGC 4102 are excluded}. The lower panel of
Figure~\ref{fig-bat-nh} displays the observed histogram of the \nh\
distribution. Even if this sample is selected by hard X-rays above 10
keV having strong penetrating power, there still remains detection
biases against obscuration at large column densities due to the
suppression of the hard X-ray flux (see Section~\ref{sec-template}),
which becomes particularly significant in CTK AGNs. The presence of this
bias in hard X-ray surveys is discussed by 
\citet[e.g.,][]{mal09,bur11}.
Importantly, the small observed fraction of CTK AGNs
($5/84\approx$6\% in our sample) does not mean that they are a minor
population.

To derive the absorption function by correcting for these
effects, we perform an ML fit of the absorption function with
the same approach as described in Section 4.1 of U03. In this analysis,
the likelihood estimator $L'$ to be minimized through fitting is defined as
\begin{equation}
\label{eq-ml1}
L' = -2 \sum_{i} {\rm ln}
\frac{f(L_{{\rm X}i}, z_i; N_{{\rm H}i}) \sum_{j} A_{\rm j} (N_{{\rm H}i},\Gamma
_i,L_{{\rm X}i},z_i)}
{\int f(L_{{\rm X}i}, z_i; N_{{\rm H}}) \sum_{j} A_{\rm j} (N_{{\rm H}},\Gamma_i
,L_{{\rm X}i},z_i) {\rm d log} N_{\rm H}}.
\end{equation}
The symbol $i$ represents each object, and 
$j$ represents each survey (only one in this case).
Here $A_{\rm j}$ gives the survey area for a count rate expected from a
source with the absorption \nh, photon index $\Gamma$, intrinsic
luminosity \lx , and redshift $z$. The count rate is calculated through
the detector response and luminosity distance by assuming the template
spectra of AGN presented in Section~\ref{sec-template}. 
The minimization is carried out on the MINUIT software package. In the
ML fit, the 1$\sigma$ statistical error of a free parameter is derived
as a deviation from the best-fit value when the $L'$-value is increased
by 1.0 from its minimum value.

Since the number of sources in this sample is limited, we only make
$\psi_{43.75}^0$ as a free parameter, which represents the fraction of
absorbed CTN AGNs in all CTN AGNs at log \lx\ =
43.75. The other parameters on the absorption function are fixed as
$\epsilon=1.7$, $\beta=0.24$, $a1=0.48$ (the best fit value from the
whole sample, see Section~\ref{sec-lf}), and \fct\ = 1.0. We thus obtain
$\psi_{43.75}^0 = 0.43\pm0.03$, which will be adopted in the following
analysis. This is in perfect agreement with the result by \citet{ued11}
based on the \maxi\ sample, who obtain the absorption fraction at log
\lx\ = 44 of $\psi_{44}=0.37\pm0.05$, corresponding to
$\psi_{43.75}^0=0.43\pm0.05$ with $\beta=0.24$.

The best-fit shape of the absorption function calculated at \lx\ = 43.5,
the averaged value from the \swift/BAT sample, is plotted in the upper
panel of Figure~\ref{fig-bat-nh} by lines (red). The data points in the
same panel give a bias-corrected \nh\ distribution calculated in the
following way. First we make a normalized detection efficiency curve as
a function of \nh\ that is proportional to $A_{\rm j} (N_{{\rm
H}},\Gamma_i ,L_{{\rm X}i}, z_i)$ for each object. Then, the observed
histogram of \nh\ is divided by the sum of the detection efficiency
curves, and is finally normalized to unity within the range between log
\nh\ = 20--24. A good agreement with the best-fit model is noticed,
justifying the choice of the fixed parameters of $\epsilon=1.7$ and
\fct$=1.0$. All parameters of the absorption function are summarized in
Table~\ref{tab-abs}.

\subsection{Evolution of Absorbed-AGN Fraction}
\label{sec-abs-evolv}

The dependence of the 
absorbed-AGN fraction on redshift are studied in
depth by \citet{has08}, following previous works by \citet{laf05},
\citet{bal06}, and \citet{tre06}, who all report a positive evolution in
the sense that more obscured AGNs toward higher redshifts. Here we
pursue this issue by utilizing our new sample from the SXDS. It is not
straightforward, however, to estimate a true intrinsic absorption
fraction that is subject to detection biases as well as statistical
errors in the measured \nh\ values, which become very large for faint
AGNs detected in deep survey data, unlike the case of the local AGN
sample analyzed in the previous subsection. To take into account the
effects of statistical fluctuation in the hardness ratio and hence that
in \nh\ in the ML fit, we introduce the ``\nh\ response matrix
function'' that gives the probability of finding an observed value of
\nh\ from its true value for each object as described in Section 4.1 of
U03. A similar approach is adopted in \citet{hir12} as well to estimate
the absorption fraction at $z>3$ from the SXDS sample.

In this study, we utilize the samples of \swift/BAT, AMSS, and SXDS. To
correctly calculate the \nh\ response matrix function, we need to have a
hardness ratio and its statistical error. Although the absorptions are
measured on the basis of spectral fitting in some samples, the resultant
parameters are inevitably subject to non-negligible statistical errors
much larger than those in the \swift/BAT sample. Unlike the simple
method to estimate \nh\ only from the hardness ratio, it is difficult to
quantify the effects of statistical fluctuation in individual
spectral fit. We find that many faint \chandra\ sources have a very few
number of photons in a given energy band. This makes the correction
method more complicated because the statistics cannot be approximated by
a Gaussian distribution. Thus, we decide to use only the AMSS sample and
the SXDS hard-band sample, which have well-defined the hardness ratio in
two energy bands with enough photon statistics, in addition to the
\swift /BAT sample. In this stage, for simplicity, we adopt the
absorption, photon index, and luminosity calculated from the observed
count rate and hardness ratio according to the procedure described in
Section~\ref{sec-sample-luminosity}. The \maxi\ sample is not included to
avoid duplication with the \swift /BAT AGNs. The parent sample thus
consists of 751 AGNs with high identification completeness (99\%).

\begin{figure}
\epsscale{1.0}
\begin{center}
\includegraphics[angle=0,scale=0.8]{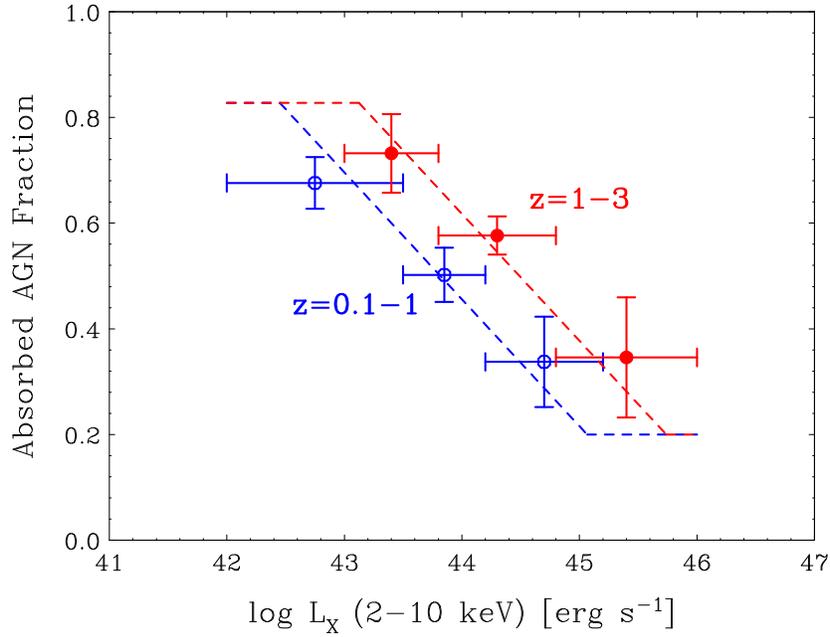}
\end{center}
\caption{
Absorbed fraction of AGNs plotted against luminosity at $z=0.1-1$
(blue) and $z=1-3$ (red) as determined from the \swift/BAT, AMSS, and
SXDS (hard band) samples. The dashed lines represent the best-fit
models calculated at the mean redshifts for the $z=0.1-1$ and $z=1-3$
samples.
}
\label{fig-absfrac}
\end{figure}

Using the list of \nh, $\Gamma$, \lx, and $z$ in the combined sample,
we perform ML fitting of the absorption function on the basis of
Equation~(\ref{eq-ml1}). Since the main purpose is to investigate the
luminosity and redshift dependence of absorption, we set $\beta=0$ and
$a1=0.0$ (i.e., constant) and make the $\psi_{43.75}^0$ parameter free
by limiting to narrow ranges of \lx\ and $z$ (shown in
Figure~\ref{fig-absfrac}). Only the region of log \nh\ = 20--24 is
used in the ML fit, and thus the eight identified CTK AGNs in the
\swift/BAT sample are excluded.  Since \nh\ is simply converted from
the hardness ratio on the basis of a single absorbed continuum model
for the AMSS and SXDS samples, it is assumed that there are no CTK
AGNs in them as mentioned in Section~\ref{sec-sample-luminosity}. This
simplification would be justified because the fraction of CTK AGNs is
expected to be small at the flux limits of the AMSS and the SXDS
according to our population synthesis model (Section~\ref{sec-model}).

The results of the $\psi$ parameter are plotted in
Figure~\ref{fig-absfrac}. As noticed, we confirm the strong
anti-correlation between the absorption fraction and luminosity. In
addition, a redshift dependence is clearly seen that the absorption
fraction becomes larger at higher redshift by keeping the similar
anti-correlation with luminosity. This is fully consistent with
\citet{has08}. These results support that the choice of $\beta=0.24$
that is constant against $z$ is appropriate. The $a1$ parameter
representing the evolution with $z$ will be determined from the main
analysis presented in Section~\ref{sec-lf}, where the whole sample is
utilized. The obtained best-fit curves calculated at the mean redshifts
for the $z=0.1-1$ and $z=1-3$ samples are overplotted by dashed lines
in the Figure.

In the above analysis, we have assumed a single absorbed spectrum to
estimate the column density for the AMSS and SXDS samples. In reality,
it is known that the photon flux of an absorbed AGN re-increase towards
the lowest energy, because the absorber does not fully cover the nucleus, or
because the absorber is ionized, or because other soft components of
different origins are present. According to the systematic spectral
study of local AGNs detected in the Swift/BAT survey
\citep{win09a,ich12}, in most cases the X-ray spectra of absorbed AGNs
can be well represented by the partial covering model, where a
fraction of $f_{\rm c}$ of the total continuum is subject to
absorption. A majority of absorbed AGNs have covering fractions of
$f_{\rm c} \simgt 0.98$, which can be explained by the leakage of the
scattered component from outside the torus (see
Section~\ref{sec-template}), while $\sim$15\% of the total AGNs show
smaller covering fractions, with a medium of $f_{\rm c} \approx$0.95,
most probably due to the complex nature of the absorber or the presence
of other components. To evaluate the possible systematic errors, we
statistically take into account the distribution of these complex
spectra in calculating the ``\nh\ response matrix function'' and
repeat the analysis. We confirm that systematic errors in the
absorbed-AGN fractions are much smaller than the statistical ones at any
luminosity and redshift regions, and hence our conclusions are robust.

\section{Template Model Spectra of AGNs}
\label{sec-template}

In our main analysis described in Section~\ref{sec-lf}, we
simultaneously treat the results from the surveys performed in different
energy bands spanning from 0.5 keV to 195 keV. Thus, it is critical to
make consistent analysis on the basis of adequate assumptions on the
broad band X-ray spectra of AGNs. In this section, we describe
``template spectra'' of AGNs adopted in this paper. They are based on
extensive studies of broad-band spectra of nearby AGNs in the
literature.
As for the intrinsic continuum spectrum, we adopt a cutoff power law
with $E_{\rm c}=$300 keV, an averaged value of bright Seyfert galaxies
in the local universe reported by \citet{dad08}. To achieve a physically
consistent picture, two reflection components from optically thick
matter are considered, one from the accretion disk and the other from
the torus. The solar abundances are assumed in both models. 
We also add
an unabsorbed scattered component from the surrounding gas located
outside the torus. For simplicity, the
spectrum is assumed to be a pure Compton scattered one of the continuum
without any emission lines, by taking into account the Klein-Nishina
cross-section. Its intensity is proportional to the opening solid angle
of the torus, which is normalized to be $1\%$ when the half sky
($\Omega=2\pi$) is covered by it \citep{egu11}.

We model the reflection component from the accretion disk with the
pexrav code \citep{mag95} by assuming that the disk is not ionized. The
parameters are the inclination angle of the line-of-sight with respect
to the normal axis of the disk, $\theta_{\rm inc}$, and the reflection
strength \rdisk, for which we adopt \rdisk\ = $0.5$ as a default
value. This value is chosen to roughly explain an averaged reflection
strength found in local Seyfert galaxies, $R_{\rm tot} \sim 1$
\citep[e.g.,][]{zdz95,dad08}, after the contribution from the
torus-reflection component is added. Its possible dependences on \lx\
and on $z$ are ignored, for which no consensus has been established
yet. The inclination $\theta_{\rm inc}$ is determined separately for
type-1 and type-2 AGNs as a function of \lx\ and $z$ according to the
description below.

To reproduce realistic spectra from AGN tori even in CTK cases, we adopt
the numerical model calculated by \citet{bri11a} based on Monte Carlo
simulations, where the absorbing torus is approximated by a uniform
sphere with bi-polar conical openings, rather than donut-shaped. The
model parameters are the photon index $\Gamma$, the column density
$N_{\rm H}^{\rm torus}$, the opening angle $\theta_{\rm oa}$ (or
$\theta_{\rm tor}$), and the inclination angle $\theta_{\rm i}$ (see
Figure~1 of \citealt{bri11a}). The spectrum consists of the transmitted
emission (unabsorbed when $\theta_{\rm i} < \theta_{\rm oa}$ and vice
versa), its reflected spectra from the torus including fluorescence
lines from abundant metals (such as iron, nickel, etc). Self-absorption
of the reprocessed emission by the torus itself is properly taken into
account for a given geometry.

Throughout our work, we assume the ``luminosity and redshift dependent unified
scheme'' of AGNs; the absorbed fraction of AGNs is simply
determined by the covering factor of the torus whose geometry depends on
both \lx\ and $z$. The torus opening angle can be then related to the
fraction of total absorbed AGNs (including CTK ones) among all
AGNs, $\psi' (L_{{\rm X}}, z)$, as
\begin{eqnarray}
\cos (\theta_{\rm oa}) 
 &=&\psi' (L_{\rm X}, z) \nonumber \\
 &=&\frac{(1+f_{\rm CTK}) \psi(L_{\rm X}, z)}{1+f_{\rm CTK} \psi(L_{\rm X}, z)}.
\end{eqnarray}
We recall that the absorbed AGNs are defined as those with log \nh\ $\geq
22$. This is based on the idea that absorptions smaller than log \nh\ =
22 are not originated from the torus, but most probably from
interstellar medium in the host galaxy (see e.g.,
\citealt{fuk11}). Hence, we multiply the absorption to all the spectral
components for type-1 AGNs (i.e., with log \nh\ $< 22$), while we ignore
other origins than the tori for the absorptions in type-2 AGNs. In this
formulation, we assume that CTK AGNs are just an extension of
CTN ones with the same \lx\ and $z$ toward larger column
densities. Using a type-1 AGN sample in the local universe, \citet{ric13}
find that the luminosity-dependent unified scheme can explain the
so-called X-ray Baldwin effect \citep{iwa93}, the anti-correlation
between the equivalent width of iron-K line and X-ray luminosity,
although the definition of the absorbed fraction is slightly different from
ours; when they refer to the result by \citet{has08}, AGNs with log \nh\
$\geq$ 21.5 are counted as absorbed ones by neglecting CTK AGNs.

Once $\theta_{\rm oa}$ is known, we can calculate a solid-angle averaged
inclination angle separately for type 1 and type 2 AGNs; 
\begin{equation}
\theta_{\rm inc} = \left\{
\begin{array}{ll}
\arccos(1-(1-\cos\theta_{\rm oa})/2)) & {\rm [for \;\;type-1]}\\
\arccos(\cos\theta_{\rm oa}/2)) & {\rm [for \;\;type-2]}\\
\end{array} \right.
\end{equation}
According to the torus geometry in the \citet{bri11a} model, the torus column
density $N_{\rm H}^{\rm torus}$ is taken to be the same as the
line-of-sight column density \nh\ in type-2 AGNs. 
For type-1 AGNs we adopt log $N_{\rm H}^{\rm torus}$ = 24 
as an average value to reflect our
assumption that the number of CTK AGNs is the same as
CTN absorbed ones (i.e., \fct\ = 1.0).

\begin{figure}
\epsscale{1.0}
\begin{center}
\includegraphics[angle=0,scale=0.6]{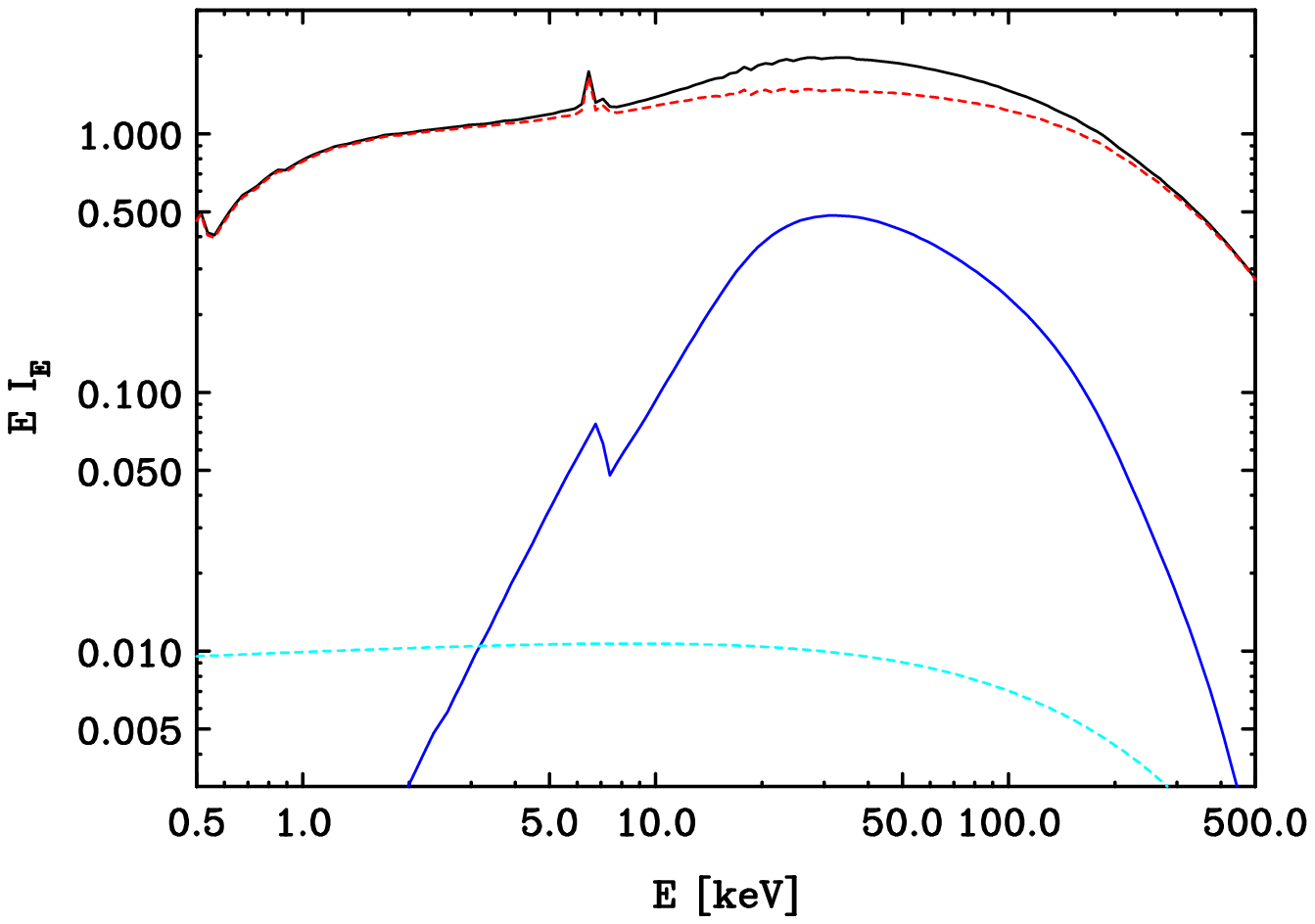}
\includegraphics[angle=0,scale=0.6]{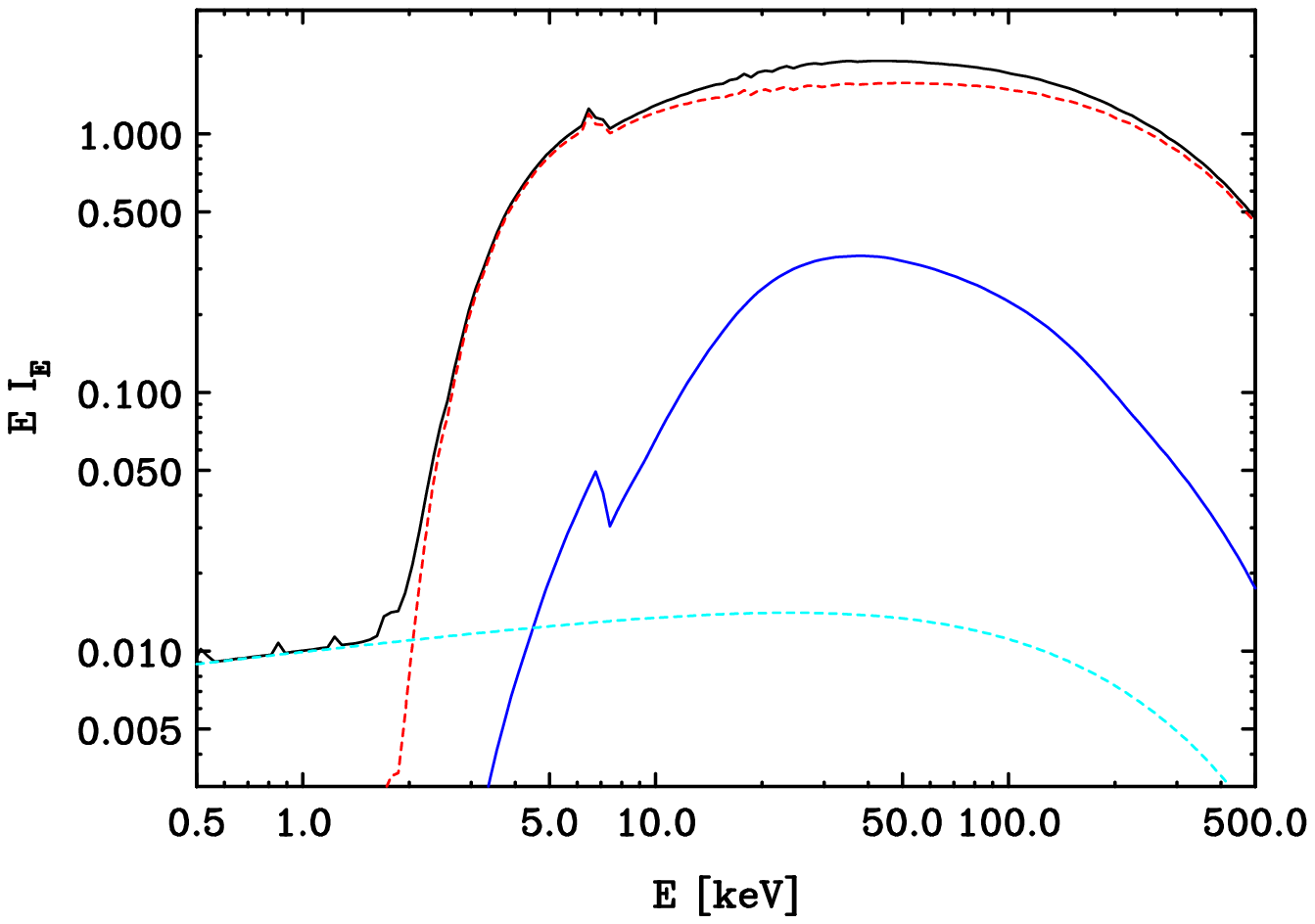}
\end{center}
\caption{
Examples of the template broad band X-ray spectra 
(in $E I(E)$, where $E$ is energy and $I(E)$ is the energy flux)
for AGNs with log
$N_{\rm H}$ = 21 and $\Gamma=1.94$ (left) and log $N_{\rm H}$ = 23 and
 $\Gamma=1.84$ (right). Units are arbitrary.
The black curve denotes the total spectrum, the red curve the
transmitted continuum plus its reflection component from the torus, the
blue curve the reflection component from the accretion disk, and the
cyan curve the scattered component.
}
\label{fig-agnspec-lnh}
\end{figure}

\begin{figure}
\epsscale{1.0}
\begin{center}
\includegraphics[angle=0,scale=0.8]{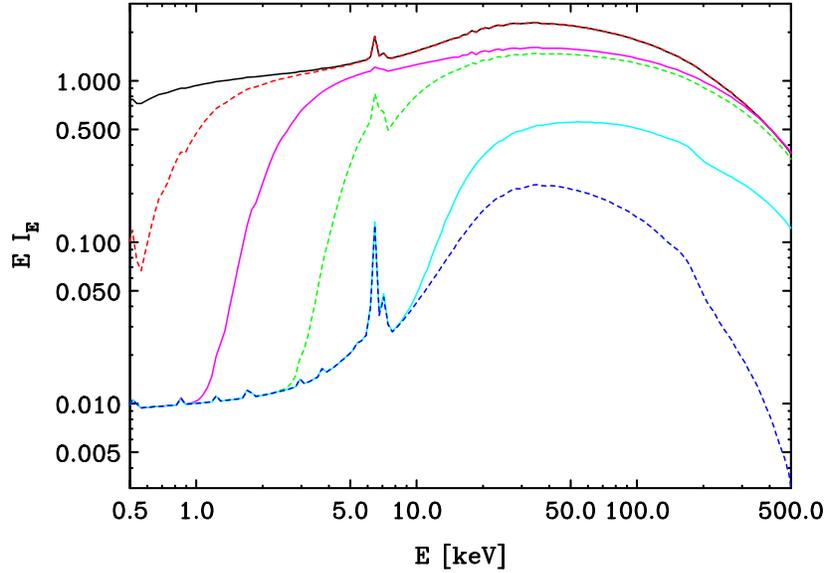}
\end{center}
\caption{
Model spectra of AGNs with
different absorptions (from top to bottom, log $N_{\rm H}$ = 20.5, 21.5,
22.5, 23.5, 24.5, 25.5). 
Units are arbitrary in $E I(E)$.
A same photon index ($\Gamma=1.9$) and a
normalization are assumed for the intrinsic cutoff power-law continuum
in all the spectra.
} \label{fig-agnspec-all}
\end{figure}

Figure~\ref{fig-agnspec-lnh} (left) and (right) plot the template model
spectra in the 0.5--500 keV band for an AGN with log \nh\ = 21
and 23, respectively. Here
we adopt $\Gamma=1.94$ for the former and $\Gamma=1.84$ for the latter
(see Section~\ref{sec-idx}).
The
disk-reflection and scattered components are separately shown.
Figure~\ref{fig-agnspec-all} shows those for log \nh\ = 20.5, 21.5,
22.5, 23.5, 24.5 and 25.5. 
For easy comparison, we adopt $\Gamma=1.9$ for all the spectra in this
Figure. The suppression of the hard X-ray flux in heavily CTK AGNs is
noticed. In both Figures, $\theta_{\rm oa}=60^\circ$ is assumed.

\section{Photon Index Distribution}
\label{sec-idx}

To estimate the intrinsic photon index ($\Gamma$) distribution of AGNs
in the local universe, we analyze the \swift/BAT spectra in the 14--195
keV band averaged for 58 months, which are available for all AGNs in the
\swift/BAT 9 month catalog. This energy band is suitable to estimate the
$\Gamma$ value of an individual CTN AGN as a first order
approximation without invoking complex, source dependent spectral
analysis covering the full 0.5--195 keV band, which is beyond the scope
of this paper. We systematically perform spectral fitting to the 14--195
keV spectra of the \swift/BAT sample with the ``template spectra''. The
parameters of the reflection and scattered components are fixed
according to our best estimate of the $\psi(L_{{\rm X}}, z)$ values
(Section~\ref{sec-abs}), and only the photon index and overall
normalization are free parameters.

\begin{figure}
\epsscale{1.0}
\begin{center}
\includegraphics[angle=0,scale=0.5]{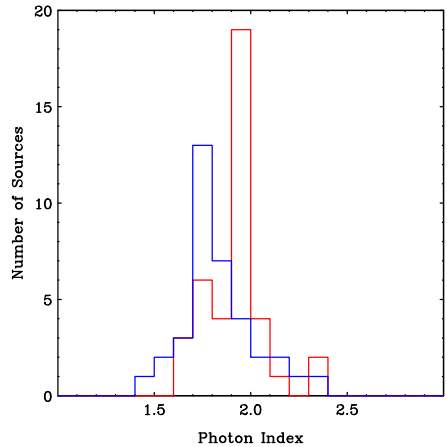}
\end{center}
\caption{
Distribution of photon index for type-1 (red) and type-2 (blue)
AGNs as determined from the \swift/BAT survey.
}
\label{fig-bat-idx}
\end{figure}

Figure~\ref{fig-bat-idx} displays the histograms of the best-fit photon
index, plotted separately for type-1 (red) and type-2 AGNs (blue). Here
we exclude the results for four objects out of 79 CTN AGNs in
the original sample for which the goodness of the fit is found to be
poor. Since it composes only a minor fraction, we regard the rest of the
sample as a representative one. As noticed from the Figure, we confirm
the trend reported previously \citep[e.g.,][]{zdz00,tue08,bur11} that
the average slope of type-1 AGNs is larger than that of type-2 AGNs,
even after properly taking into account the reflection components in the
spectra. 
This can also explain the suggestion by
\citet{ued11} that the averaged photon index in the 4--195 keV band
is larger in a high luminosity range where type-1 AGNs dominate their
sample.
From these histograms, we obtain the average and standard
deviation (with their $1\sigma$ errors) of
$\overline{\Gamma_1}=1.94\pm0.03$ and $\Delta\Gamma_1=0.09\pm0.05$ for
type-1 AGNs, and $\overline{\Gamma_2}=1.84\pm0.04$ and
$\Delta\Gamma_2=0.15\pm0.06$ for type-2 AGNs, as summarized in
Table~\ref{tab-idx}. Here $\Delta\Gamma_{1,2}$ presents the intrinsic
scatter of the photon index after subtracting that caused by the
statistical errors in the spectral fits.

To quantitatively incorporate the intrinsic scatter of photon index, we
introduce the ``photon index function'', $g (N_{\rm H}, L_{\rm X}, z;
\Gamma)$, similarly to the absorption function, which gives the probability of
finding $\Gamma$ per unit $\Gamma$ space from an AGN with \nh\ and \lx\ 
located at $z$. Specifically, we model it by a normalized Gaussian function as 
\begin{equation}
g (L_{\rm X}, z, N_{\rm H}; \Gamma) = 
\frac{1}{\sqrt{2\pi} \Delta \Gamma_{1,2}} 
\exp [-\frac{(\Gamma-\overline{\Gamma_{1,2}})^2}{2 (\Delta\Gamma_{1,2})^2}]
\end{equation}
where $\overline{\Gamma_1}=1.94$ and $\Delta\Gamma_1=0.09$ 
for log \nh\ $< 22$, 
and $\overline{\Gamma_2}=1.84$ and $\Delta\Gamma_2=0.15$ for log \nh\ $\geq 22$.
In our paper, we ignore the \lx\ and $z$ dependences, which have not
been established yet. The effects by changing the $\overline{\Gamma_{1,2}}$ or
$\Delta\Gamma_{1,2}$ parameters onto our final results will be examined in
Section~\ref{sec-model-dependence}.

\section{Luminosity Function}
\label{sec-lf}

\subsection{Analysis Method}
\label{sec-lf-method}

In this section we describe the main part of analysis where the XLF is
determined by performing an ML fit to our sample at $z=0.002-5.0$.
We define the
XLF of CTN AGNs so that 
\begin{equation}
\frac{d \Phi^{\rm CTN}_{\rm X} (L_{\rm X}, z)}{d{\rm log} L_{\rm X}}
\end{equation} 
represents the number density per unit co-moving volume per log \lx\
as a function of \lx\ and $z$ in units of Mpc $^{-3}$ dex$^{-1}$.
The ML fit to unbinned data is a standard method to determine the model
parameters when the sample size is limited. Unlike in many previous
works, however, we do not use the list of \lx\ and $z$ as the input
data. This is because, as already mentioned in
Section~\ref{sec-sample-luminosity}, we cannot avoid uncertainties in
the determination of \lx\ for which an accurate measurement of
absorption is required, while spectral information is limited due to
poor statistics in faint sources. In particular when we expand the \nh\
region of interest above log \nh\ $>24$ to include CTK AGNs,
the hardness ratio does not uniquely correspond to a single \nh\ value
because of the complexity of the spectrum, producing additional
systematic errors to determine \lx.

Thus, we develop a new analysis method to utilize the list of the count
rate \cx\ and $z$ obtained in each survey, the most basic observational
quantities without any corrections. This approach is based
on a ``forward method'', where comparison between the prediction and
observation is made at the final stage. Namely, we search for a set of
parameters of the XLF and absorption function that best reproduce the
count-rate versus $z$ distributions of all surveys used in the
analysis. The merit is that we can properly take into account any
complex X-ray spectra in the analysis including CTK
AGNs. Another merit is that we can treat surveys of the same field
(including all sky surveys) in different energy bands as independent
data of one another, enabling us to utilize all samples without worrying
about the overlap of objects detected in multiple surveys.

Specifically, we define a likelihood estimator as 
\begin{eqnarray}
\label{eq-ml2}
L = -2 &&\times \nonumber \\
\sum_{i} {\rm ln} && \frac{\int \int \int 
N_j(N_{\rm H}, \Gamma, \hat{L_{{\rm X}}}, z)
 d{\rm log} N_{\rm H} d \Gamma d z
}
{\sum_{j} \int \int \int \int N_j( N_{\rm H}, \Gamma, L_{{\rm X}}, z) d{\rm log} N_{\rm H}
d \Gamma d{\rm log} L_{\rm X} d z},
\end{eqnarray}
where 
$$
\hat{L_{{\rm X}}}=4\pi d_{\rm L}^2(z) a_j(N_{\rm H},\Gamma,z) C_{{\rm X}i}
$$
in the integrand of the numerator.
The suffixes $i$ and $j$ represent each independent detection and
survey, respectively. Here $d_{\rm L}(z)$ is the luminosity distance,
$a_j(N_{\rm H},\Gamma,z)$ is the conversion factor
from the count rate in the $j$-th survey into the de-absorbed flux 
in the observer's frame $2/(1+z)- 10/(1+z)$ keV band, and $C_{{\rm X}i}$ is the
observed count rate of the $i$-th detection. The term $N_j(N_{\rm H},
\Gamma, L_{{\rm X}}, z)$ represents the expected number from the $j$-th
survey
calculated as
\begin{eqnarray}
N_j(N_{\rm H}, \Gamma, L_{{\rm X}}, z) && =  f(L_{\rm X}, z; N_{\rm H})  g(L_{\rm X}, z, N_{\rm H}; \Gamma) \times \nonumber \\
  \frac{d \Phi^{\rm CTN}_{\rm X} (L_{\rm X}, z)}{d{\rm log} L_{\rm X}} &&
d_{\rm A}(z)^2 (1+z)^3 c \frac{d\tau}{dz}(z) A_j (N_{\rm H},\Gamma,L_{\rm X},z) 
\label{eq-nj}
\end{eqnarray}
where $d_{\rm A}(z)$ is the angular distance, $d\tau/d z$ the
differential look back time, and $A_j (N_{\rm H},\Gamma,L_{\rm X},z)$ is
the area of the $j$-th survey expected from an AGN with \nh , $\Gamma$,
\lx , and $z$. 

In principle, it is possible to simultaneously constrain the XLF,
absorption function $f(L_{\rm X}, z; N_{\rm H})$, and photon index
function $g(L_{\rm X}, z, N_{\rm H}; \Gamma)$ through an ML fit. In
practice, however, to avoid strong parameter coupling we only make the
index of the evolution factor $a1$ in the absorption function
(Equation~\ref{eq-a1}) as a free parameter, besides those in the XLF. In
our baseline model, the other parameters of the absorption function and
photon index function are all fixed at the values presented in
Sections~\ref{sec-abs} and \ref{sec-idx}, respectively. Since this ML
fit does not constrain the normalization of the XLF, we determine it so
that the expected number of the total detections agrees with the
observed one, $N_{\rm tot} = 4039$,
and basically estimate the uncertainty from its Poisson error (but see below).
In ML fits, the minimized value of the likelihood estimator itself
cannot be utilized to evaluate the absolute goodness of the fit. Thus,
we verify it by comparing the flux and redshift distribution between the
model prediction and observation on the basis of $\chi^2$ test.

In a normal ML fit performed to a completely independent dataset (like
the analysis presented in Section~\ref{sec-abs-local}), the 1$\sigma$
error for a single parameter is defined as the deviation from the
best-fit when the $L$-value is increased by $\Delta-L = 1.0$ from its
minimum value. In our case, however, we utilize multiple-band data from
the same objects (i.e., at common redshifts) for a significant fraction
of the sample, which would work to underestimate the true errors in the
XLF parameters. Hence, here we conservatively estimate their 1$\sigma$
errors by adopting $\Delta-L = 2.0$ instead of $\Delta-L = 1.0$, to 
take into account the ``double counting'' effects.
For the same reason, we estimate the relative uncertainty in the
normalization of the XLF as $1/\sqrt{N_{\rm tot}/2}$ instead of 
$1/\sqrt{N_{\rm tot}}$.

In our analysis, we neglect the effects of AGN variability in
determination of the XLF. Many X-ray surveys utilized in our study are,
however, based on observations with a typical exposure of $\sim$ a day,
except for the \swift /BAT and \maxi\ surveys and very deep ones like
the CDFN and CDFS. This means that we measure an instantaneous flux (or
luminosity) of an AGN, which may well be different from its ``true''
flux averaged over a much longer period. For instance, \citet{pao04}
report that most of AGNs in the CDFS posses intrinsic X-ray variability
on timescales ranging from a day to a year. They find that the
fractional variability is $\simlt 0.2$ for 90\% of the AGN. To check the
possible systematic effects, we thus perform an ML fit with the same XLF
model as described in Section~\ref{sec-lf-results} by taking into
account variability of each AGN in Equation~(\ref{eq-ml2}) except for
the \swift /BAT, \maxi, CDFN, and CDFS samples. Here the distribution of
the observed flux relative to the intrinsic one is assumed to be a
Gaussian with a standard deviation of 0.2, which is adopted as the
maximum value regardless of the luminosity. The result verifies
that the best-fit XLF parameters are not affected by the time
variability over the statistical uncertainties.

\subsection{Results}
\label{sec-lf-results}

We model the luminosity function in the local universe
by a smoothly-connected double
power law model that has slopes $\gamma_1$ and $\gamma_2$ below and
above the break luminosity $L_{*}$, respectively:
\begin{equation}
\frac{d \Phi^{\rm CTN}_{\rm X} (L_{\rm X}, z=0)}{d{\rm log} L_{\rm X}} 
= A [(L_{\rm X}/L_{*})^{\gamma_1} + (L_{\rm X}/L_{*})^{\gamma_2}]^{-1}.
\end{equation}
Many previous works based on a sample covering a sufficiently wide \lx\
and $z$ range have revealed that the evolution of the XLF is more
complex than that approximated by a simple model like the pure density
evolution (PDE) or the pure luminosity evolution (PLE). Here we adopt
the luminosity dependent density evolution (LDDE), which is found to be
give a good representation of the XLF of AGN in a number of studies
based on hard X-ray ($>$2 keV) selected samples
\citep{ued03,sil08,ebr09,yen09} and on soft X-ray ($<$2 keV) selected
samples \citep{miy00,has05}.

We basically follow the formulation of the XLF in U03 with a few
additional modifications. The XLF at a given $z$ is calculated by
multiplying a luminosity-dependent evolution factor $e(z, L_{\rm X})$ to
the local one: 
\begin{equation}
\frac{ d \Phi^{\rm CTN}_{\rm X} (L_{\rm X}, z)}{ d{\rm log} L_{\rm X}} 
= \frac{ d \Phi^{\rm CTN}_{\rm X} (L_{\rm X}, 0)}{ d{\rm log} L_{\rm X}} e(z, L_{\rm X}).
\end{equation}
Recent studies based on large area surveys like COSMOS and SXDS with
high completeness have established a decay of the comoving number
density of luminous AGNs with \lx\ $\simgt 44$ toward higher redshift
above $z \simgt 3$ \citep{bru09,civ11,hir12}. The similar trend was
suggested by a previous work by \citet{sil05} using a \chandra\
serendipitous survey called CHAMP, where the completeness correction was
made in the X-ray and optical flux plane. Hence, we take into account
the decline in the evolution factor by introducing another
(luminosity-dependent) cutoff redshift above which the decline of $d
\Phi^{\rm CTN}_{\rm X} (L_{\rm X}, z) / d {\rm log} L_{\rm X}$ with $z$ starts to appear.

The evolution factor as a function of $z$ and \lx\ is thus
represented as
\begin{eqnarray}
  &&e(z, L_{\rm X} ) = \nonumber \\
&& \left\{ \begin{array}{ll}
     (1 + z)^{p1} & [z \le z_{c1}(L_{\rm X})] \\
     (1 + z_{c1})^{p1} 
     \left(\frac{ 1 + z}{ 1 + z_{c1}}\right)^{p2} & [z_{c1}(L_{\rm X}) < z \le z_{c2}] \\
     (1 + z_{c1})^{p1} \left(\frac{ 1 + z_{c2}}{ 1 + z_{c1}}\right)^{p2} \left(\frac{ 1 + z}{ 1 + z_{c2}}\right)^{p3} & [z > z_{c2}] \\
  \end{array} \right.
\end{eqnarray}
Here $z_{\rm c1}$ and $z_{\rm c1}$ represent two cutoff redshifts where
the evolution index changes from $p1$ to $p2$ and from $p2$ to $p3$,
respectively. We adopt $p2=-1.5$, the same value as adopted in U03, 
and $p3=-6.2$, based on the result by \citet{hir12}.
Following H05, we consider the luminosity dependence for the
$p1$ parameter as 
\begin{equation}
p1(L_{\rm X})=p1^* + \beta_1 ({\rm log} L_{\rm X} - {\rm log} L_{\rm p}),
\end{equation}
where we set ${\rm log} L_{\rm p} = 44$.

Both cutoff redshifts are given by power law functions of
\lx\ with indices of $\alpha1$ and $\alpha2$ below luminosity
thresholds of $L_{\rm a1}$ and $L_{\rm a2}$, respectively;
\begin{equation}
z_{\rm c1}(L_{\rm X}) = \left\{ \begin{array}{ll}
z_{\rm c1}^* (L_{\rm X}/L_{\rm a1})^{\alpha1} & [L_{\rm X} \le L_{\rm a1}] \\
z_{\rm c1}^* & [L_{\rm X} > L_{\rm a1}] \\
  \end{array} \right.
\end{equation}
and
\begin{equation}
z_{\rm c2}(L_{\rm X}) = \left\{ \begin{array}{ll}
z_{\rm c2}^* (L_{\rm X}/L_{\rm a2})^{\alpha2} & [L_{\rm X} \le L_{\rm a2}] \\
z_{\rm c2}^* & [L_{\rm X} > L_{\rm a2}] \\
  \end{array} \right.
\end{equation}
We fix $z_{\rm c2}^* = 3.0$, log $L_{\rm a2} = 44$, and $\alpha2 = -0.1$, which
well represent our XLF at $z>3$ and are also consistent with the results
by \citet{fio12} based on multiwavelength studies in the CDFS field.

\begin{figure}
\epsscale{1.0}
\begin{center}
\includegraphics[angle=0,scale=0.5]{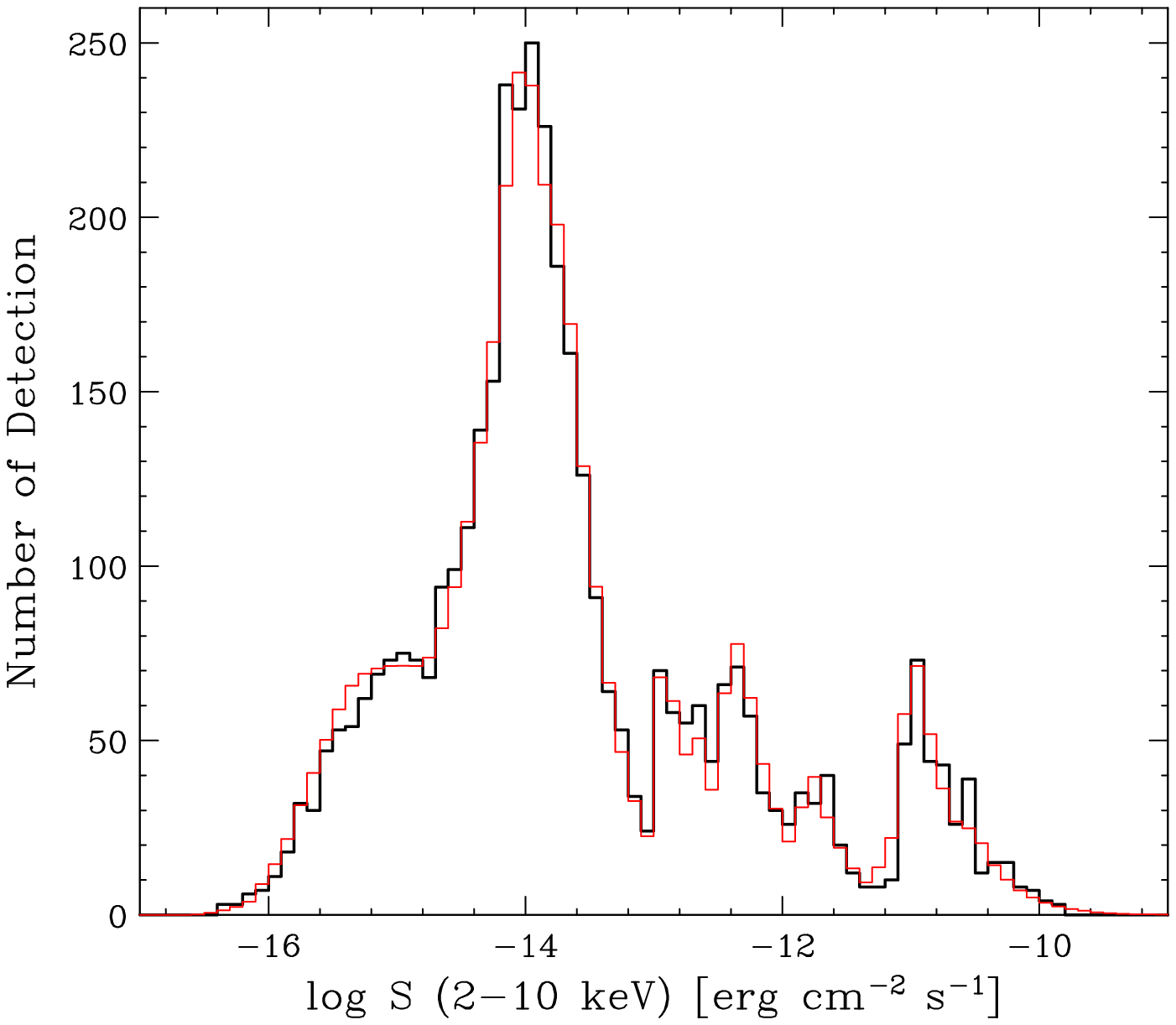}
\includegraphics[angle=0,scale=0.5]{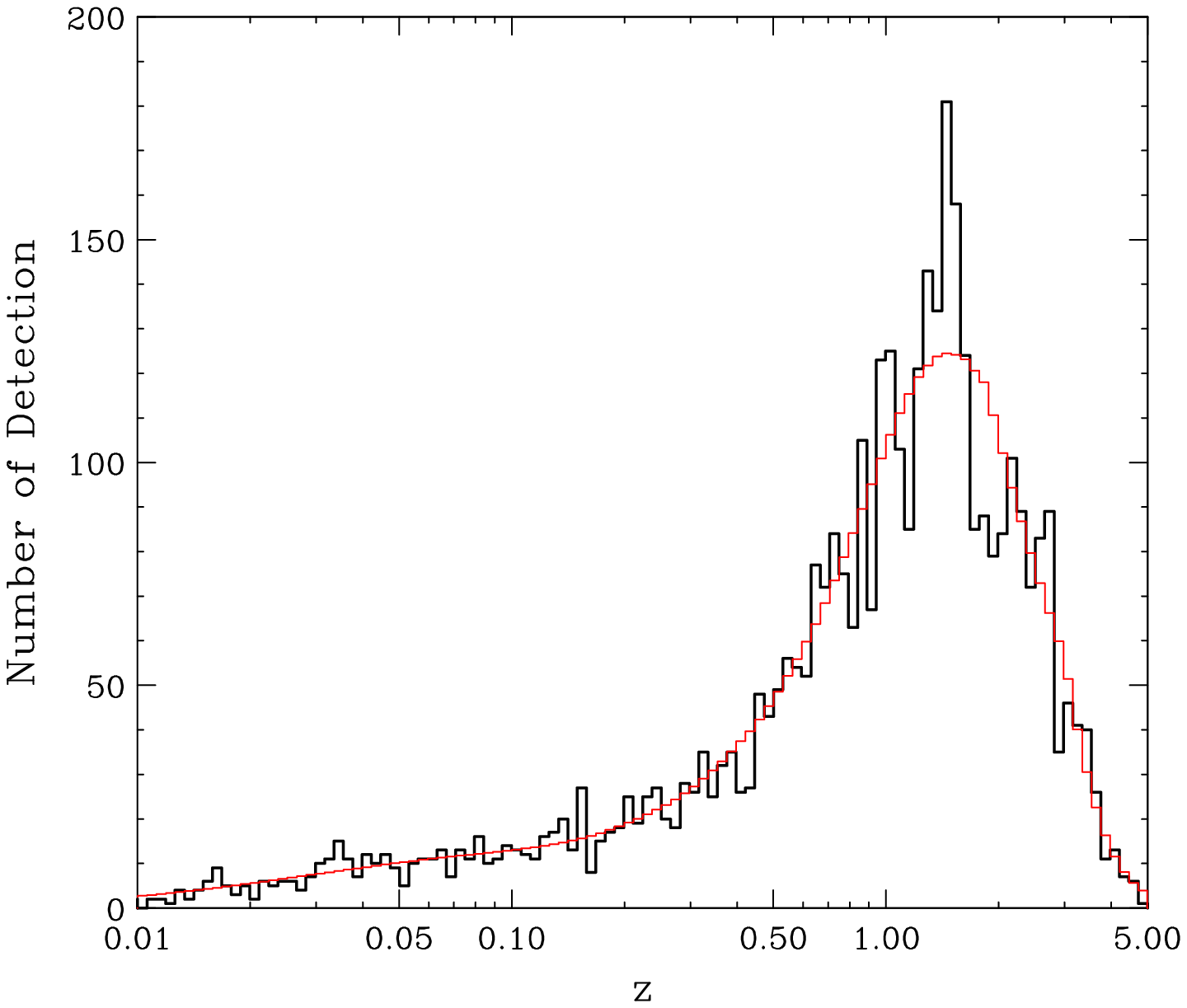}
\end{center}
\caption{
Observed histograms (thick, black)
of flux (left) and redshift (right) of our sample
compared with model predictions (thin, red).
}
\label{fig-dist}
\end{figure}

Adopting the LDDE model for the XLF along with the absorption and photon
index functions described in Sections~\ref{sec-abs} and \ref{sec-idx}, we
perform an ML fit to the whole sample consisting of 4039 detections. The
evolution index $a1$ in the absorption function and all parameters of
the XLF except for those mentioned as fixed above are left to be free
parameters. The best-fit parameters and the 1$\sigma$ errors are
summarized in Table~\ref{tab-abs} ($a1$) and Table~\ref{tab-lf}
(XLF). To verify the absolute goodness of the fit, we calculate the
2-dimensional histogram of flux and redshift predicted by the best-fit
model. The count rates in each survey are converted to the 2--10 keV
flux by assuming a power law index of 1.4 so that we can combine the
results from the multiple surveys. The histogram has 10 and 17
logarithmic bins in the flux range between $10^{-9}$--$10^{-17}$ and
redshift range between 0.002--5.0, respectively. To compare it with the
observed histogram made in the same way, the $\chi^2$ value between the
two histograms is calculated by adopting the 1$\sigma$ error of
$1+\sqrt{N+0.75}$ in each bin of the observed histogram where $N$ is the
number of sources \citep{geh86}. We obtain $\chi^2=102.7$ with a degree
of freedom (d.o.f.) of 114, indicating that the model is
acceptable. Figures~\ref{fig-dist} (left) and (right) show the projected
histograms onto the flux and redshift axes, respectively, together with
the model predictions (curve). Good agreements between the data and
model are seen, although there is a peak feature in the observed
redshift distribution around $z\approx 1.5$ related to the large scale
structure in the SXDS field \citep{aki14}.

\begin{figure}
\epsscale{1.0}
\begin{center}
\includegraphics[angle=0,scale=0.30]{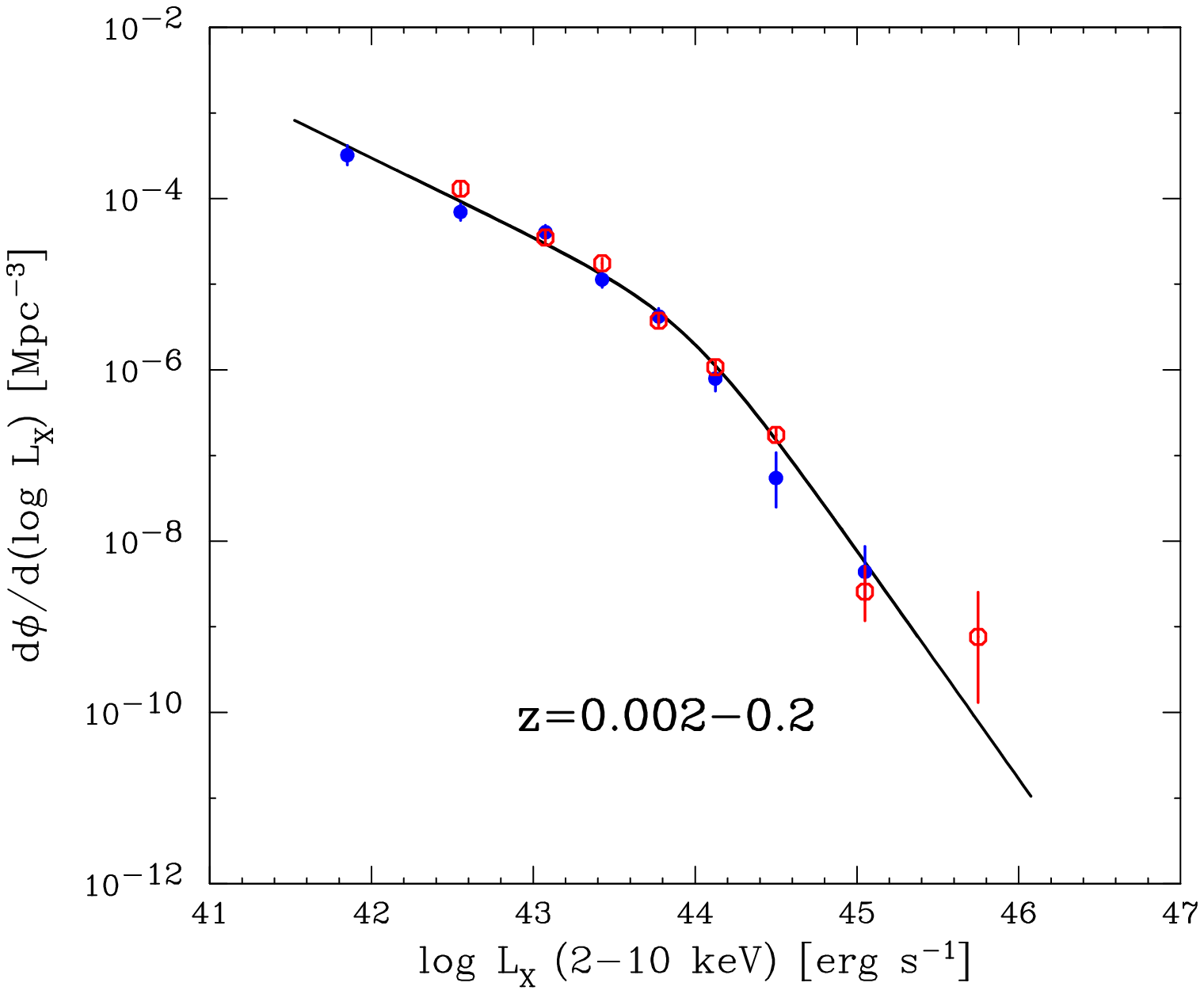}
\includegraphics[angle=0,scale=0.30]{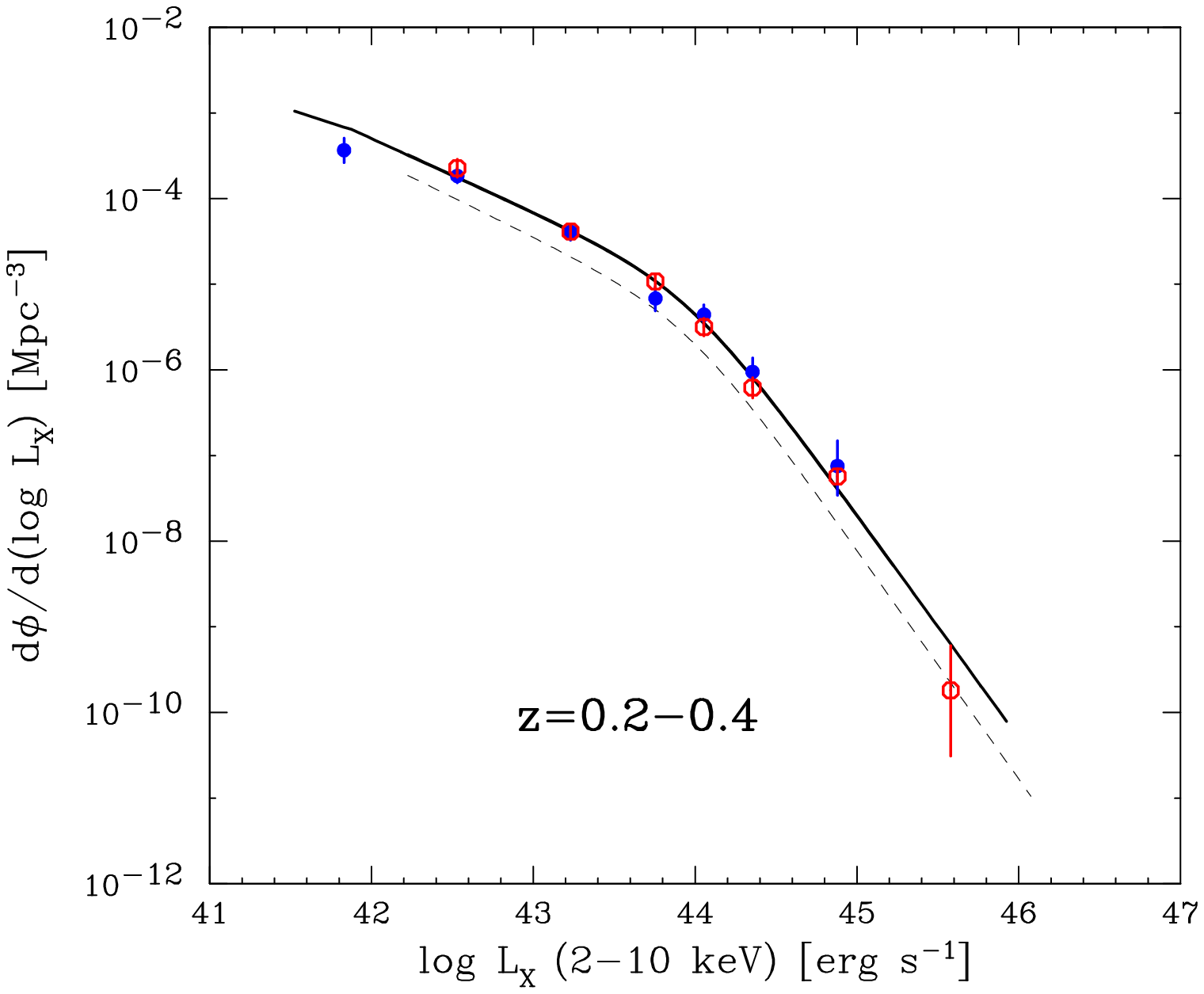}
\includegraphics[angle=0,scale=0.30]{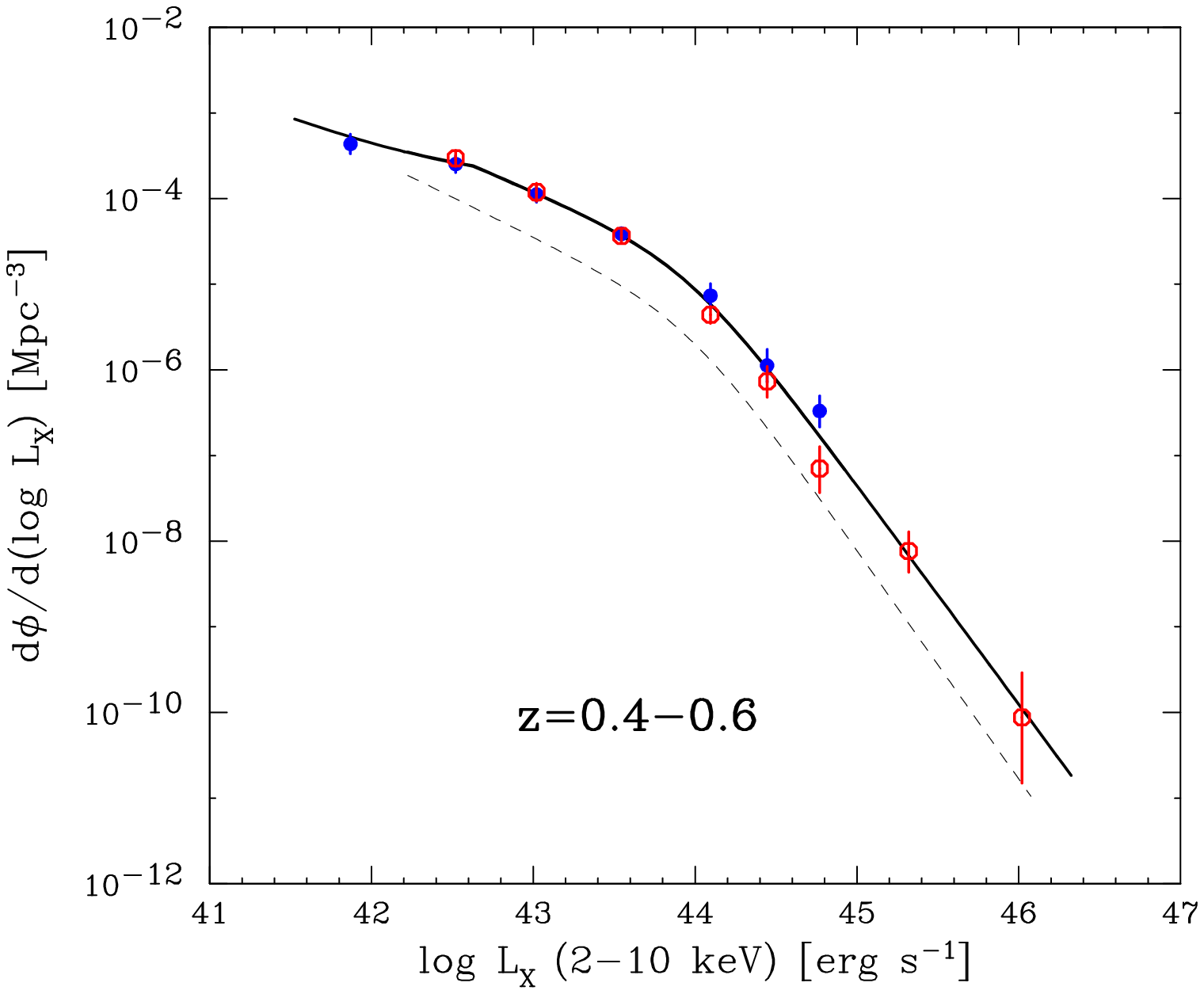}
\includegraphics[angle=0,scale=0.30]{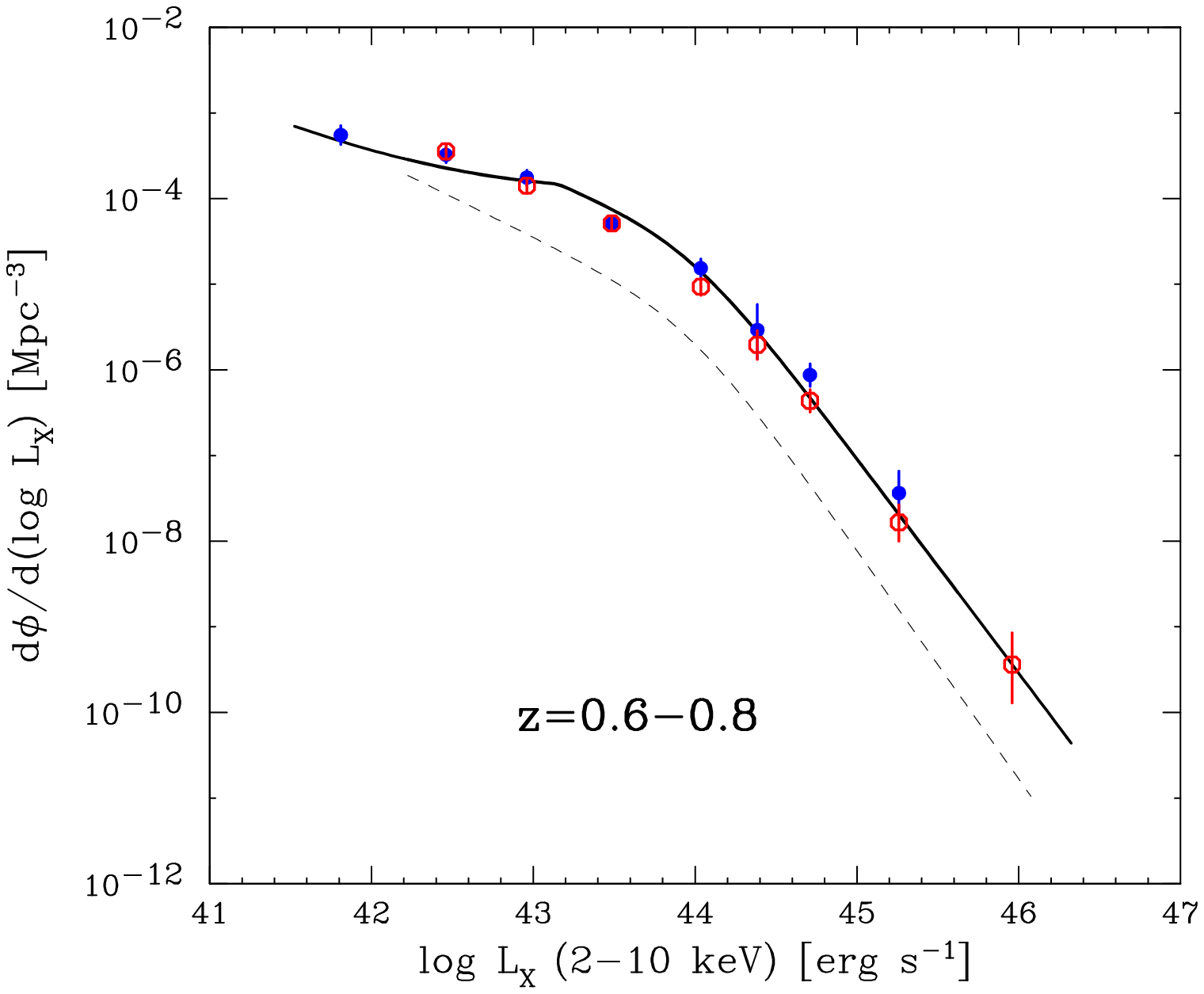}
\includegraphics[angle=0,scale=0.30]{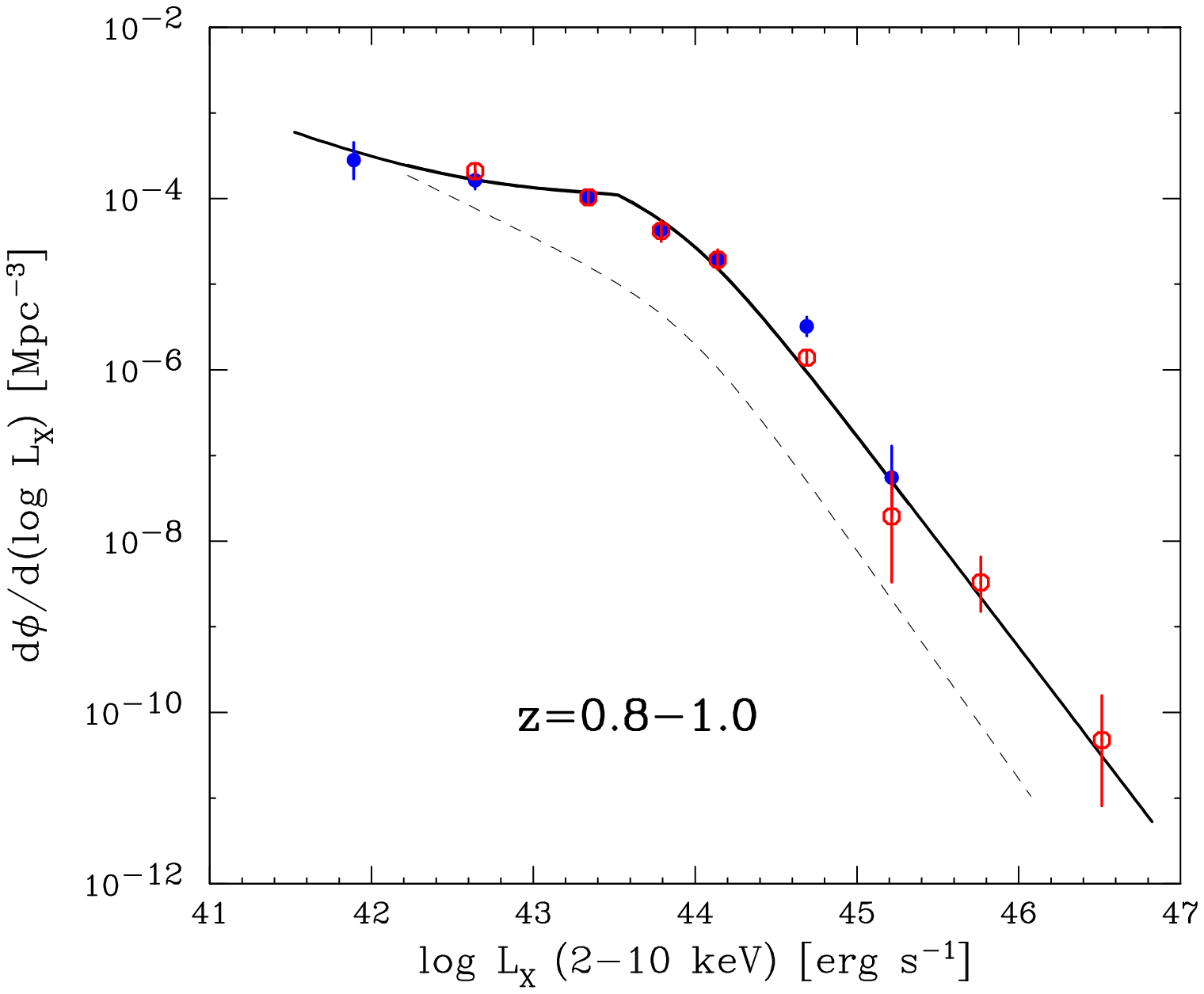}
\includegraphics[angle=0,scale=0.30]{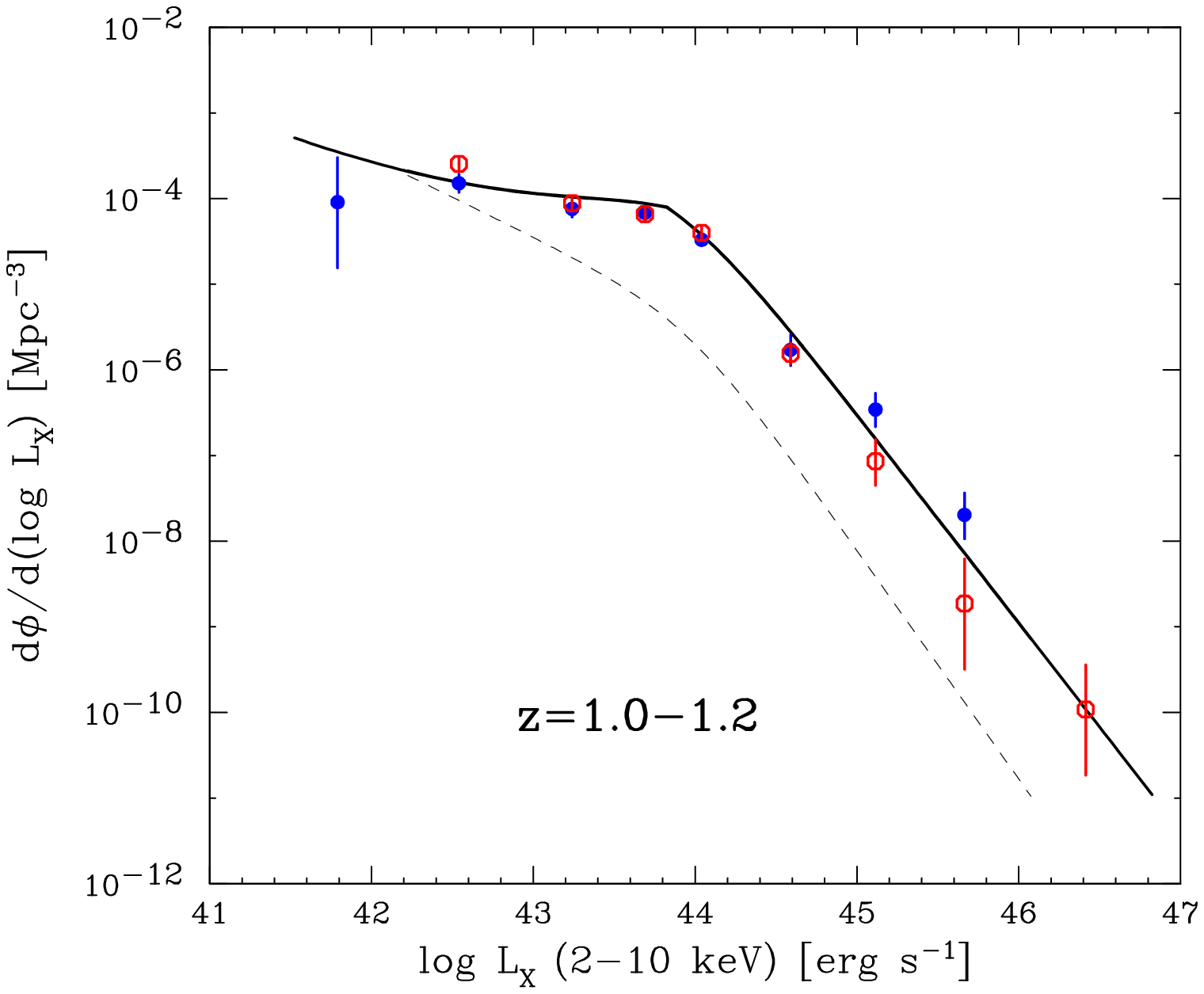}
\includegraphics[angle=0,scale=0.30]{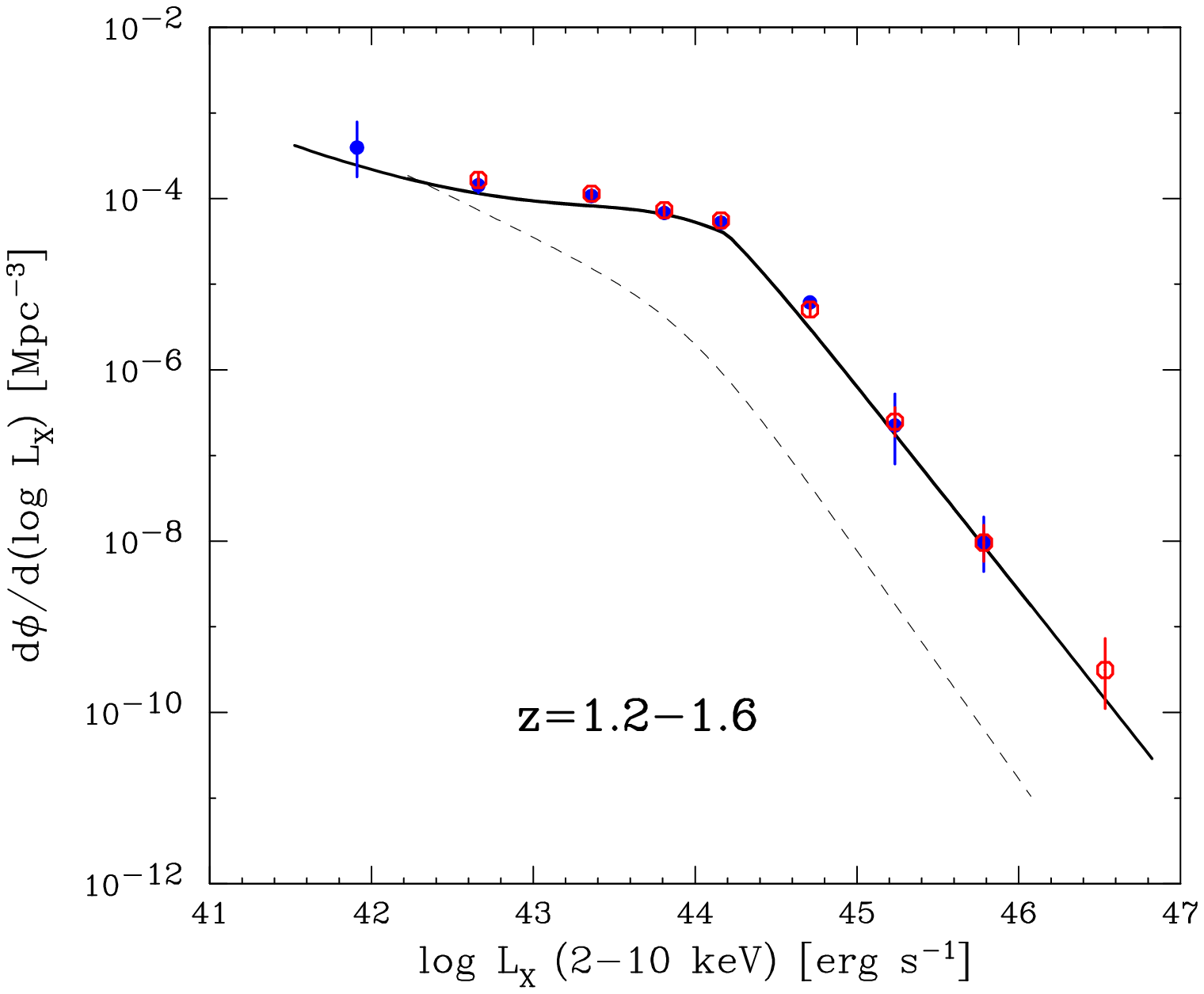}
\includegraphics[angle=0,scale=0.30]{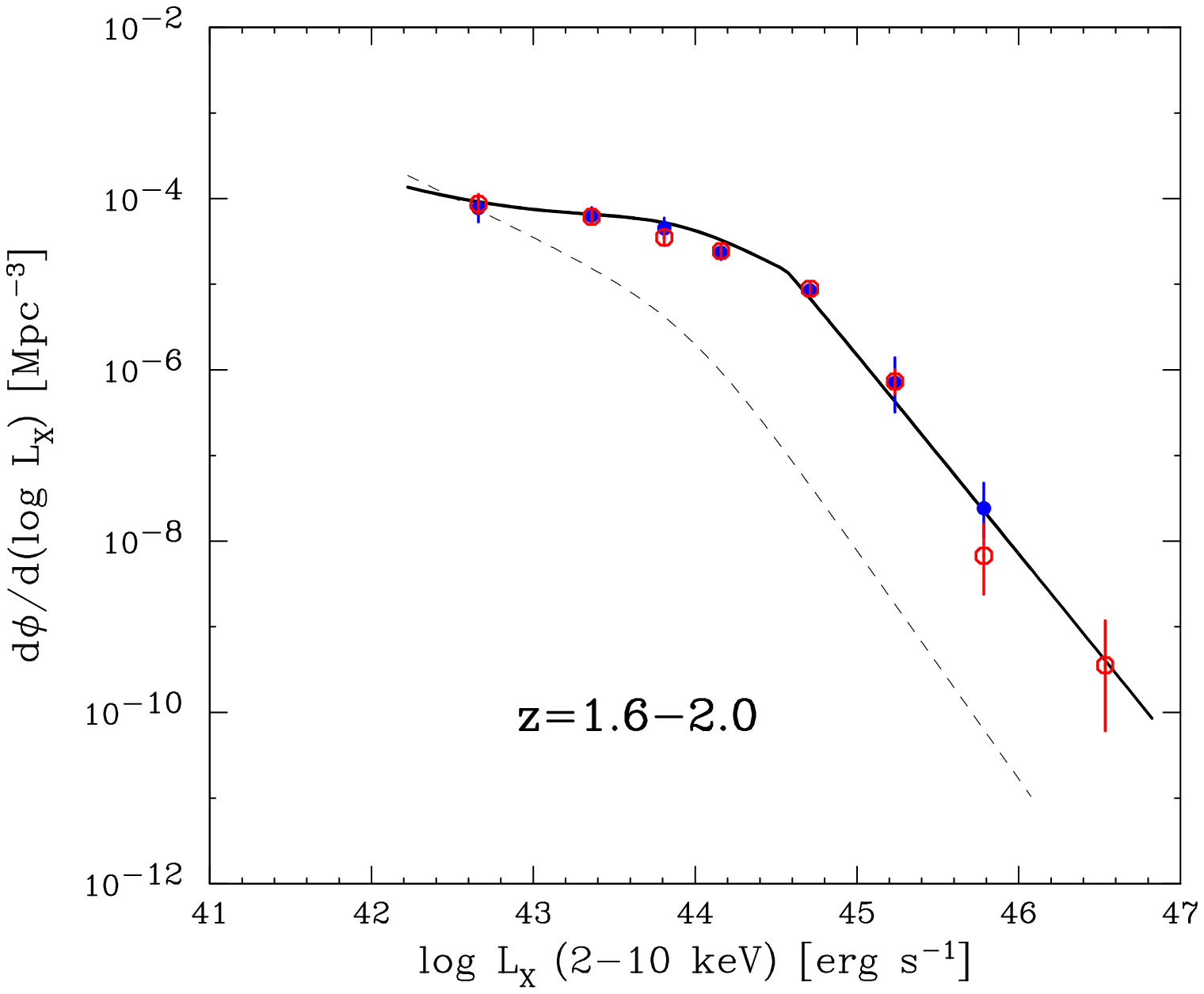}
\includegraphics[angle=0,scale=0.30]{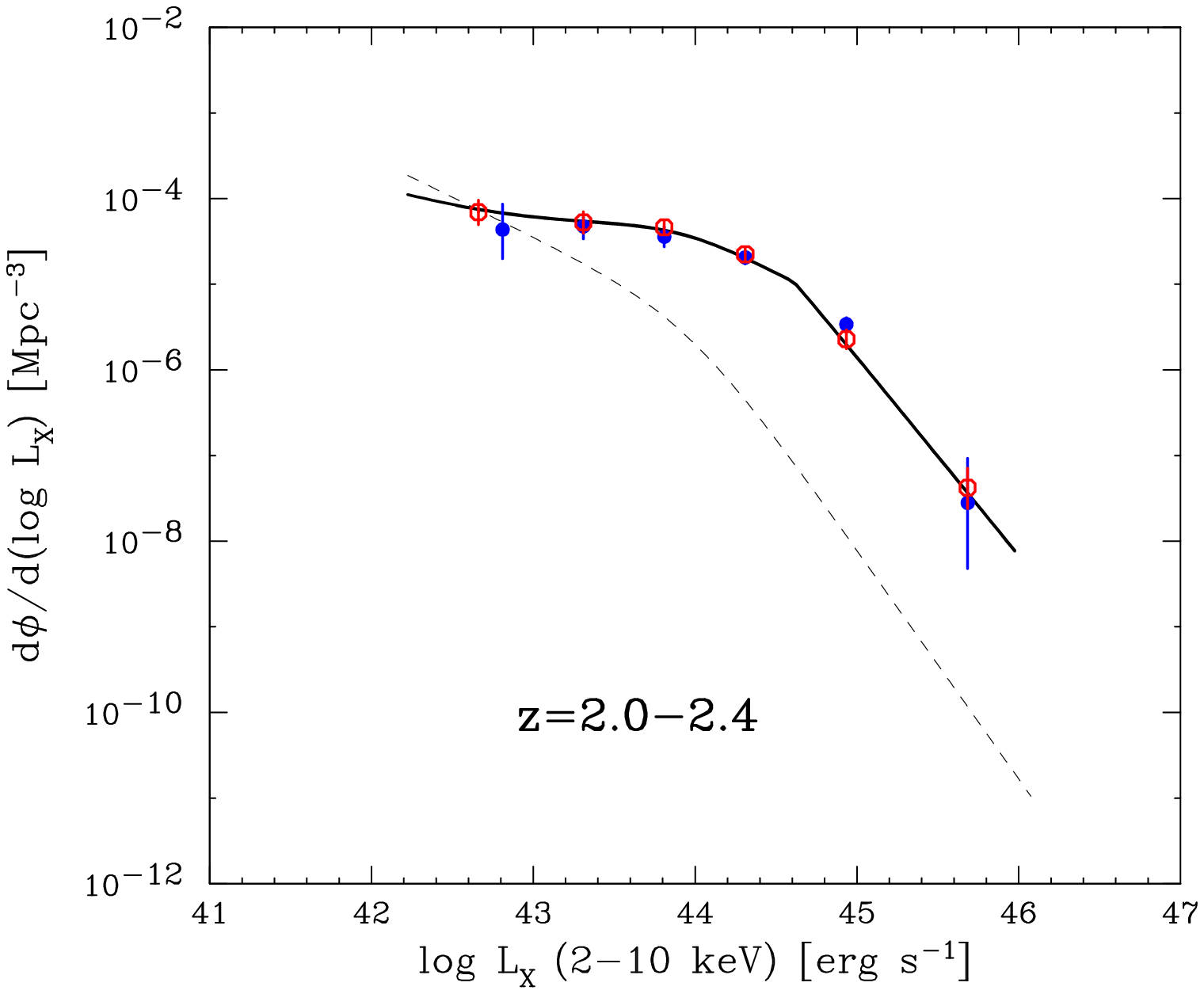}
\includegraphics[angle=0,scale=0.30]{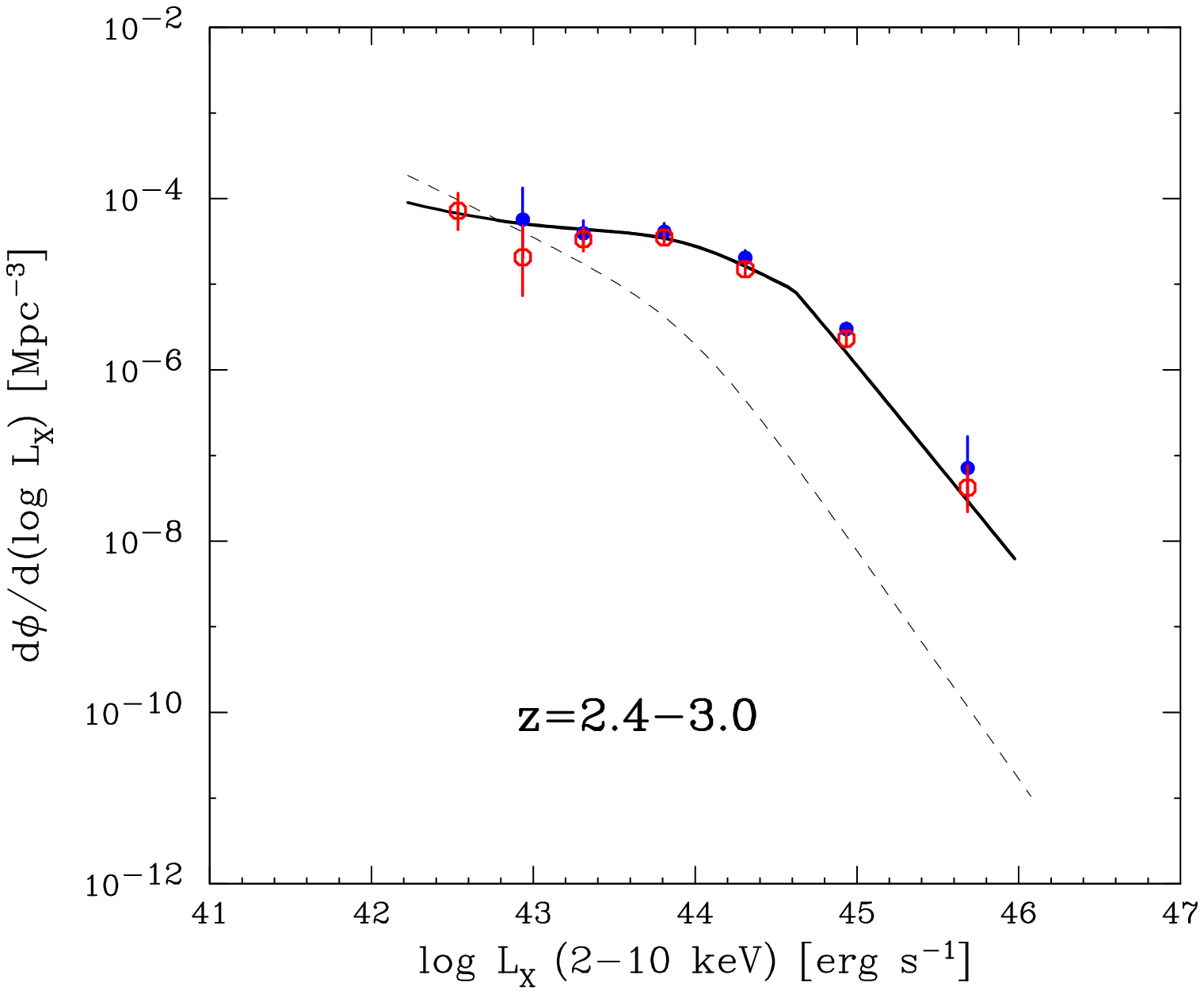}
\includegraphics[angle=0,scale=0.30]{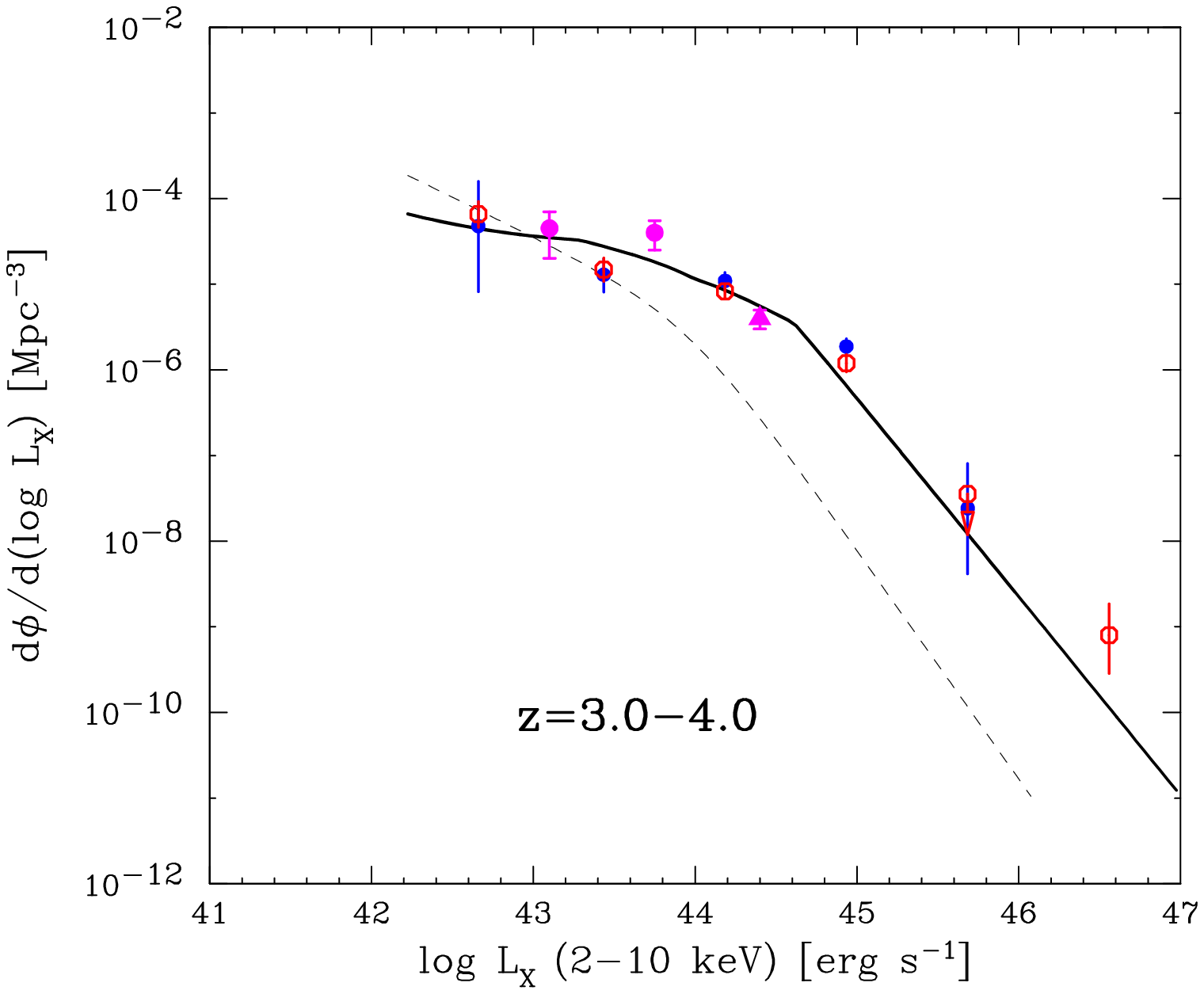}
\includegraphics[angle=0,scale=0.30]{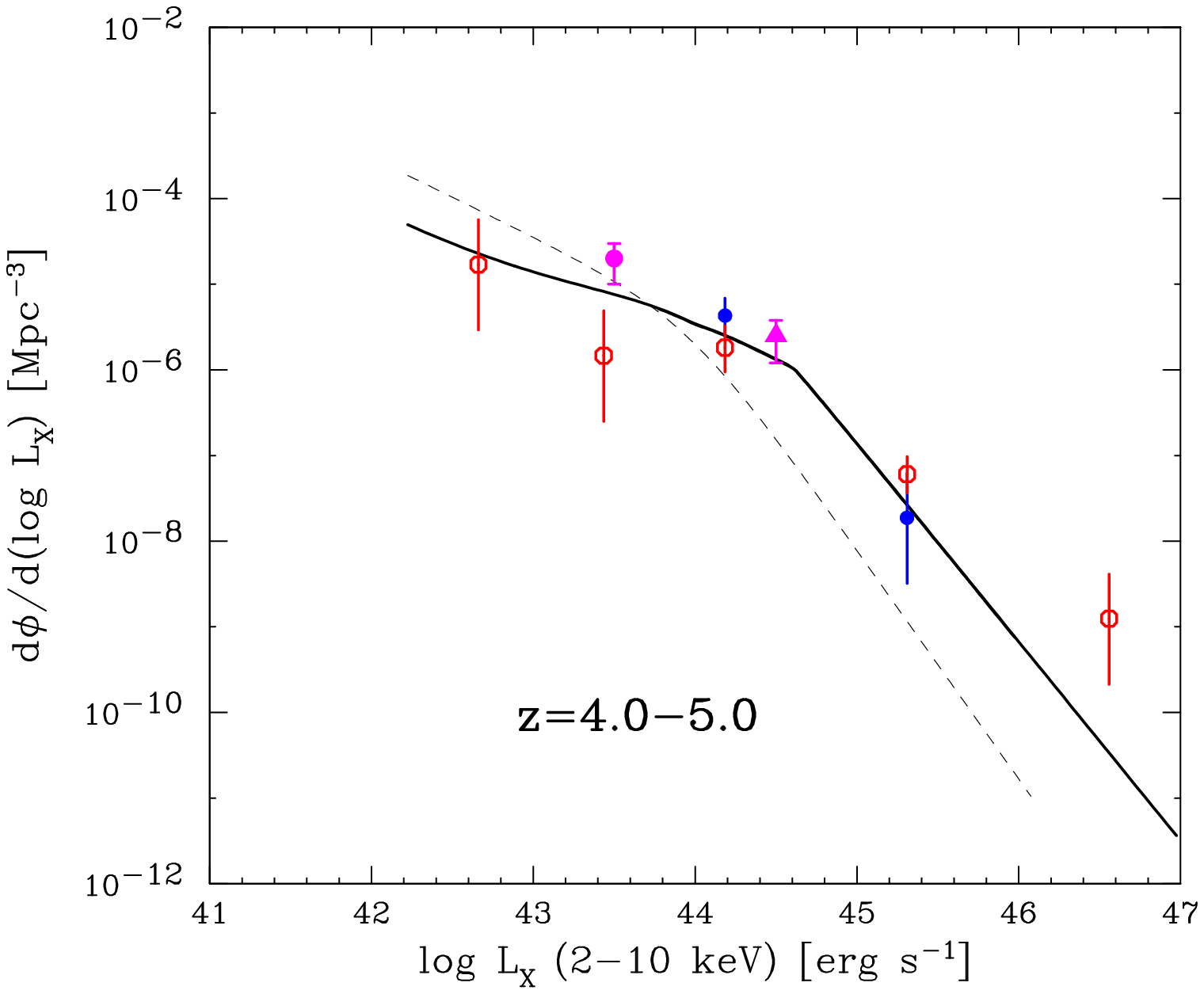}
\end{center}
\caption{
De-absorbed, rest frame 2--10 keV XLF of AGNs at different redshift
ranges (CTN AGNs only).
The solid curve represents the best-fit XLF at the central redshift in
each $z$ bin. The dashed curve is that in the local universe. Blue (red)
data points are plotted according to the ``$N^{\rm obs}/N^{\rm mdl}$
method'' with 1$\sigma$ Poisson errors by using the hard (soft) band
sample. The magenta points in the $z=3.0-4.0$ and $z=4.0-5.0$ panels are
taken from \citet{fio12}.  } \label{fig-lf}
\end{figure}

\begin{figure}
\epsscale{1.0}
\begin{center}
\includegraphics[angle=0,scale=0.8]{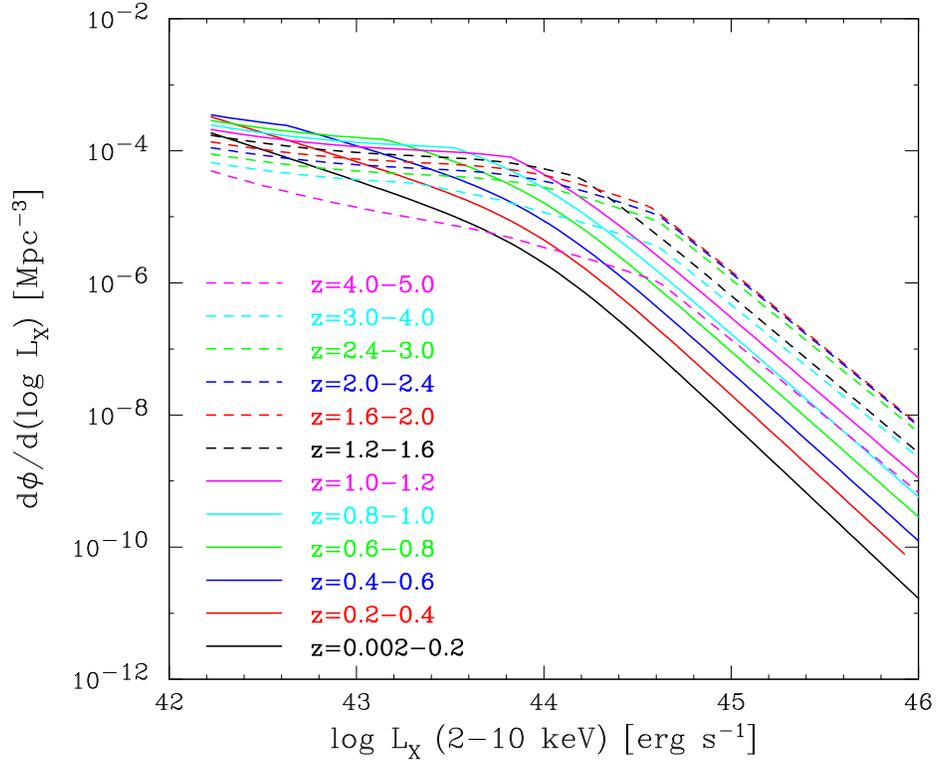}
\end{center}
\caption{
Comparison of the best-fit XLF shape between different redshifts (CTN AGNs only).
}
\label{fig-lf-nodata}
\end{figure}

Figure~\ref{fig-lf} displays the best-fit XLFs of CTN AGN,
$d \Phi^{\rm CTN}_{\rm X} (L_{\rm X}, z=0)/d{\rm log} L_{\rm X}$, in 12 different redshift
bins covering from $z=0.002$ to $z=5.0$. The shape of the XLF at the
central redshift of each bin is represented by the curves. The data
points are calculated on the basis of the ``$N^{\rm obs}/N^{\rm mdl}$
method'' \citep{miy01}; for a given luminosity bin, we plot the model at the
logarithmic center of \lx\ multiplied by the ratio between the number of
observed sources and that of the model prediction. Here we utilize the
\lx\ value assigned to each object according to the procedures described
in Section~\ref{sec-sample-luminosity}. Thus, the plotted data should be
regarded as an approximation by considering the uncertainties in
calculating \lx, in particular for CTK AGNs. This would become
an issue only for faintest AGNs detected in the hard band,
$\sim$10--20\% of which could be CTK AGNs at fluxes of
$S=10^{-15}-10^{-16}$ \ergs\ (2--10 keV) according to our best-fit model
(see Section~\ref{sec-model-predictions}). The data points are independently
calculated from the hard band ($>2$ keV) and soft band ($<2$ keV)
samples, which are marked by filled circles (blue) and open circles
(red), respectively. Here the \maxi\ sample is not included due to its
significant overlap with the \swift/BAT sources. The error bars reflect
the relative Poissonian 1$\sigma$ errors in the observed number of
sources based on the formula of \citet{geh86}. The arrows denote the
90\% confidence upper limits when no object is found in that luminosity
bin. To show the redshift dependence of the XLF, we plot the best-fit
model computed at different redshifts in Figure~\ref{fig-lf-nodata}.

In Figure~\ref{fig-lf} ($z=3.0-4.0$ and $z=4.0-5.0$), we also plot the
luminosity function derived by \citet{fio12}. They adopt a fainter flux
limit than that in \citet{xue11} by utilizing the positional information
of $z>3$ galaxies based on optical and mid-infrared catalogs. 
Our
best-fit XLF is in good agreement with their results; the maximum 
deviation of the data points is $<2\sigma$ statistical error.
Note that \citet{vit13} analyze the $z>3$ AGNs in
the CDFS by adopting rather conservative selection criteria. The \logn\
of their sample is smaller than that of \citet{fio12} by a factor of
$\sim$2. The discrepancy could be explained by incompleteness. The same
problem could be present in our sample, too. However, assuming the
extreme case that all the unidentified AGNs detected in the soft band
are $z>3$ AGNs, we find that the XLF normalization at $z>3$ becomes only
1.5 times larger than the present data points in average, which is still
consistent with the best fit model within the errors.

\begin{figure}
\epsscale{1.0}
\begin{center}
\includegraphics[angle=0,scale=0.8]{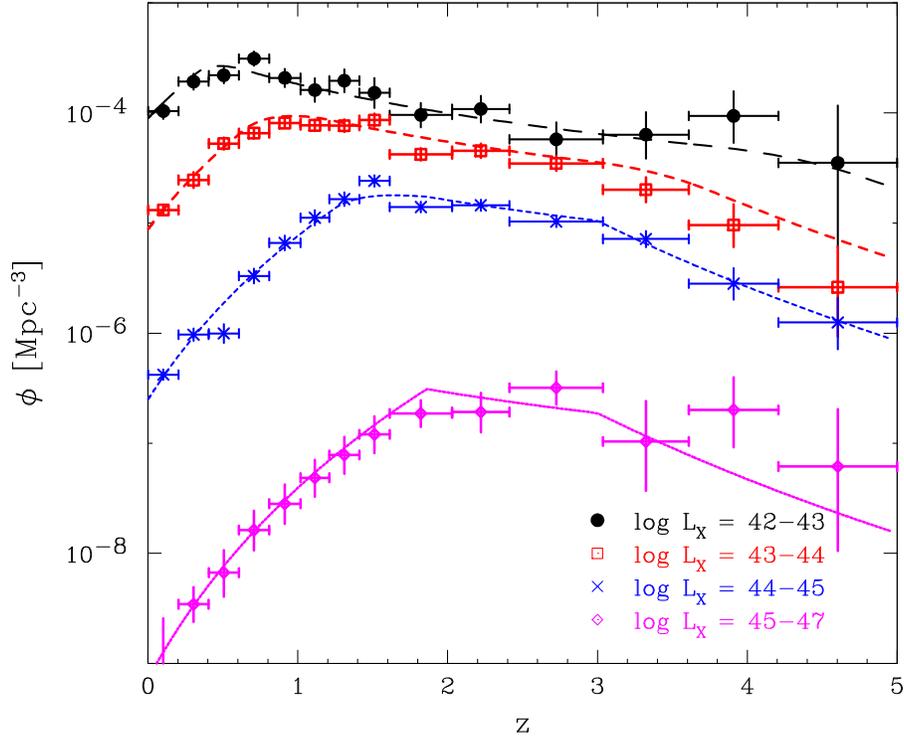}
\end{center}
\caption{
Comoving number density of AGNs plotted against redshift 
in different luminosity bins 
(CTN AGNs only). The curves are the best-fit model, and the data points 
are calculated from either the soft or hard band sample (see Section~\ref{sec-lf}).
}
\label{fig-zf}
\end{figure}

Figure~\ref{fig-zf} plot the co-moving space number density of
CTN AGNs as a function of redshift integrated in different
luminosity bins, log \lx\ = 42--43, 43--44, 44--45, and 45--47. The
curves show that of the best-fit model, while the data points are based
on the ``$N^{\rm obs}/N^{\rm mdl}$ method'' as explained above. 
In this Figure, to ensure complete independence of the plotted data, we
utilize either the hard-band or soft-band selected sample to calculate
the data points in each redshift and luminosity bin. Specifically, we
adopt the hard-band samples (without the \maxi\ one) in the region of
$z<2$ and log \lx\ $< 44$, and the soft-band samples for the rest. This
is because at higher redshifts soft-band surveys become more efficient
even for obscured AGNs thanks to the K-correction effect. Also, at large
\lx\ ranges the majority of AGNs are unobscured populations
(Section~\ref{sec-abs}), for which wide area surveys with \rosat\
provide a large number of sources.

From Figure~\ref{fig-zf}, one can clearly confirm the global
``down-sizing'' evolution, where more luminous AGNs have their number
density peak at higher redshifts compared with less luminous ones. We
note, however, that when we only focus attention on the high redshift
range of $z \simgt 3$, our LDDE model with $\alpha2 = -0.1$ indicates an
``up-sizing'' evolution instead (i.e., the number ratio of less luminous
AGNs to more luminous ones is larger at earlier epochs). This is what is
expected from the hierarchical structure formation in the early
universe. Thus, the SMBH growth must be correctly described by {\it
``up-down sizing''}. To firmly establish this behavior, it is critical
to determine the space number density of all AGNs at $z \simgt 3$ in
both the lowest and highest luminosity ranges with better accuracies.

\subsection{LADE model}
\label{sec-lf-lade}

As noticed in Figures~\ref{fig-lf} and \ref{fig-lf-nodata}, the shape of
the XLF at $z=1-3$ 
is quite different from that in the local universe
in the sense that the slope at the low luminosity range is significantly
flatter than those observed at lower redshifts. This trend can be well
reproduced by the LDDE model. Here we check if the LADE model of the XLF
proposed by \citet{air10} also gives a good description of our
data. Unlike the LDDE, the LADE assumes a constant relative shape of the
XLF in the logarithmic scales over the full redshift range, and its
break luminosity and normalization is given as a function of
redshift. We perform an ML fit to the whole sample by adopting the same
formulation of the XLF as given in \citet{air10}. A chi-squared test for
the 2-dimensional histograms of flux and redshift between the best-fit
model and data yields $\chi^2=207.1$ (d.o.f=114). The LADE model is thus
rejected with a $p$ value of $< 10^{-7}$. We infer that it is difficult
to distinguish the LDDE and LADE models in \citet{air10} because of the
smaller number of sample used there; indeed \citet{air10} show that the
LDDE gives a better fit to their data than the LADE, although the
difference is not significant.

\subsection{Comparison with Previous Works}
\label{sec-lf-comparison}

The parameters of the AGN XLF are better constrained than in any of
previous works thanks to our large sample size ($\approx$15 and
$\approx$4 times larger than those used by U03 and H05, respectively).
Here we compare them with those of the LDDE model by U03
and by H05 
as representative ones.
Although the direct comparison with U03
is not trivial as the formulation of the XLF in U03 is simpler than ours
(e.g., $\beta_1=0$ is assumed in U03), the overall parameters are in
good agreement between our work and U03 except for $\gamma_2$.
The overall shape of our XLF derived for all CTN AGNs is 
almost consistent with that by H05 derived
only for type-1 AGNs (see 
their Table~5) within the errors
except for $\alpha$ (=$\alpha1$ in our paper), which is found to be 
slightly larger ($\alpha1=0.29\pm0.02$) than in H05 ($\alpha=0.21\pm0.04$).
Note that the $z_{\rm c,44} = 0.21\pm0.04$
parameter defined in H05 can be converted to $z_{\rm c} = 1.96\pm0.15$
with $\alpha=0.21$ (=$\alpha1$ in our paper), and thus agrees with our
result ($z_{\rm c} = 1.86\pm0.07$).
Our best-fit model has steeper slopes in the double power-law form for
the local XLF, 
$\gamma_1=0.96\pm0.04$ and $\gamma_2=2.71\pm0.09$,
than those obtained by H05. 
This can be explained by the luminosity dependence of the absorbed-AGN
fraction.
Our local XLF is well consistent with the 
\citet{bal14} result as determined by the ``multi-band'' fit.

We also determine the evolution of the absorption fraction with an
unprecedented accuracy, $a1 = 0.48\pm0.05$ in the form of $(1+z)^{a1}$
that is saturated above $z=2$. \citet{laf05} model the redshift
evolution of the absorption fraction by a different parameterization,
adopting a linear function of $z$ for the fraction of AGNs with log \nh\
$< 21$. According to their best-fit model (model 4), where the constant
\nh\ distribution is assumed over log \nh\ = 21--25, the fraction of
absorbed CTN AGNs (log \nh\ = 22--24) in the total CTN AGNs (log \nh\ $<
24$) at log \lx\ = 44 is 2.3 times higher at $z=2$ than at $z=0$. This
corresponds to $a1 \approx 0.75$ when modeled by
$(1+z)^{a1}$. Similarly, \citet{has08} obtain $(1+z)^{0.62\pm0.11}$ that
is saturated at $z>2$. The reason why both \citet{laf05} and
\citet{has08} obtain larger indices than ours could be the difference in
the adopted absorption fraction in the local universe. Both of them
utilize the \heao\ samples, from which somewhat smaller absorption
fractions are estimated compared with the \swift/BAT and \maxi\
results. In the \citet{laf05} model, the fraction of CTN AGNs in the
total CTN AGNs is $\approx$0.25 at log \lx\ = 44, which can be converted
to $\psi_{43.75}^0 \approx 0.31$ with $\beta=0.24$. This value is
similar to that presented in \citet{has08}, while it is smaller than our
result obtained from the \swift/BAT sample, $\psi_{43.75}^0 =
0.43\pm0.03$. The reason for the discrepancy is unclear but may be
attributed to the statistical error due to the small size of the \heao\
A2 sample \citep{pic82} and/or incompleteness of the \heao\ A1 and A3
sample \citep{gro92}. Note that our best-fit slope is larger than that
in the model by \citet{bal06}, $a1\approx0.3$, where the absorption
fraction is assumed to be saturated above $z=1.0$. \citet{tre06} obtain
a similar slope to ours, $a1 \approx 0.4\pm0.1$ without saturation up to
$z=4$, by correcting for selection biases due to the low completeness
($53\%$) in their sample.

\section{Standard Population Synthesis Model of the XRB}
\label{sec-model}

\subsection{Model Predictions}
\label{sec-model-predictions}

We have constructed a new XLF of AGNs by utilizing one of the largest
sample with a high degree of identification completeness combined from
surveys in different energy bands. We also model the absorption and
photon index functions on the basis of a hard X-ray ($>15$ keV) selected
AGN sample in the local universe for which detailed spectral information
is available. The redshift dependence of the absorption function is
taken into account, whose evolution index $a1$ is simultaneously
determined along with the XLF parameters. We consider the contribution
of CTK AGNs by assuming that their number density at a given
luminosity and a redshift is the same as that of obscured CTN
ones. The combination of the XLF, absorption function, and photon index
function with the template broad-band spectra of AGNs enables us to
establish a new population synthesis model of the XRB. In this section,
we examine the basic properties of the model.

\begin{figure}
\epsscale{1.0}
\begin{center}
\includegraphics[angle=0,scale=0.8]{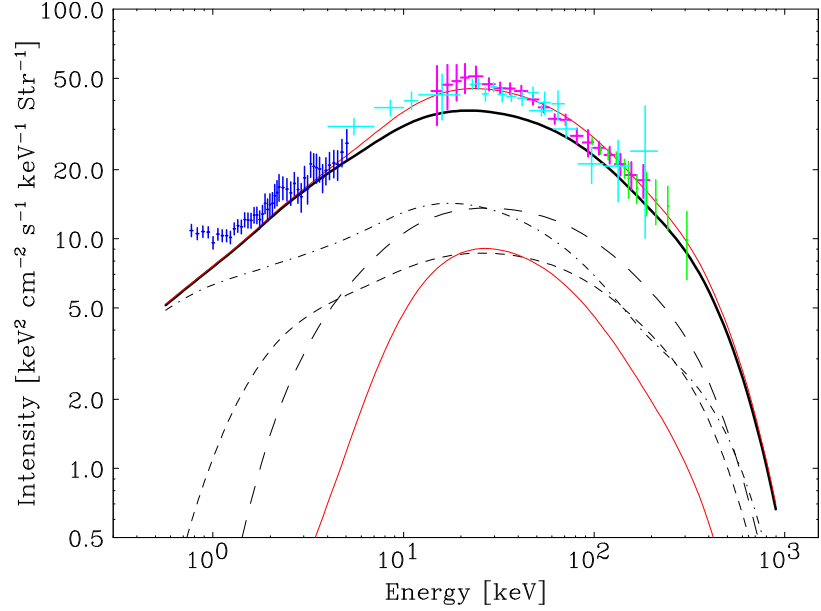}
\end{center}
\caption{
XRB spectrum calculated from our AGN population synthesis model
(upper solid curve, red)
compared with the observed data by various missions \citep{aje08}.
Middle solid curve (black): the integrated spectrum of CTN AGNs (log
\nh\ $<$24).
Lower solid curve (red): that of CTK AGNs (log \nh\ = 24--26). 
Long-dashed curve (black): that of AGNs with log \nh\ = 23--24. 
Short-dashed curve (black): that of AGNs with log \nh\ = 22--23.
Dot-dashed curve (black): that of AGNs with log \nh\ $<$22.
Data points in the 0.8--5 keV (blue), 4--215 keV (cyan), 14--195 keV
(magenta), and 100--300 keV (green) bands refer to the XRB spectra
observed with \asca/SIS \citep{gen95}, \integral\ \citep{chu07}, 
\swift/BAT \citep{aje08}, and \heao\ A4 \citep{gru99}, respectively.}
\label{fig-cxbspec}
\end{figure}

Figure~\ref{fig-cxbspec} shows the integrated broad band spectrum of the
whole AGNs at $z=0.002-5.0$ with log \lx\ = 41--47 predicted from our
model. The spectrum of each AGN is modelled by the ``template spectrum''
presented in Section~\ref{sec-template}, which is given as a function of
luminosity, column density, and photon index. The data points represent
the measurements of the XRB observed with various missions including
\heao\ A4 in the 100--300 keV band \citep{gru99}, \swift/BAT in the
14--195 keV band \citep{aje08}, and \integral\ in the 4--215 keV band
\citep{chu07}. A good agreement is confirmed between the model
prediction and the hard XRB, supporting the overall validity of our
model, including the fraction of CTK AGNs and the reflection strengths
from the accretion disk and torus based on the luminosity and redshift
dependent unified scheme (Section~\ref{sec-template}). Effects by
changing these model parameters will be examined in
Section~\ref{sec-model-dependence}.

There are discrepancies in the absolute flux measurements of the XRB
between different missions, most probably due to calibration
uncertainties. These issues are discussed in detail by e.g.,
\citet{bar00} for the XRB below 10 keV and by \citet{aje08} above 10
keV. In Figure~\ref{fig-cxbspec}, for clarity, we only plot the \asca\
result obtained by \citet{gen95} as the representative data of the XRB
in the 0.8--5 keV band. The XRB spectrum obtained by the \heao\ A2
experiment gives systematically smaller fluxes in the energy range below
10 keV than most of more recent missions. The maximum flux is reported
by \citet{delu04} with \xmm, which is 40\% higher than that of the
original \heao\ A2 result \citep{mar80}. The reasons are yet unclear. In
addition, we do not include the emission from populations other than
AGNs in our model. For instance, clusters of galaxies could contribute
to the XRB by $\sim$10\% at 1 keV level. With these reasons, we mainly
discuss our population synthesis model on the basis of the hard XRB
above 10 keV, where the contribution from AGNs is dominant.

\begin{figure}
\epsscale{1.0}
\begin{center}
\includegraphics[angle=0,scale=0.4]{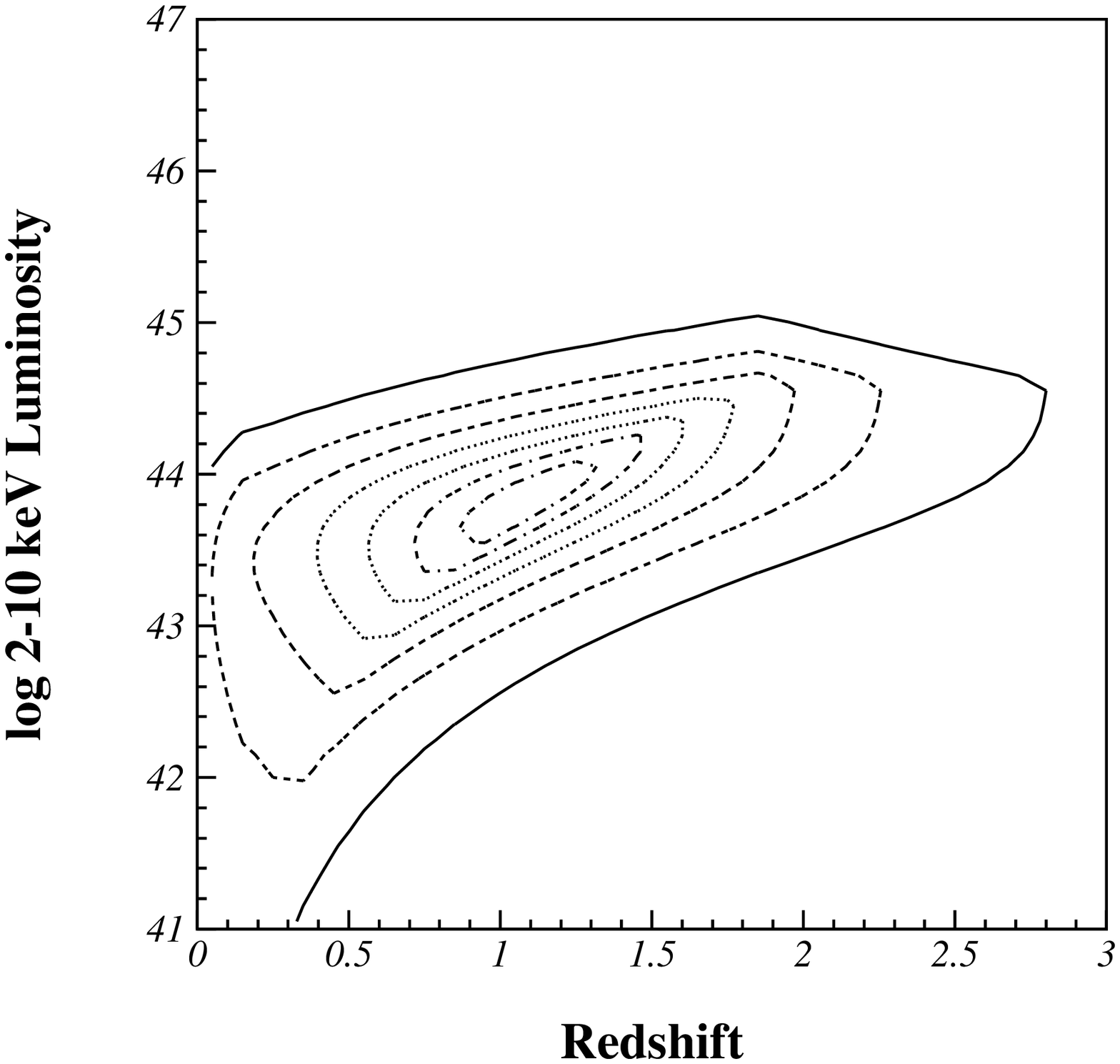}
\includegraphics[angle=0,scale=0.4]{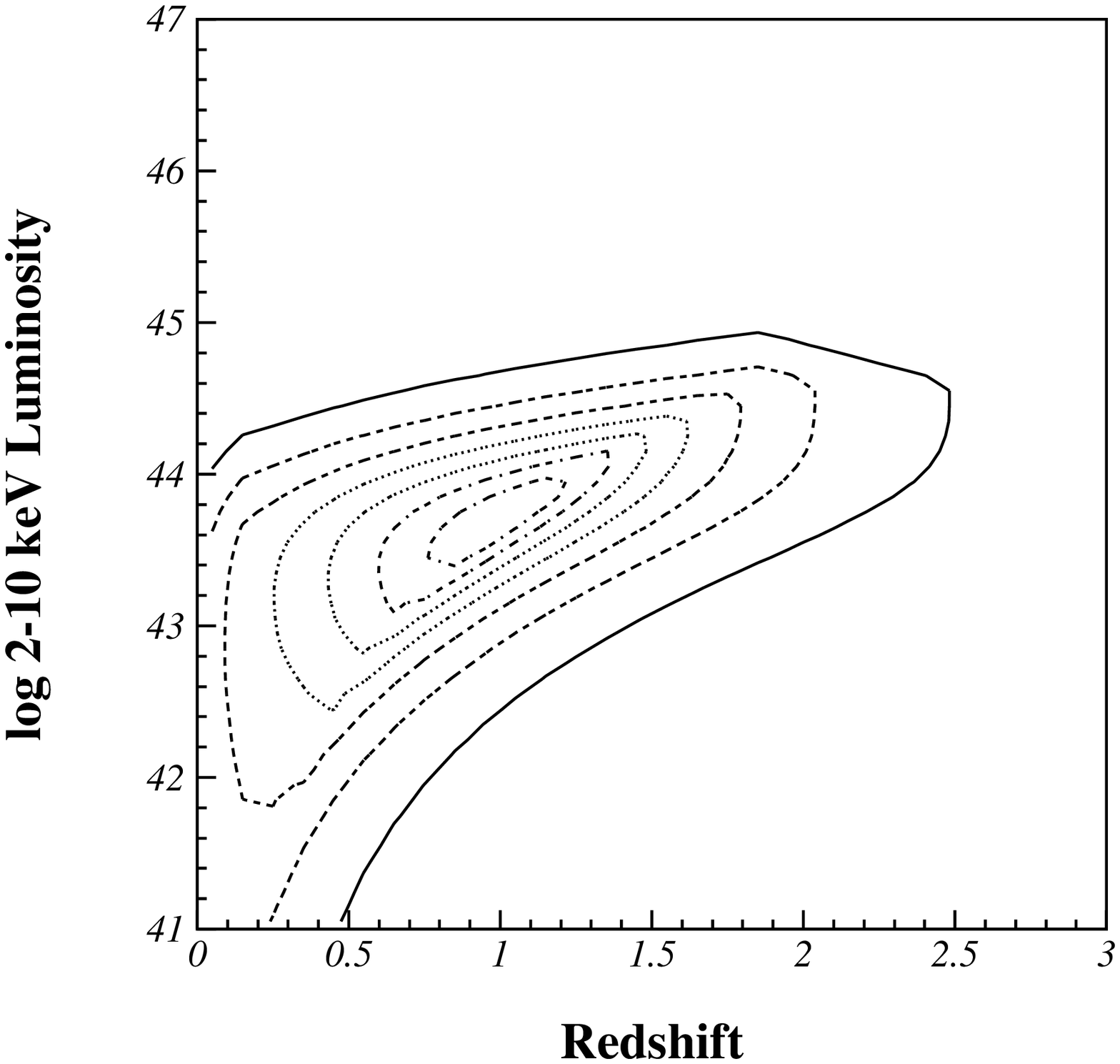}
\end{center}
\caption{
{\it Left}: contour plot showing the contribution of all (CTN+CTK) AGNs
per unit $z$ and log \lx\ to the 2--10 keV XRB. The intervals are
constant in linear scale. 
{\it Right}: the same but for the 10--40 keV XRB.}
\label{fig-cxbcontour}
\end{figure}

\begin{figure}
\epsscale{1.0}
\begin{center}
\includegraphics[angle=0,scale=0.6]{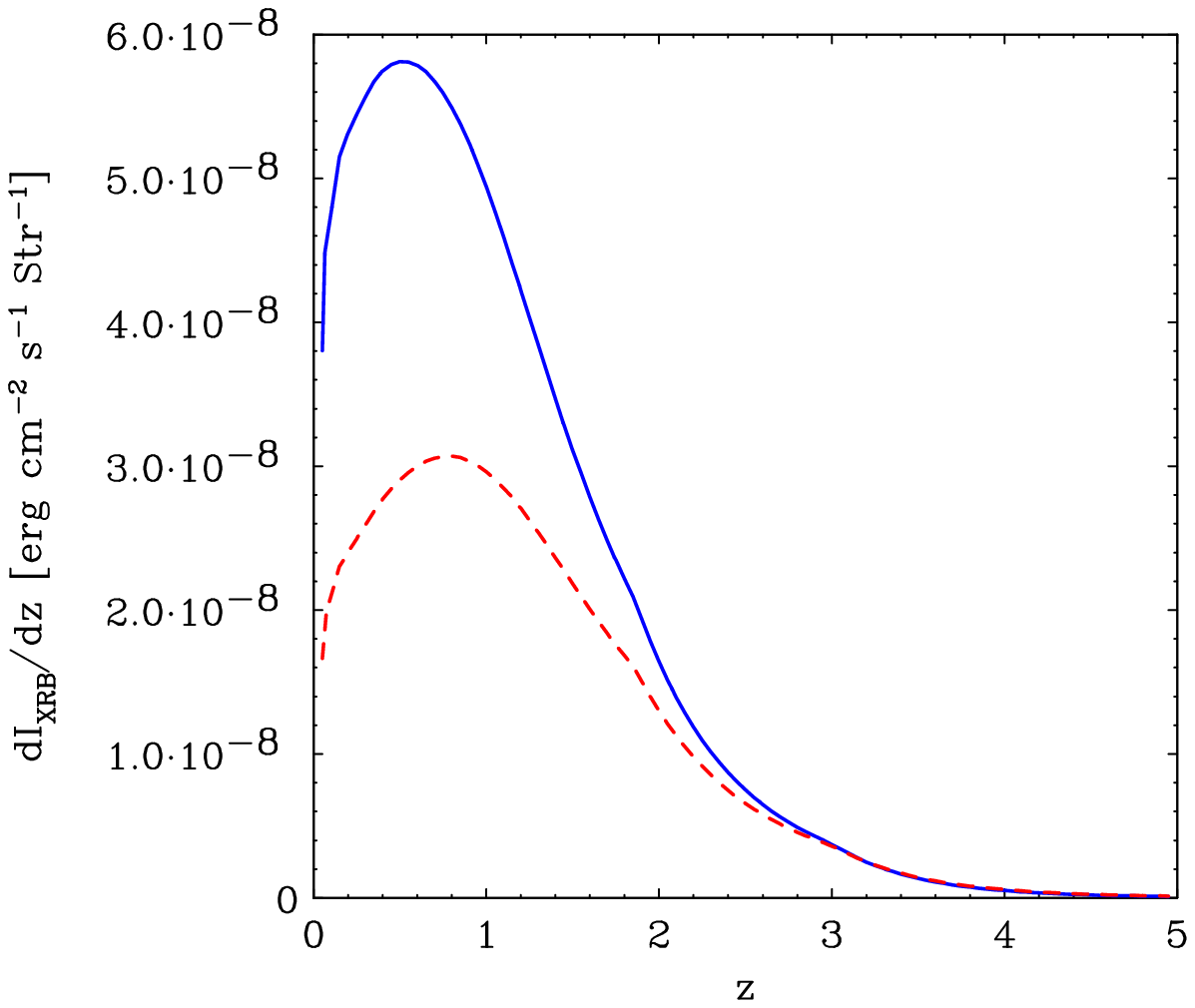}
\includegraphics[angle=0,scale=0.6]{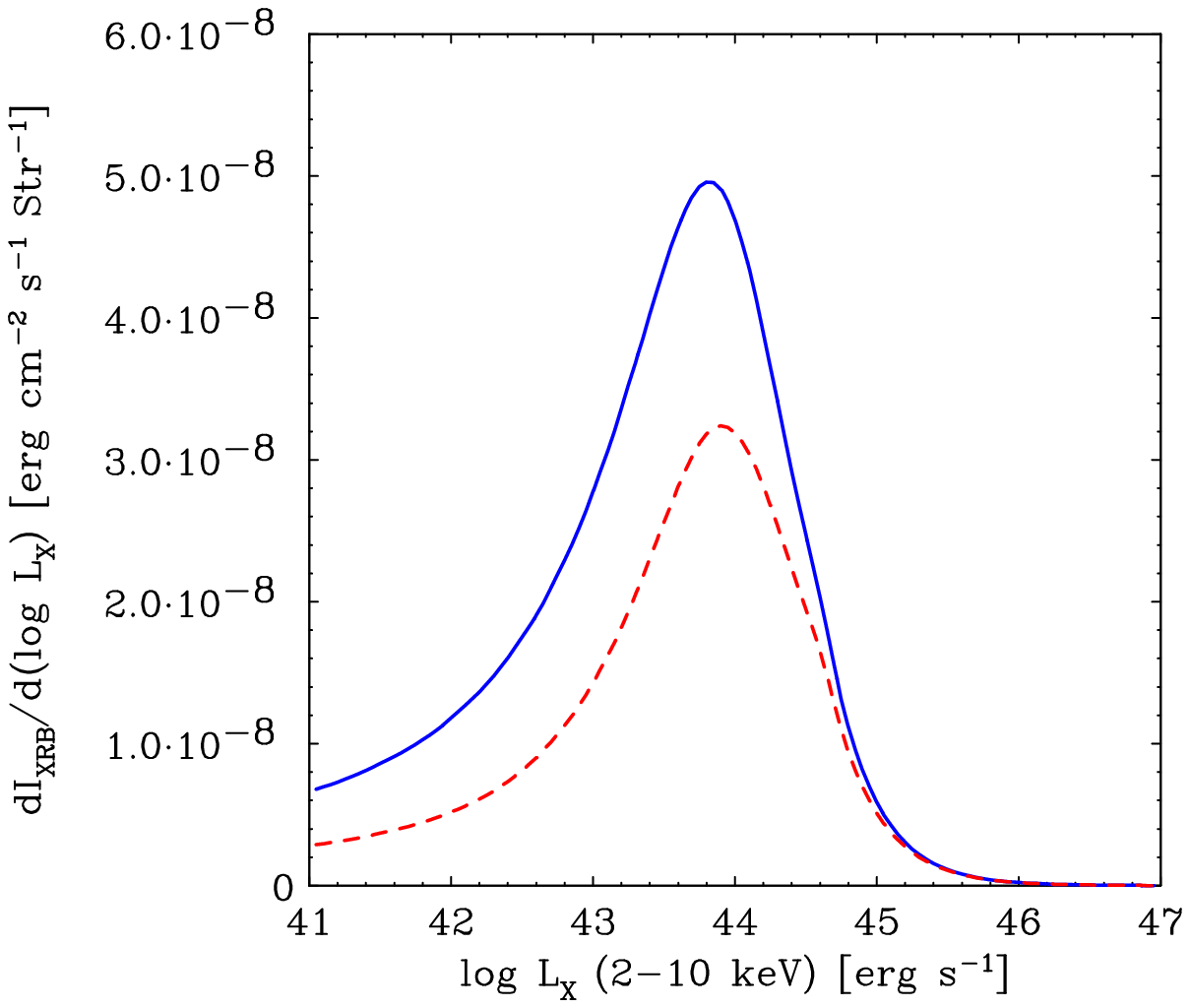}
\end{center}
\caption{
{\it Left}: 
 Differential contribution to the 2--10 keV (red dashed) 
or 10--40 keV (blue solid) XRB intensity as a function of redshift (in
 units of per $z$) from all (CTN+CTK) AGNs with log \lx\ = 41--47.
{\it Right}: that as a function of luminosity (in units of per log \lx )
from all AGNs at $z=0.002-5$.
}
\label{fig-cxbdist}
\end{figure}

The contributions from all (CTN+CTK) AGNs 
per unit $z$ per unit log \lx\ 
to
the XRB flux in the 2--10 keV and 10--40 keV bands are shown by the
contours in Figures~\ref{fig-cxbcontour} (left) and (right),
respectively. As noticed from the figures, AGNs with log \lx\
$\approx$43.8 ($\approx$43.7) at $z \approx$1.1 ($\approx$1.0) make
the largest contribution to the XRB in the 2--10 keV (10--40 keV)
band. Figures~\ref{fig-cxbdist} (left) and (right) plot the differential
XRB intensity per unit log \lx\ 
in a redshift region of
$z=0.002-5$, and that per unit $z$ 
in a luminosity region of
log \lx\ = 41--47, respectively.

\begin{figure}
\epsscale{1.0}
\begin{center}
\includegraphics[angle=0,scale=0.6]{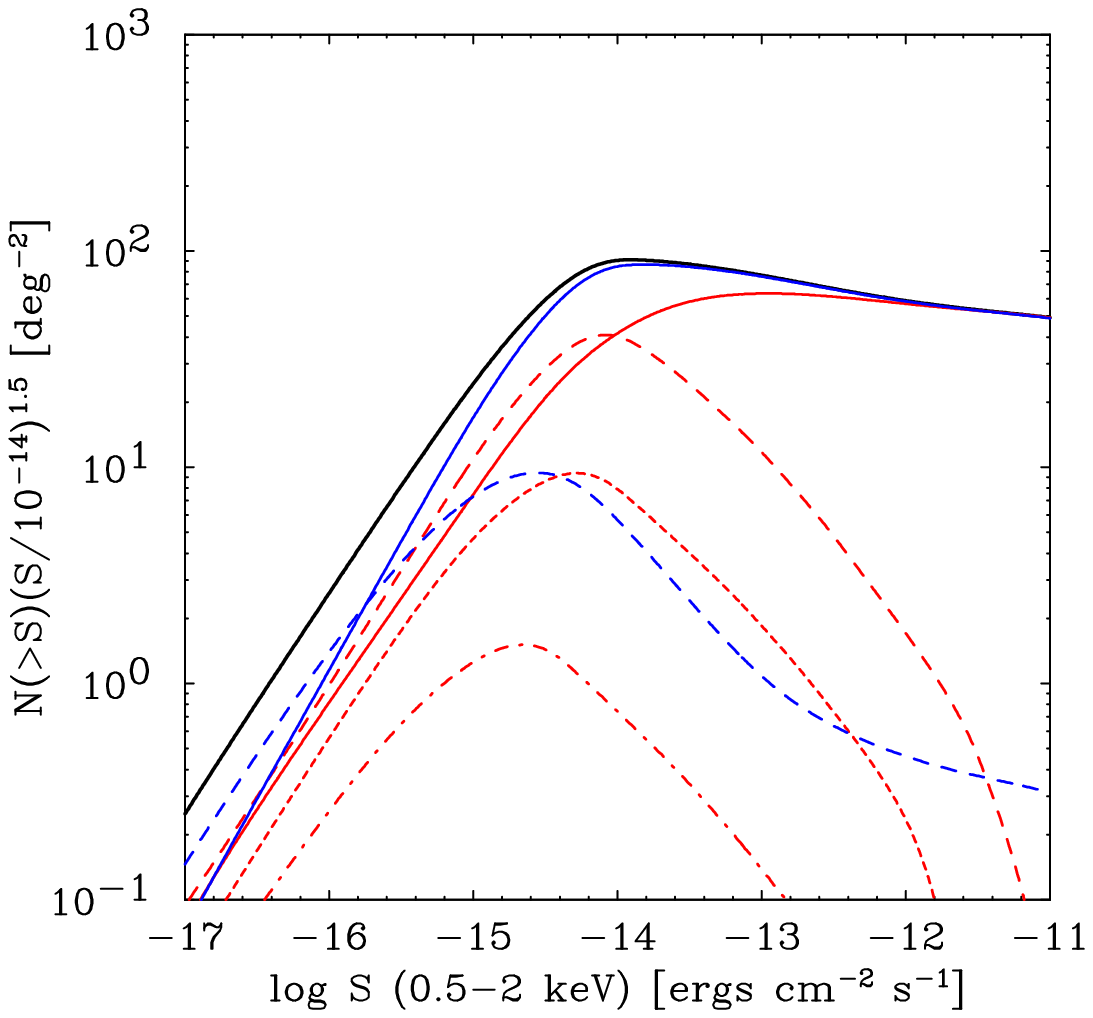}
\includegraphics[angle=0,scale=0.6]{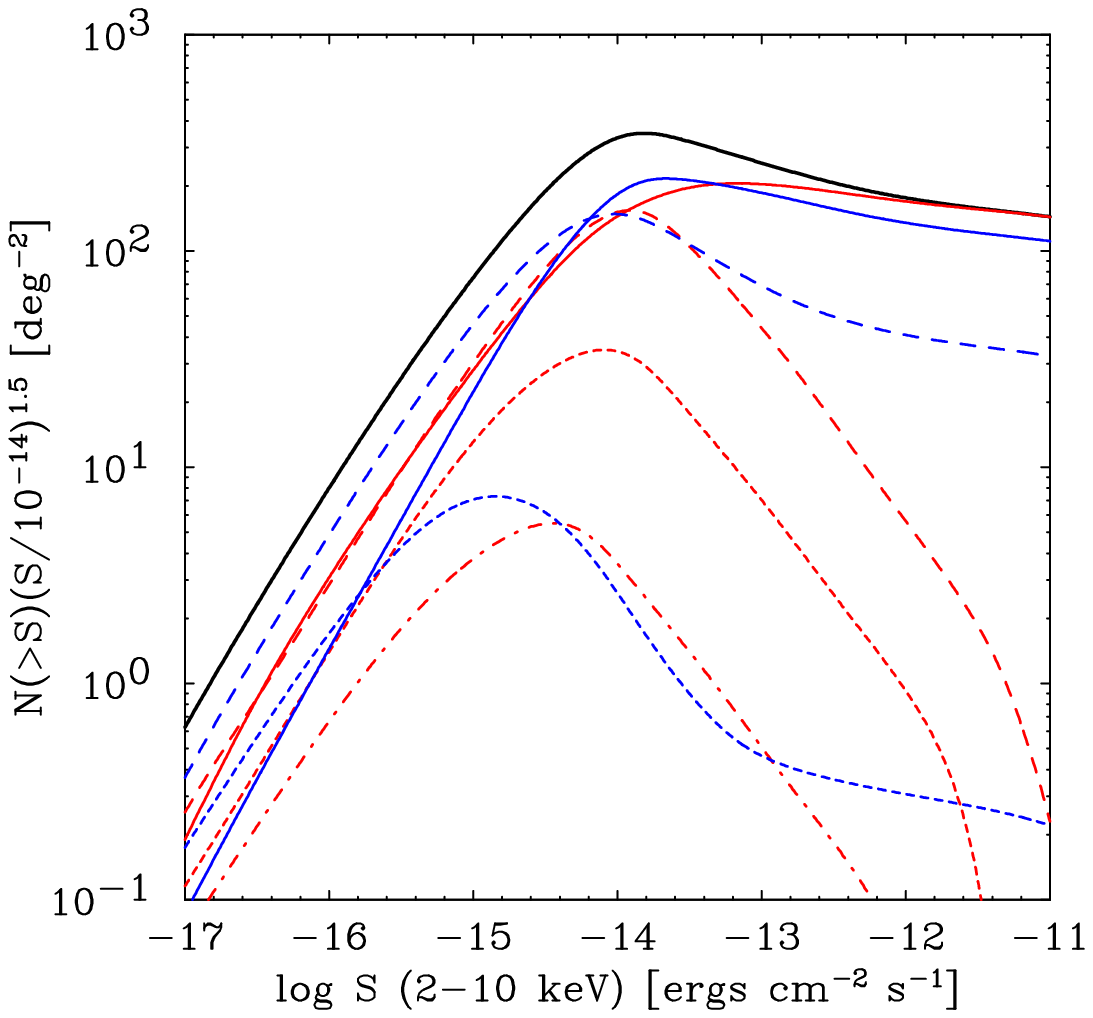}
\includegraphics[angle=0,scale=0.6]{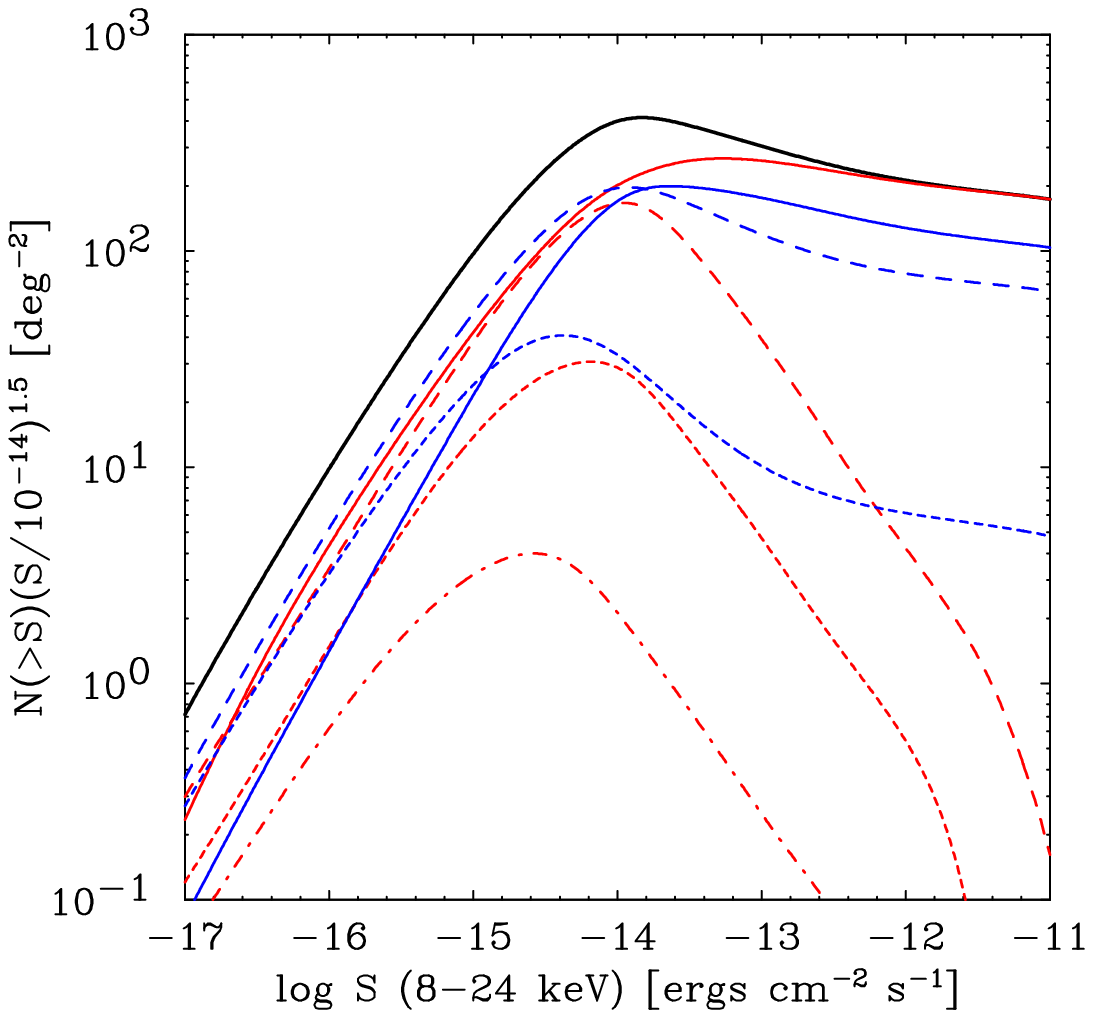}
\includegraphics[angle=0,scale=0.6]{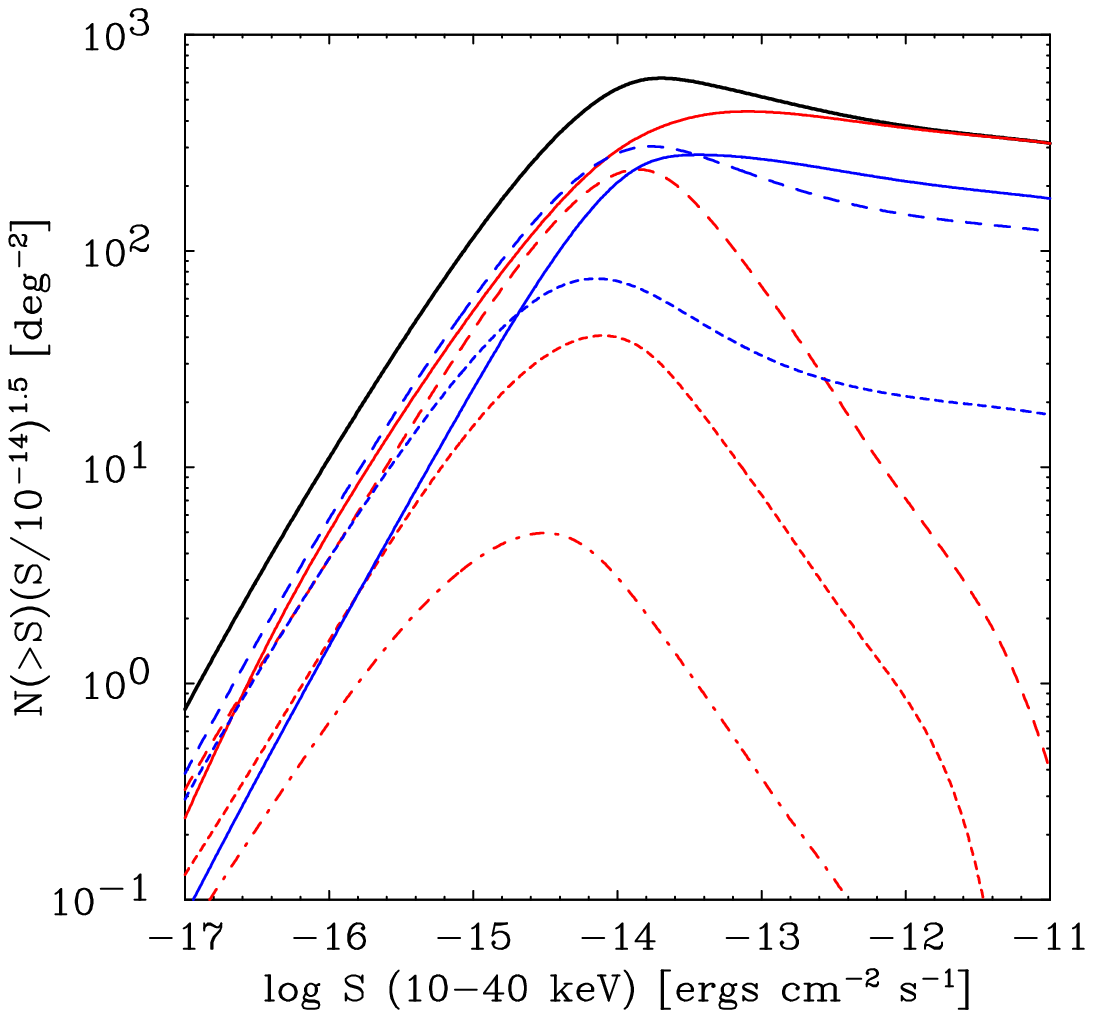}
\end{center}
\caption{
{\it upper left}: predicted integrated \logn s in the 0.5--2 keV band from our baseline model
normalized as $N(>S) (S/10^{-14})^{1.5}$. The red curves correspond to
those in different redshifts (solid: $z<1$, long-dashed: $z=1-2$,
short-dashed: $z=2-3$, dot-dashed $z=3-5$), and blue curves to those 
with different absorptions (solid: log \nh\ = 20--22, long-dashed: log
 \nh\ = 22--24, short-dashed: log \nh\ = 24--26).
{\it top right}: the same but for the 2--10 keV band.
{\it lower left}: the same but for the 8--24 keV band.
{\it lower right}: the same but for the 10--40 keV band.
}
\label{fig-logn-model}
\end{figure}

\begin{figure}
\epsscale{1.0}
\begin{center}
\includegraphics[angle=0,scale=0.7]{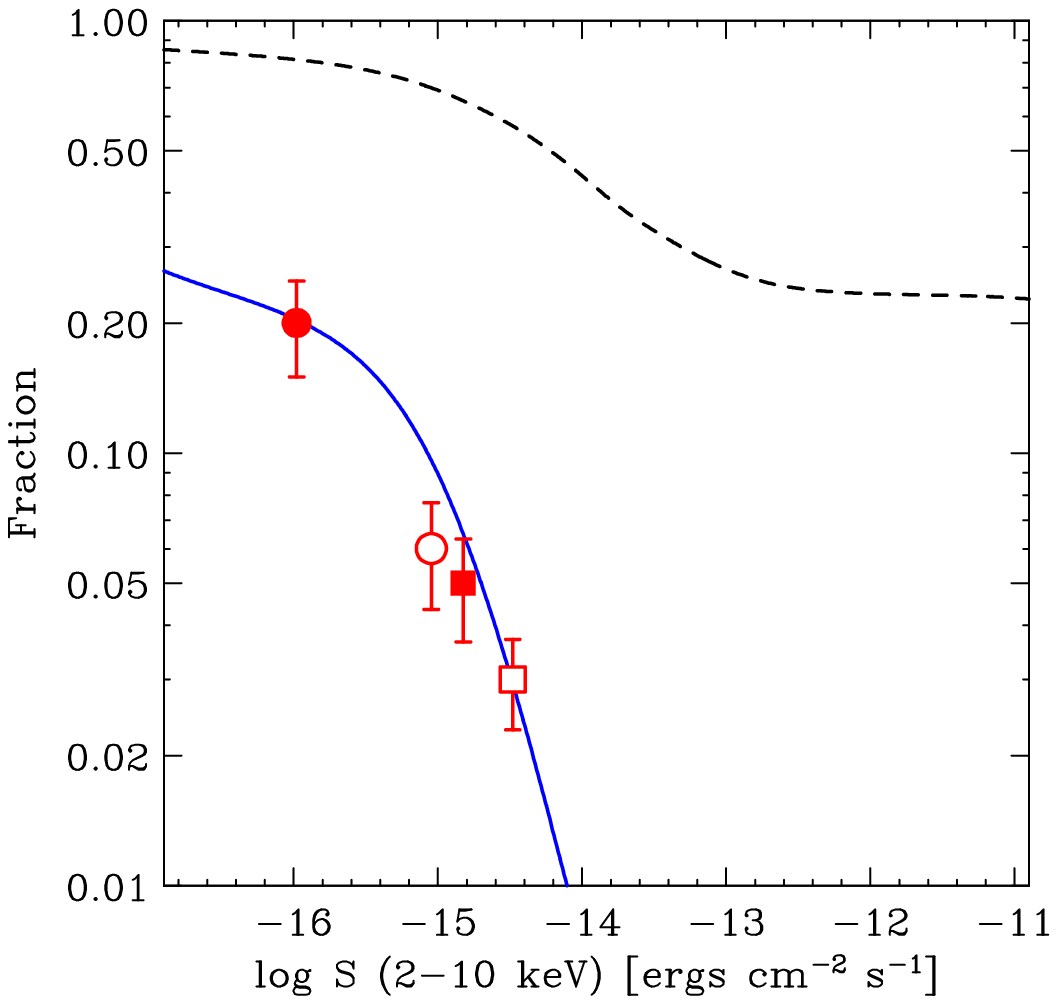}
\includegraphics[angle=0,scale=0.7]{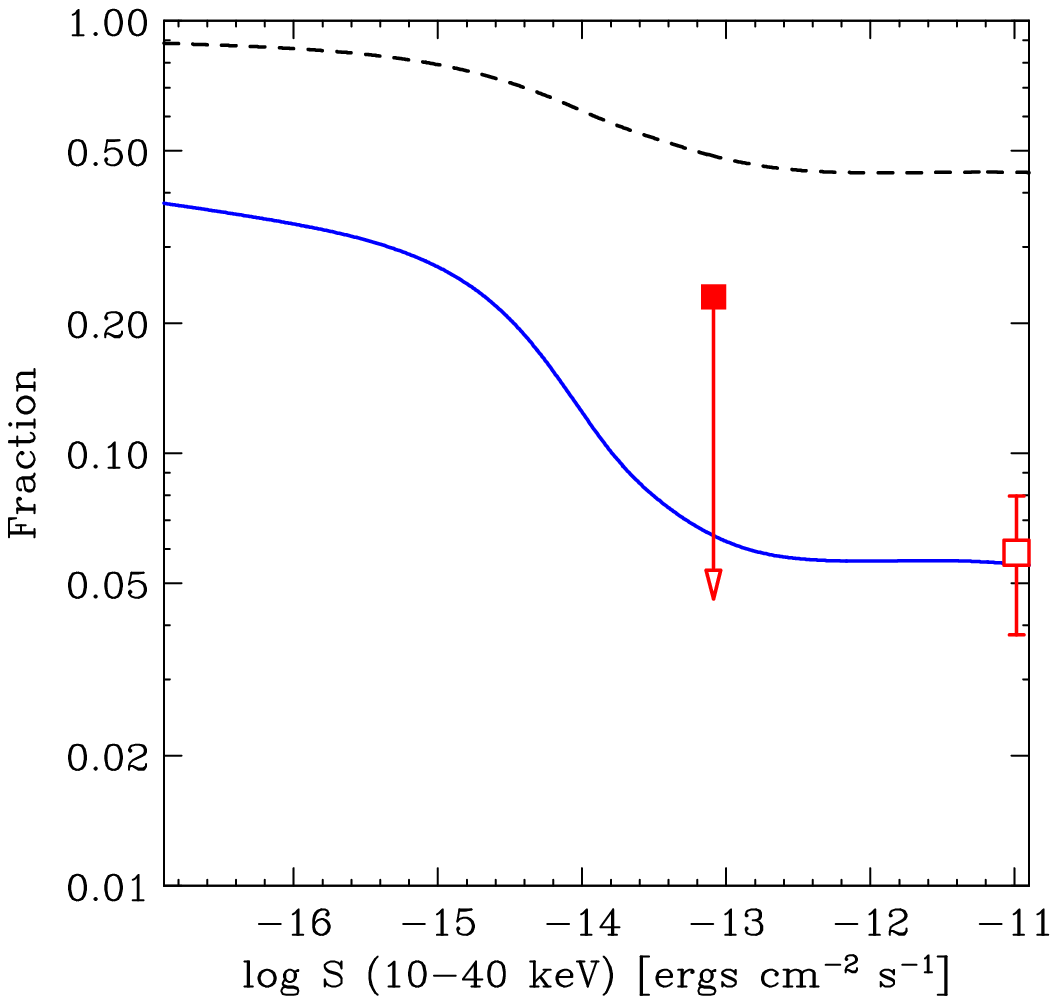}
\end{center}
\caption{
{\it Left}: fractions of CTK AGNs (log \nh\ = 24--26, solid blue) and obscured
AGNs (log \nh\ = 22--26, dashed black) in the total AGNs given as a
function of flux in the 2--10 keV band, predicted from our baseline
model. The data points correspond to the observed CTK AGN fractions by
\citet{bri12} (filled circle), \citet{bru08} (open circle),
 \citet{toz06} (filled square), and \citet{has07} (open square), from
 left to right. Here the result by \citet{bri12} is plotted by
converting the 0.5--8 keV flux to the 2--10 keV one by assuming a photon
index of 1.4.
{\it Right}: the same but for the 10--40 keV flux. 
The arrow denotes the 90\% confidence upper limit of the CTK fraction
obtained by {\it NuSTAR} in the 8--24 keV band \citep{ale13}, and the
right data point is that from the \swift /BAT 9-month survey in the
14--195 keV band \citep{tue08,ich12}. The fluxes are converted into the
10--40 keV band by assuming a photon index of 1.8.
} 
\label{fig-ctfrac}
\end{figure}

The predicted \logn\ of AGNs in the 0.5--2 keV, 2--10 keV, 8--24 keV,
and 10--40 keV bands are plotted in Figure~\ref{fig-logn-model}.
We separately plot the contributions from AGNs at different
redshift ranges ($z<1$, $z$=1--2, $z$=2--3, and $z$=3--5) and from those
with different absorptions (log \nh\ = 20--22, 22--24,
24--26). Figure~\ref{fig-ctfrac} shows the fractions of CTK AGNs
(log \nh\ = 24--26) and obscured AGNs (log \nh\ = 22--26) in the total
AGNs (log \nh\ $\leq$ 26 ) as a function of flux predicted from surveys
in the 2--10 keV (left) and 10--40 keV (right) bands. The CTK AGN fraction reaches
$\approx$20\% at $S \sim 10^{-16}$ \ergs\ in the 2--10 keV band, the
flux limit of \chandra\ deep surveys. We find that the observed CTK AGN
fractions at various flux limits in the 2--10 keV (or 0.5--8 keV) band
reported by \citet{toz06}, \citet{has07}, \citet{bru08}, and
\citet{bri12} are generally in good agreements with the model prediction. 
In the 10--40 keV band, our model is consistent with the observed CTK
fraction at $S \sim 10^{-11}$ \ergs\ observed by the \swift /BAT 9-month
survey performed in the 14--195 keV band \citep{tue08,ich12}, and with
the upper limit ($<$0.23 at a 90\% confidence level) obtained from the
first {\it NuSTAR} extragalactic survey in the 8--24 keV band
\citep{ale13}. In our baseline model, the intrincic fraction of CTK AGNs
among the whole AGNs at log \lx\ = 43.75 is $30\pm2$\% at $z=0$,
$37\pm2$\% at $z=1$, and $42\pm2$\% at $z\geq2$, which are calculated as
$f_{\rm CTK}\psi_{43.75}(z)/[1+f_{\rm CTK }\psi_{43.75}(z)]$. They are
fully consistent with the results obtained by \citet{bri12} from the
CDFS data at $z>0.1$. Note that using the \swift/BAT 3-year survey,
\citet{bur11} report a slightly smaller CTK fraction of $20^{+9}_{-6}$\%
than the above value, though within the errors, because they do not
include heavily CTK AGNs with log \nh\ $>25$.

\begin{figure}
\epsscale{1.0}
\begin{center}
\includegraphics[angle=0,scale=0.8]{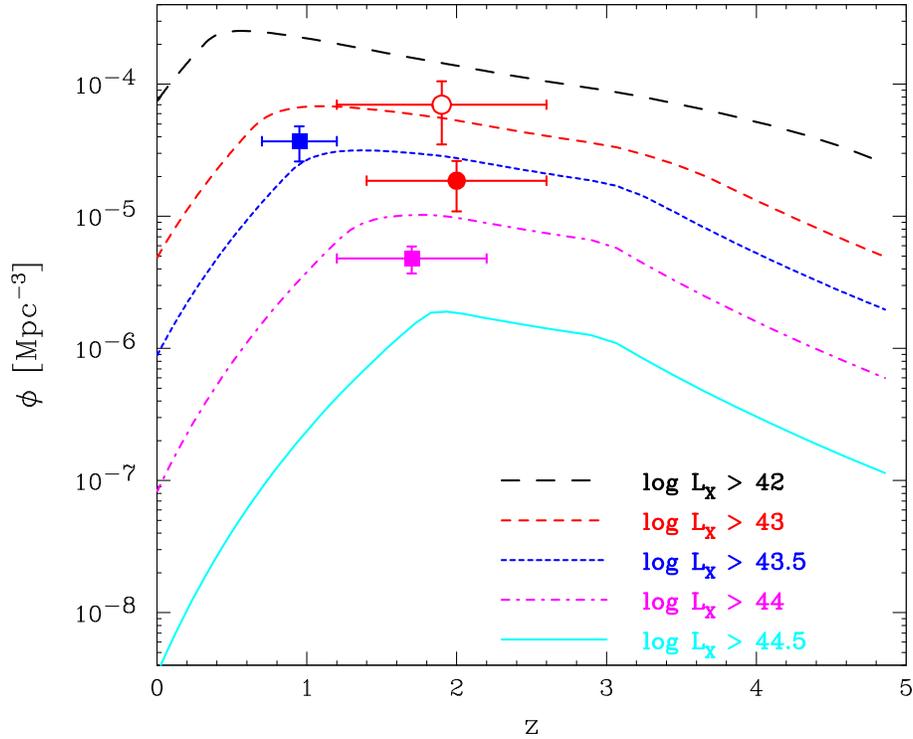}
\end{center}
\caption{
Comoving number density of CTK AGNs with different lower limits for
the X-ray luminosity predicted from our baseline model. The data points
represent the estimates by \citet{ale11} for log \lx\ $\simgt 43$
(filled circle, red), \citet{fio08} for log \lx\ $>43$ (open circle,
red), and \citet{fio09} for log \lx\ $>43.5$ (filled square, blue) and
for log \lx\ $>44$ (filled square, magenta).
} \label{fig-ctzf}
\end{figure}

Figure~\ref{fig-ctzf} plots the comoving number density of CTK AGNs with
different lower luminosity limits as a function of redshift predicted
from the baseline model. For comparison, the estimates from X-ray
stacking analyses obtained by \citet{fio08,fio09} are over-plotted.
The result by \citet{fio08} at $z=1.2-2.6$ 
for log \lx\ $> 43$ (open circle)
well agrees with our model.
More recent results reported by \citet{fio09} from the COSMOS data
(filled squares) at $z=0.7-1.2$ (log \lx\ $> 43.5$) and at $z=1.2-2.2$
(log \lx\ $> 44$) are within a factor of $\sim$2 from our baseline
model, which would be acceptable by considering possible uncertainties
in the luminosity range of the samples.
In the figure, we also plot the estimate based on X-ray detected heavily
obscured AGNs in the CDFS at $z=1.4-2.6$ with log \lx\ $\simgt 43$
obtained by \citet{ale11} (filled circle), who updated the \citet{dad07}
result using deeper X-ray data and new analyses. Our prediction is by a
factor of $\sim$3 higher than their result, which should be regarded as
a conservative lower limit \citep{ale11}. 

\subsection{Constraints on Compton-Thick AGN Fraction}
\label{sec-model-dependence}

As desribed above, our population synthesis model has the following
parameters that are fixed in the main analysis of Section~\ref{sec-lf}:
(1) the fraction of CTK AGNs \fct\ = 1.0, (2) the strength of the
reflection component from the accretion disk \rdisk\ = 0.5, (3) mean
photon index and its scatter of $\overline{\Gamma_1}=1.94$ and
$\Delta\Gamma_1=0.09$ for type-1 AGNs and $\overline{\Gamma_2}=1.84$ and
$\Delta\Gamma_2=0.15$ for type-2 AGNs. In this subsection, we evaluate
the dependence of model predictions on these fixed parameters and
discuss constraints on the fraction of CTK AGNs.
As the boundary condition that must be reproduced from the XRB model, we use
the XRB intensity integrated in the 20--50 keV band, \IXRB. Considering
the systematic uncertainties between different missions (see Table~2 of
\citealt{aje08}), we conservatively adopt \IXRB\ =
$(5.7-6.7)\times10^{-8}$ \ergss\ as the constraint; the minimum and
maximum values are obtained by \sax\ \citep{fro07} and \integral\
\citep{chu07}, respectively, when we adopt \citet{gru99} as the \heao 's
result. We also check the AGN source counts in the 2--8 keV band at a
representative flux of $S=2.7\times10^{-16}$ \ergs, which sensitively
depends on the assumed \fct\ parameter, to be compared with the
\chandra\ result obtained by \citet{leh12}, $N(>S=2.7\times10^{-16}) =
4290\pm240$ deg$^{-2}$ (for AGN only). At fainter fluxes, the contribution of
normal galaxies becomes more important, which are difficult to be
unambiguously distinguished from AGNs \citep{xue11}.

\begin{figure}
\epsscale{1.0}
\begin{center}
\includegraphics[angle=0,scale=0.8]{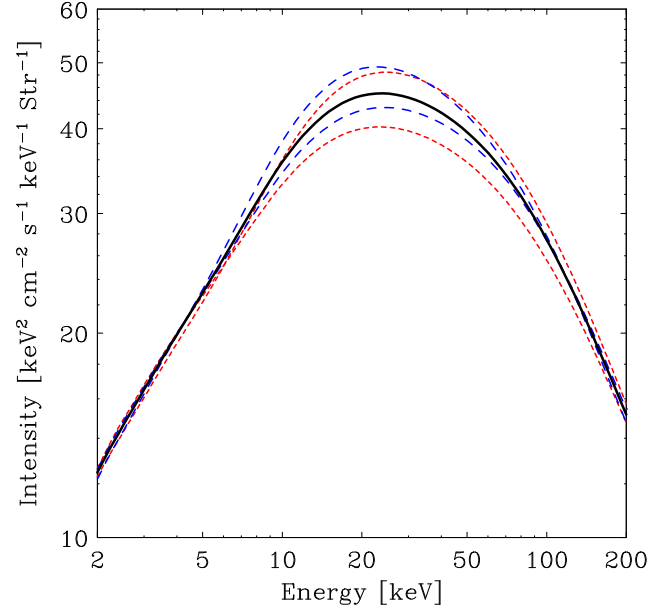}
\end{center}
\caption{
The predicted XRB spectra by assuming different fractions of CTK AGN
relative to obscured CTN AGNs (short-dashed, red; \fct\ = 2 and 0.5 from
upper to lower), or different reflection strengths from the accretion
disk (long-dashed, blue; \rdisk\ = 1.0 and 0.25 from upper to
lower). The baseline model (\fct\ = 1.0 and \rdisk\ = 0.5) is plotted by
the solid line (solid, black).  } 
\label{fig-cxbspec-comp}
\end{figure}

Since these fixed parameters affect the fitting results of the XLF and
absorption function, we repeat ML fit to the list of our AGN sample by
replacing the default parameters with different values. For simplicity,
only one set of parameters (i.e., either \fct, \rdisk,
$\overline{\Gamma_{1,2}}$, or $\Delta\Gamma_{1,2}$) is changed from the
default values. This enables us to etimate the error for a single
parameter by ignoring the coupling between
them. Table~\ref{tab-predictions} summarizes the results obtained for
different values of the fixed parameters. Since we find the XLF
parameters are not significantly different over the statistical errors,
we only show the evolution index in the absorption function $a1$. The
predicted XRB intensity in the 20--50 keV band and the 2--8 keV source
count at $S=2.7\times10^{-16}$ \ergs\ are
listed. Figure~\ref{fig-cxbspec-comp} compares the integrated spectra
for the cases of \fct\ = 0.5 and \fct\ = 2.0 (short-dashed, red) and
\rdisk\ = 0.25 and \rdisk\ = 1.0 (long-dashed, blue) with our baseline
model (black). Taking the 20--50 keV XRB intensity \IXRB\ =
$(5.7-6.7)\times10^{-8}$ \ergss\ as the observational constraint, 
we constrain that \fct\ = 0.5--1.6 in
 the case of \rdisk\ = 0.5; for this range of \fct, we confirm that the
predicted source count at $S=2.7\times10^{-16}$ in the 2--8 keV band is
consistent with the observed one, $N(>S=2.7\times10^{-16}) =4290\pm240 $
deg$^{-2}$. As discussed in many previous works, there are degeneracies
between the estimate of CTK AGN fraction and the strength of Compton
reflection components in order to reproduce the XRB spectrum.  From
results listed in Table~\ref{tab-predictions}, we can roughly estimate
that the best-estimate of \fct\ will be changed by $+50\%$ and $-50\%$
when we assume \rdisk\ = 0.25 and 1.0, respectively, although the choice
of \rdisk\ = 0.5 in our baseline model is the most reasonable from
observations of local AGNs (Section~\ref{sec-template}).

\subsection{Comparison with Previous XRB Models}
\label{sec-model-comparison}

We compare our new population synthesis model of the XRB with major
previous works published after 2003: U03, \citet{bal06}, \citet{gil07},
\citet{tre09}, and \citet{aky12}. All these models, including ours, assume
that the CTK AGNs follow the same cosmological evolution of CTN AGNs and
introduce the \fct\ (or its equivalent)
parameter. Table~\ref{tab-models} summarizes the details of the
ingredients in each model: the XLF, the evolution index of the
absorption fraction $a1$ ($a1=0$ if no evolution), \fct\ with the range
of column density of CTK AGNs, spectral parameters ($\overline{\Gamma}$
and $\Delta\Gamma$), reflection strength, high energy cutoff), and the
predicted XRB intensity at 25 keV. \citet{aky12} intensively explore the
degeneracies between these parameters by fitting the XRB spectra with
the model predictions. In Table~\ref{tab-models}, only a representative
set of the parameters that fits the XRB data are listed (taken from
their Figure~1).

All these works except \citet{gil07} utilize the 2--10 keV XLF of the
whole CTN AGNs obtained by U03. \citet{gil07} basically adopt the 0.5--2
keV XLF of type-1 AGNs derived by H05, and determine a
luminosity-dependent (but redshift independent) ratio between obscured
and unobscured AGNs by fitting the data points of the 2--10 keV XLF
obtained by U03 and \citet{laf05}. \citet{bal06} and \citet{tre09}
take into account the evolution of the absorption fraction. The scatter
in the photon index distribution is considered in the \citet{gil07} and
\citet{aky12} models. All these authors adopt slightly different values
of the high energy cutoff and Compton reflection strength.

A validity of the models can be checked by the predicted XRB intensity.
\citet{bal06} model significantly overproduces the XRB intensity at 25
keV when compared with recent measurements by \swift/BAT and
\integral. The same problem existing in the earlier model by
\citet{tre05} is corrected in \citet{tre09}, where a very small CTK
AGN fraction (\fct\ = 0.17) is assumed on the
basis of hard X-ray ($>$10 keV) surveys in the local universe. However,
after correcting for biases against detecting heavily CTK AGNs even in
the hard X-ray band 10 keV, the intrinsic fraction of CTK AGNs could be
much larger than the value assumed in \citet{tre09}
(Section~\ref{sec-abs}; see also \citealt{bur11}). 

Our model thus supersedes the older models, and may be regarded as a
standard population synthesis model of the XRB at the current stage.
The biggest advantage is that it utilizes the most precise XLF and
absorption function that depends both on luminosity and redshift.  Our
model also takes into account the broad band spectra including the
reflection components from the tori based on the ``luminosity and
redshift dependent unified scheme'' as well as the photon index
distributions that are different between type-1 and type-2 AGNs. The
whole analysis has been performed self-consistently on the basis of
these assumptions. Compared with \citet{gil07}, we predict a higher
fraction of obscured and CTK AGNs at faint fluxes due to the inclusion
of the redshift evolution in the absorption fraction
(Figure~\ref{fig-ctfrac}). We note that \citet{dra09} report a possible
contribution of blazars to the XRB, which is ignored in our
model. According to their model with an X-ray duty cycle of 13\%, the
integrated emission of blazars can account for $\sim$2\% of the XRB at
20 keV, which is much smaller than the current uncertainties in its
absolute intensity. The estimate should be largely uncertain, however,
as the model significantly overpredicts the blazar source counts
obtained with \swift/BAT \citep{dra09}.

\section{Bolometric Luminosity Function of AGNs and Growth History of SMBH}
\label{sec-blf}

\subsection{Bolometric Luminosity Function}

The luminosity function of the whole AGN populations provides a basis
for understanding the growth history of SMBHs in galactic
centers. While hard X-rays are an ideal energy band for a complete
survey of AGNs with little contamination, they represent only a
limited fraction of the total radiation energy emitted from an AGN,
whose SED has a peak around the ultra-violet band. Thus,
it is very convenient to determine the ``bolometric'' luminosity
function (BLF) of all AGNs (including both CTN and CTK AGNs) based on
the XLF. The BLF is defined as a function of bolometric luminosity
\lbol\ (instead of \lx ) and $z$ so that
\begin{equation} 
\frac{d \Phi_{\rm bol} (L, z)}{{d \rm log} L}
\end{equation}
gives the comoving spatial number density per ${\rm log} L$.
\citet{hop07} derived a BLF of AGNs by simultaneously analyzing multiple
AGN surveys performed in the X-ray, optical, and mid infrared bands.
In this section, we derive the AGN BLF directly from our new XLF by
taking into account the luminosity dependence bolometric correction
and its scatter that are estimated by \citet{hop07}.
The full revision of the work of \citet{hop07} by utilizing
AGN multi-band luminosity functions
other than the XLF is beyond the scope of this
paper. To combine multiwavelength data sets, it is crucial to evaluate
the selection function against obscuration, which is not necessarily
trivial given the complex situation of each survey.

\begin{figure}
\epsscale{1.0}
\begin{center}
\includegraphics[angle=0,scale=0.5]{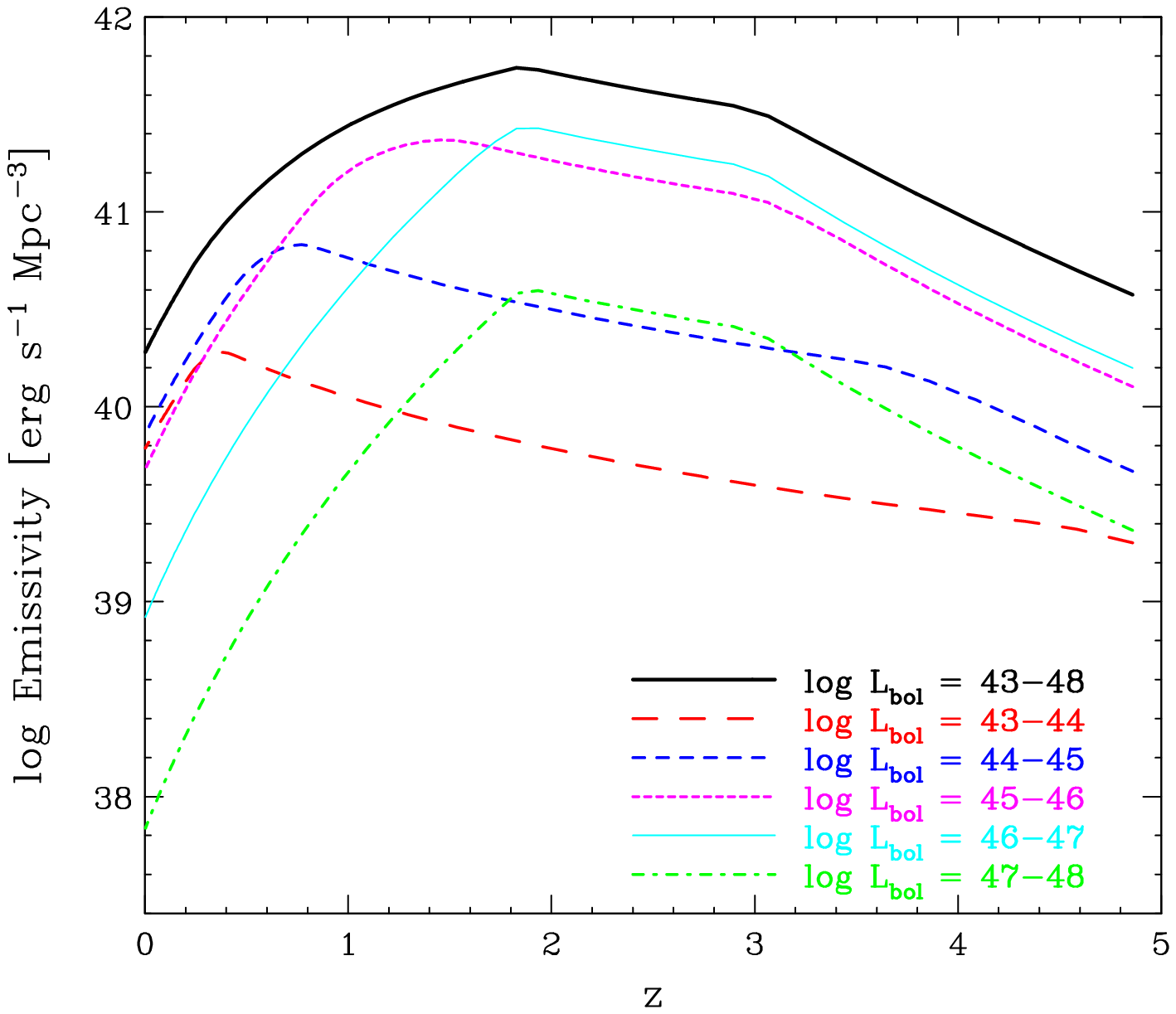}
\includegraphics[angle=0,scale=0.5]{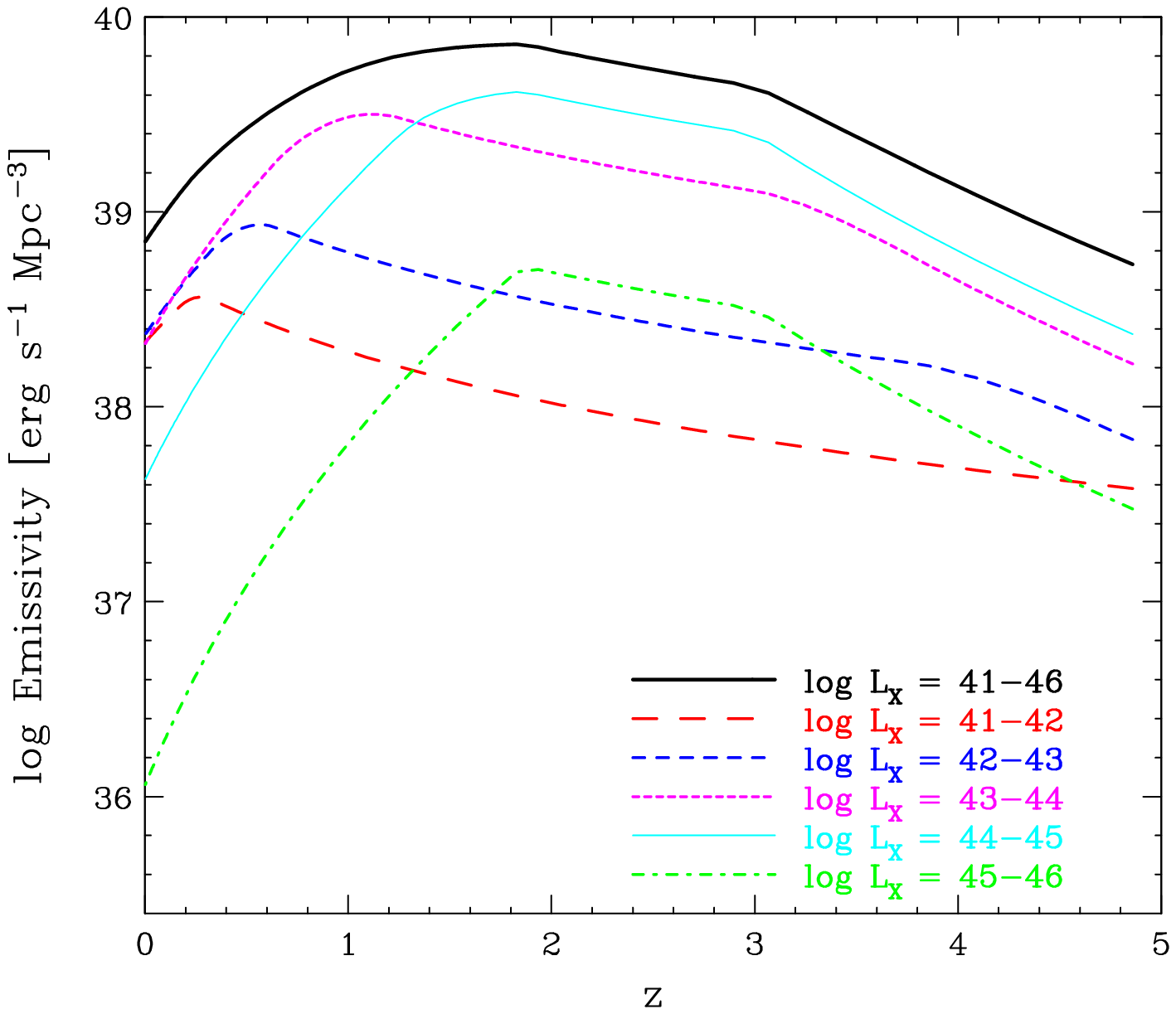}
\end{center}
\caption{
{\it Left}: comoving bolometric luminosity density (emissivity) in different
luminosity ranges calculated from our baseline model. 
{\it Right}: that in the 2--10 keV band. }
\label{fig-bhmfld}
\end{figure}

We define the bolometric correction factor $k \equiv L/L_{\rm X}$ 
to convert from an X-ray luminosity into a bolometric one.
According to \citet{hop07}, 
its average $\overline{k}$ is a function of \lbol\ and 
is represented as 
\begin{equation}
\label{eq-bolcor}
\overline{k}(L) = 10.83 (\frac{L}{10^{10}\solarlum})^{0.28} + 6.08 (\frac{L}{10^{10}\solarlum})^{-0.020}.
\end{equation}
The standard deviation in log $k$ is also given as a function of \lbol;
\begin{equation}
\sigma_{{\rm log}k} (L) = 0.06 (L/10^9 \solarlum)^{0.10} + 0.08.
\end{equation}
To determine the BLF from our data, we take an approximated approach
instead of performing detailed calculations as done in Section~\ref{sec-lf}.
A BLF of all AGNs can be converted into the XLF of CTN AGNs by
assuming that the logarithm of the bolometric correction factor has a
Gaussian distribution;
\begin{eqnarray}
&&\frac{d\Phi^{\rm CTN}_{\rm X} (L_{\rm X}, z)}{d {\rm log} L_{\rm X}} \nonumber
\\
 = &&\int 
\frac{d\Phi_{\rm bol}(L, z)}{d {\rm log} L}
\frac{1}{\sqrt{2\pi}\sigma_{{\rm log}k}}
\exp [-\frac{({\rm log}(k/\overline{k}(L)))^2}{2\sigma_{{\rm log}k}^2}] d {\rm log} k.
\end{eqnarray}
Once the XLF is obtained, we calculate the predicted number of
detectable AGNs in our surveys as a function of \lx\ and $z$ by simply
correcting for the ratio of the XLF value to the best-fit one
presented in Section~\ref{sec-lf-results}, 
\begin{equation}
N (L_{\rm X},z) = 
N_{\rm best}(L_{\rm X},z) 
\frac{d\Phi^{\rm CTN}_{\rm X}(L_{\rm X}, z)/d {\rm log} L_{\rm X}}
{d\Phi^{\rm CTN}_{\rm X,best}(L_{\rm X}, z)/d {\rm log} L_{\rm X}},
\end{equation}
which can be compared with the observed number of AGNs. 
By dividing the \lx\ and $z$ plane within the range of \lx\ =
$10^{41}-10^{47}$ and $z = 0.002-5.0$ into 120$\times$136 logarithmic
bins, respectively, we perform the Poisson maximum likelihood fit to
the binned data because the numbers of sources in each pixel are small,
often less than 10. Here we adopt the same analytic form for the BLF as
for the XLF by setting log $L_{\rm p}$ = log $L_{\rm a2}$ = 45.67, a
bolometric luminosity that corresponds to log \lx\ = 44 with the
conversion given in Equation~(\ref{eq-bolcor}).
In the fit, we fix $z^{*}_{\rm c1} = 1.86$ and $\alpha_1=0.29$, the
best-fit values of the XLF, while the other parameters are left
free. The resultant best-fit parameters of the BLF are summarized in
Table~\ref{table-lf}. 
Figure~\ref{fig-bhmfld} (left) plots the bolometric
luminosity density (i.e., emissivity) of all AGNs $\int L (d \Phi_{\rm
bol} (L,z)/d {\rm log} L) d {\rm log} L$ as a function of redshift
integrated in different luminosity ranges. For reference, those in the
2--10 keV band based on the XLF are plotted in Figure~\ref{fig-bhmfld}
(left). In both Figures, the integrated emissivity has a peak around $z
\sim 2$, where AGNs with log $L = 46-47$ or log \lx\ $ = 44-45$ make the
largest contribution. We note that the peak redshift of the integrated
emissivity is significantly larger than $z \approx 1.2$ predicted from
the LADE model by \citet{air10} (see their Figure~11). This reflects the
fact that our LDDE model gives a larger number of luminous AGNs with log
\lx\ $\simgt$ 44 than the LADE model at $z \simgt 1$.

\subsection{Evolution of Mass Function of SMBHs}

As mentioned in Section~\ref{sec-intro}, an AGN is the process where a
SMBH gains its mass by accretion and hence the AGN luminosity 
function records the growth
history of SMBHs. A bolometric luminosity $L$ can be related to the mass
accretion rate onto a SMBH, $\dot{M}_{\rm acc}$, as $L = \eta
\dot{M}_{\rm acc} c^2$, where $\eta$ is the mass-to-energy conversion
factor (or radiation efficiency). The $\eta$ value is predicted to be
0.054 for a standard disk around a non-rotating black hole and becomes as large
as 0.42 for that with a maximum spin. In radiatively inefficient
accretion flows (RIAFs), it could be significantly smaller. Hence, in
general, the averaged value of $\eta$ could depend on parameters like
black hole mass $M$ and $z$.

\begin{figure}
\epsscale{1.0}
\begin{center}
\includegraphics[angle=0,scale=0.8]{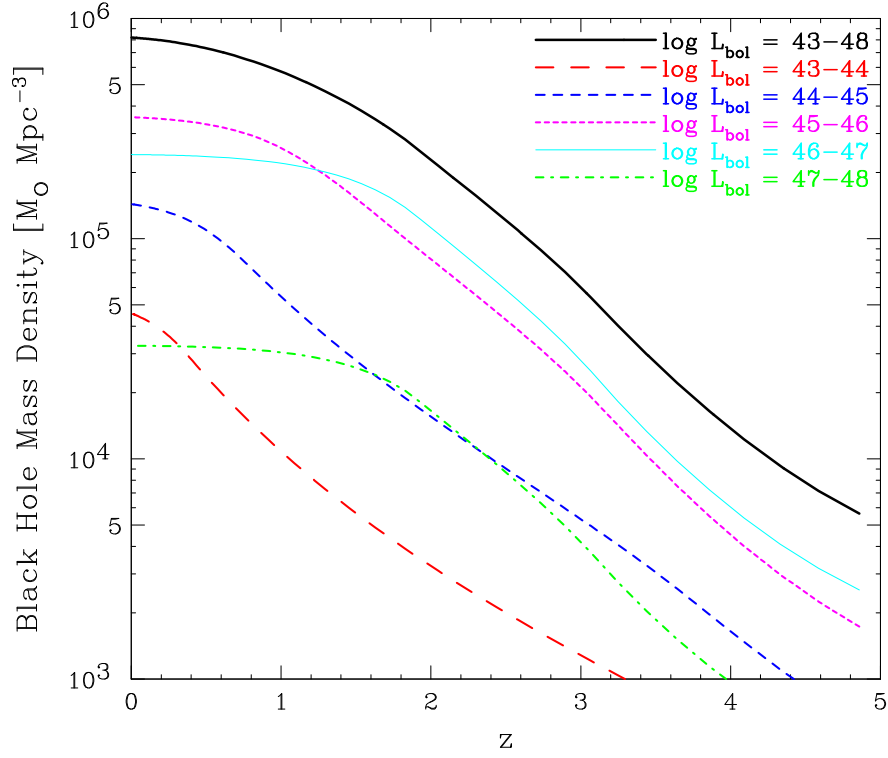}
 \end{center}
\caption{
Comoving mass density of all SMBHs plotted against redshift
(uppermost solid curve, black). Those calculated within limited
luminosity ranges are separately shown. The averaged radiation efficiency of 
$\overline{\eta}=0.05$ is assumed.
}
\label{fig-bhmfrho}
\end{figure}

On the basis of Soltan's argument \citep{sol82}, one can estimate the total
mass density of SMBHs $\rho (z)$ as a function of redshift once the
BLF of AGN is known by using the following equation:
\begin{equation}
\rho (z) = \rho (z_{\rm s}) + \frac{1-\overline{\eta}}{\overline{\eta} c^2}
\int_{z}^{z_{\rm s}} dz\frac{dt}{dz}
\int_{L_{\rm min}}^{L_{\rm max}} L \frac{d \Phi_{\rm bol} (L,z)}{d {\rm log}
L} d {\rm log} L.
\end{equation}
Here $\rho (z_{\rm s})$ gives the initial mass density at $z=z_{\rm
s}$ from which the time integration starts, and $\overline{\eta}$
represents an averaged radiation efficiency, which 
is assumed to be independent of $z$ and $L$.
A detailed calculation using our model indicates that 
$\approx$74\% ($\approx$37\%) of
the total energy emitted by whole AGNs in the history of universe 
(hence the total mass of all SMBHs at $z=0$) was produced by obscured
accretion with log \nh\ = 22--26 (log \nh\ = 24--26).
The mass density of SMBHs in the local universe 
can be independently estimated
from the empirical relation between SMBH mass and
host-spheroid luminosity (or mass).
For instance, if we adopt the result by \citet{vik09} $\rho^{\rm obs}
(z=0) = (4.9\pm0.7)\times10^5 \solarmass$ Mpc$^{-3}$,
$\overline{\eta}=0.080^{+0.013}_{-0.009}$ is suggested. This confirms
earlier works based on the hard XLF of AGNs \citep[e.g.,][]{mar04,li12}.
As described below, however, \citet{kor13} have recently reported
that the SMBH masses of classical bulges and elliptical galaxies should be
revised upward by a factor of $\sim$2--4, which would lead 
to a reduction of $\overline{\eta}$ by a similar factor 
(see also \citealt{nov14}).
Figure~\ref{fig-bhmfrho} plots the results calculated with
$\overline{\eta}=0.05$ in different luminosity ranges, log \lbol\ =
43--48, 43--44, 44--45, 45--46, 46--47, and 47--48. 
We adopt $z_{\rm s}=5$
and estimate $\rho (z_{\rm s})$ by assuming that all SMBHs were AGNs
with a mean Eddington ratio of $10^{-1.1}$ (see below).

As studied by many authors \citep[e.g.,][]{sma92,yu02,mar04,sha04,tam06,
cao08,sha09,li12,sha13}, 
it is also possible to trace
the cosmological evolution of the mass function (MF) of all SMBHs
including both active (i.e., AGN) and non-active ones from the AGN 
luminosity function
(``AGN relic MF''). Here we define the MF of all SMBHs and that of
only AGNs as $N(z,M)$ and $N_{\rm AGN}(z,M)$, respectively, which
represent their comoving spatial number density per unit mass 
at redshift $z$. The Eddington ratio is given as $\lambda
\equiv L/L_{\rm Edd}$, where $L_{\rm Edd} = 1.25\times 10^{38}
(M/\solarmass) $ erg s$^{-1}$ is the Eddington limit for a mass of
$M$. Under the assumption that the merging of SMBHs can be ignored, we
can introduce the continuity equation of the MF of all SMBHs
\citep[e.g.,][]{sma92},
\begin{eqnarray}
\frac{\partial N(z,M)}{\partial z} \frac{dz}{dt}
 &=& - \frac{\partial}{\partial M}[N(z,M)<\dot{M}>] \\ 
 &=& - \frac{\partial}{\partial M}[ \frac{1-\eta(z,M)}{\eta(z,M)}
\frac{\overline{\lambda}(z,M) L_{\rm Edd} N_{\rm AGN}(z,M)}{c^2}],
\end{eqnarray}
where $<\dot{M}>$ is the averaged black hole growth rate of {\it all SMBHs}
and $\overline{\lambda}(z,M)$ gives the averaged Eddington ratio 
of {\it AGNs} with a mass of $M$ at redshift $z$.
The MF of AGNs can be calculated as
\begin{equation}
N_{\rm AGN}(z,M) = \frac{d {\rm log} M}{d M}
\int \frac{d \Phi_{\rm bol}(z,L)}{d {\rm log} L} 
P(\lambda|L,z) d {\rm log} \lambda,
\end{equation}
where $P(\lambda|L,z)$ is the Eddington ratio distribution function
per unit log $\lambda$ at a given luminosity $L$ and redshift $z$.
The averaged AGN Eddington ratio is then given as
\begin{equation}
\overline{\lambda}(z,M) = \int \lambda \frac{\Phi_{\rm bol}(z,L)}{N_{\rm AGN} (z, M)} P(\lambda|L,z) d {\rm log} \lambda.
\end{equation}
Following \citet{li12}, we assume that 
the Eddington ratio distribution function is log-normal,
\begin{equation}
P(\lambda|L,z) = 
\frac{1}{\sqrt{2\pi}\sigma_{{\rm log}\lambda}}
\exp [-\frac{({\rm log}(\lambda/\overline{\lambda}))^2}{2\sigma_{{\rm
log}\lambda}^2}], 
\end{equation}
and is independent of luminosity and redshift.

\citet{mar04} consider the simplest case where $\eta(M,z)$ is constant
and $P(\lambda|L,z)$ is a delta function at $\overline{\lambda}$
(i.e., $\sigma_{{\rm log}\lambda} = 0.0$). Using the BLF converted
from the U03 XLF, they find that $\eta\sim0.1$ and $\lambda\sim 1.0$
to explain 
the observed MF of all SMBHs at $z=0$ 
with the AGN relic MF. \citet{tam06} show that the SMBH MFs at several
redshifts of $z=0-1.05$ derived from early-type galaxy luminosity
functions are broadly consistent with the AGN relic ones calculated with
$\eta=0.1$ and $\lambda=1.0$.

\begin{figure}
\epsscale{1.0}
\begin{center}
\includegraphics[angle=0,scale=0.5]{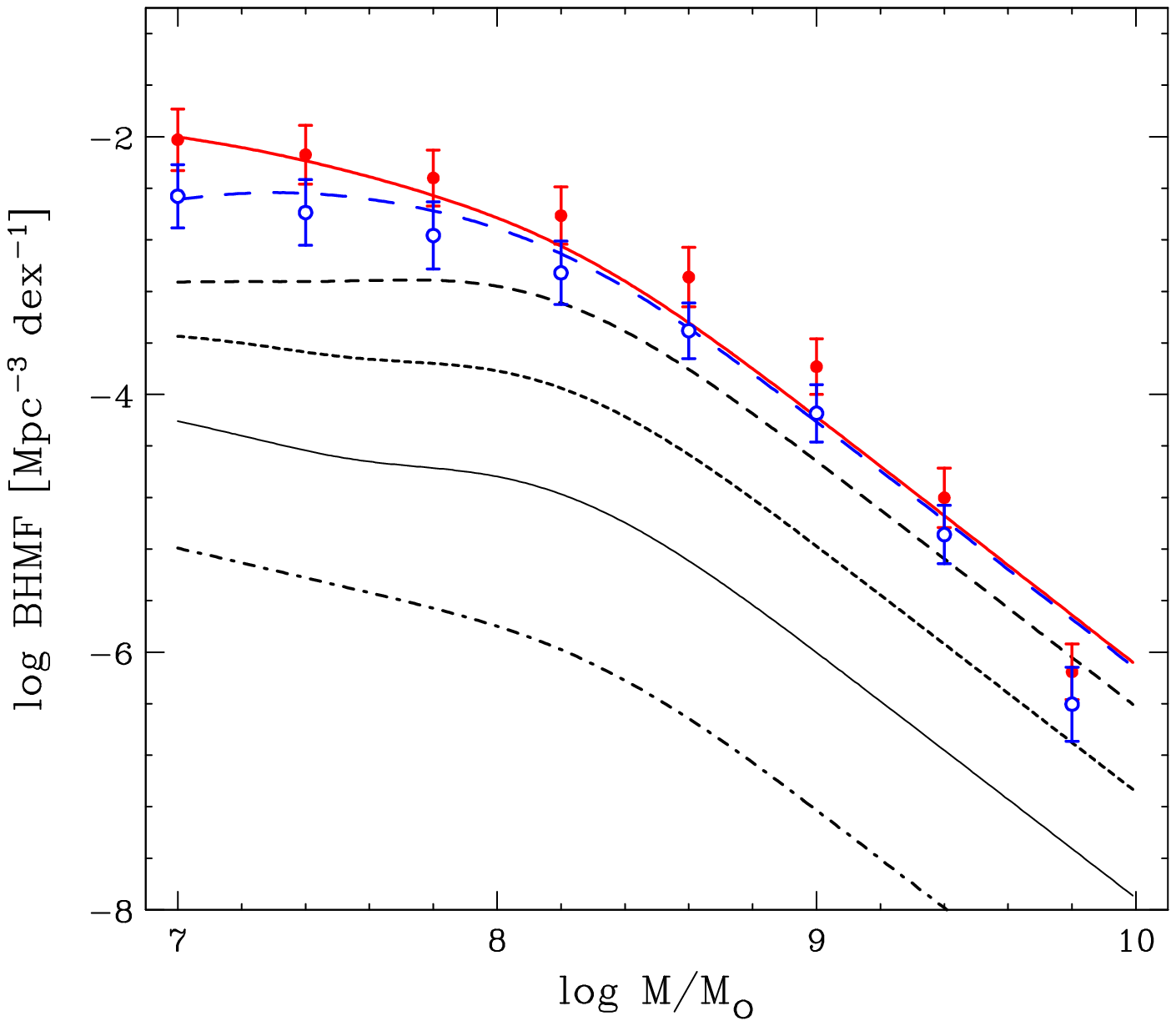}
\includegraphics[angle=0,scale=0.5]{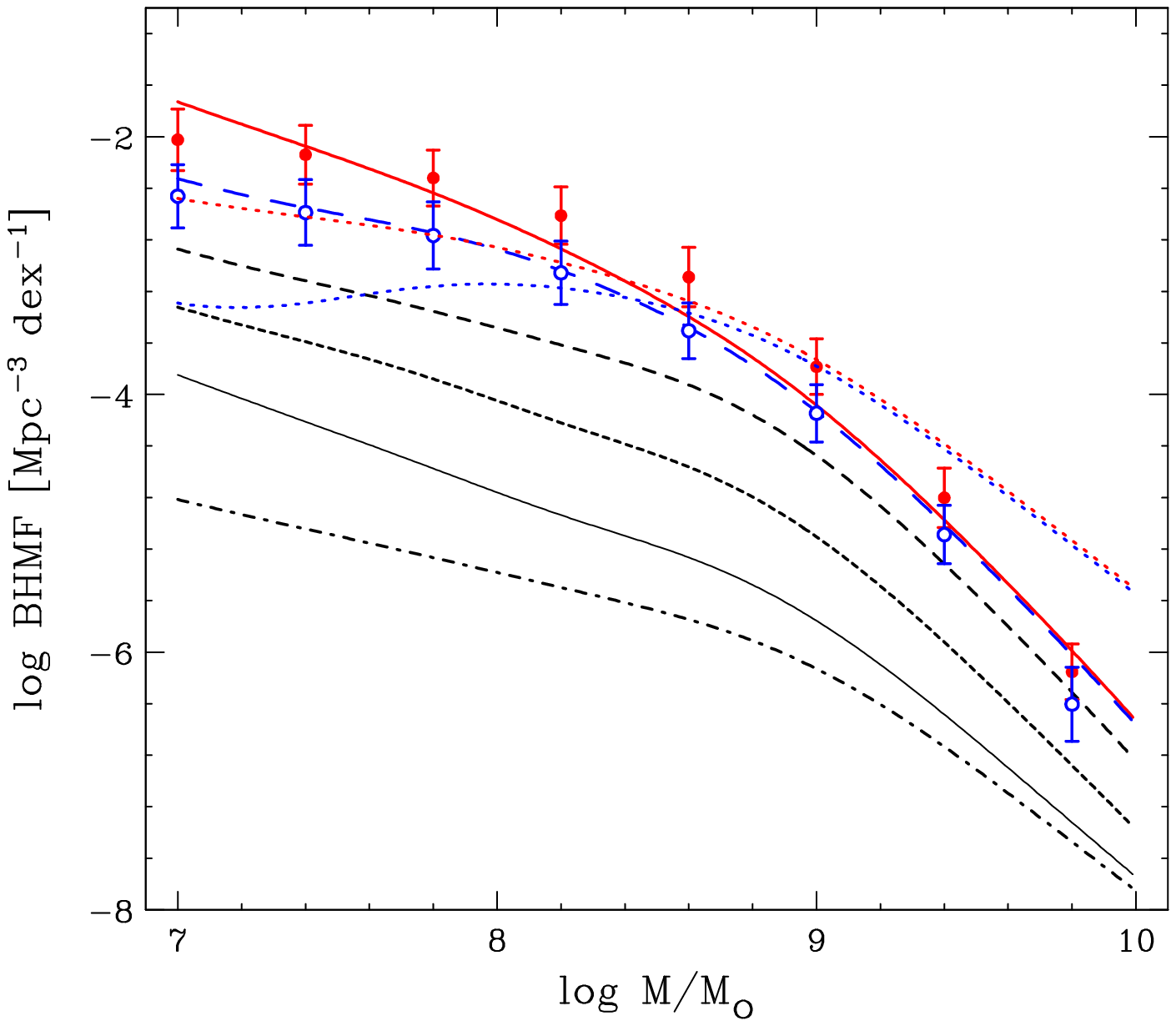}
\end{center}
\caption{
{\it Left}: the curves represent AGN relic mass functions of SMBHs calculated
from the continuity equation with a constant radiation efficiency of
$\eta = 0.091$ and an averaged Eddington ratio of log
$\overline{\lambda}=0.07$ at $z=5$ (dot-dashed, black), $z=4$ (thin
solid, black), $z=3$ (short-dashed, black), $z=2$ (med-dashed, black),
$z=1$ (long-dashed, blue), and $z=0$ (thick solid, red).
The filled circles (red) and open circles (blue) represent the observed
SMBH mass functions at $z=0$ and $z=1$, respectively, sampled from 
\citet{li11}.
{\it Right}: the same but with a mass-dependent radiation efficiency in the form of
$\eta = 0.093 (M/10^8 \solarmass)^{0.42}$ and an averaged Eddington
ratio of log $\overline{\lambda}=-0.6$. The lower (blue) and upper (red)
dotted curves represent the best-fit results with a constant radiation
efficiency and log $\overline{\lambda}=-0.6$.
}
\label{fig-bhmfmf}
\end{figure}

In a similar way, we first assume that the radiation efficiency does
not depend on black hole mass and redshift. 
We calculate AGN relic MFs
based on the new AGN BLF derived above, and compare them with the
observed SMBH MFs at $z=0$ and $z=1$
estimated by \citet{li11}.
For
each redshift, we take 8 discrete data points of log $N(z,M)$ equally
separated in a range of log $(M/\solarmass)$ = 7.0--9.6, and attach an
effective error to each point by the half difference between the
minimum and maximum allowed values indicated in Figure~4 of
\citet{li11}. We make the radiation efficiency $\eta$ and the averaged
Eddington ratio $\overline{\lambda}$ free parameters, and fix
$\sigma_{{\rm log}\lambda} = 0.3$. The initial MF is calculated $z=5$
from the BLF at the same redshift by assuming that all SMBHs were
shining (i.e., AGNs) with a constant Eddington ratio of
$\overline{\lambda}$ then. Fitting the AGN relic MFs simultaneously
to the data points at $z=0$ and $z=1$ with the $\chi^2$ algorithm, we
obtain 
$\eta=0.091^{+0.019}_{-0.016}$ 
and log $\overline{\lambda}=0.07\pm0.08$ with
$\chi^2/{\rm d.o.f.\ } = 18.7/14$ (the errors are 1$\sigma$ confidence
limits). These values are consistent with the previous result by
\citet{mar04} within the errors. The resultant AGN relic MFs at $z$=5,
4, 3, 2, 1, and 0 are plotted in Figure~\ref{fig-bhmfmf} (left) compared
with the data points of the observed SMBH MFs at $z=0$ and $z=1$.

Studies based on black hole mass measurements of optical
\citep[e.g.,][]{kol06,kel13} and X-ray \citep[e.g.,][]{nob12} selected
AGN samples indicate, however, that their averaged Eddington ratio is
significantly smaller than $\lambda = 1$. This suggests that the
apparently good reproduction of the SMBH MF by assuming constant $\lambda
\simeq 1$ and $\eta \simeq 0.1$ would not represent the actual case. A
solution to solve this contradiction is to introduce the mass dependence
of the radiation efficiency, as pointed out by \citet{cao08}. More
recently, \citet{li12} constrain the radiation efficiency as a function
of both $z$ and $M$; they find that $\eta(z,M)$ is roughly proportional
to $M^{0.5}$ at $z>1$, confirming the trend reported by \citet{cao08},
while the mass dependence becomes weaker or even inverted at lower
redshifts.

Accordingly, we empirically model $\eta$ as a power law function of
black hole mass 
\begin{equation} \eta(z,M) = \eta_8 (M/10^8 \solarmass)^\delta, 
\end{equation} although here we ignore the
redshift dependence for simplicity. We adopt log
$\overline{\lambda}=-0.6$ and $\sigma_{{\rm log}\lambda} = 0.3$
\citep{kol06}, as done in \citet{li12}. Then, performing a $\chi^2$ fit
to the observed SMBH MFs at $z=0$ and $z=1$ by \citet{li11}, we obtain
$\eta_8 = 0.093^{+0.012}_{-0.010}$ and $\delta=0.42\pm0.05$ with
$\chi^2$/{\rm d.o.f.\ } = 10.3/14. Figure~\ref{fig-bhmfmf} (right) plots the
predicted AGN relic MFs at several redshifts, which are in very good
agreements with the observed ones (data points). If the mass dependence
of $\eta$ is ignored (i.e., $\delta=0$), the AGN relic MFs significantly
under(over)-estimates the observed MFs at mass ranges lower (higher)
than log $M \approx 8$ at both $z=1$ and $z=0$; we plot this case by the
dotted lines in Figure~\ref{fig-bhmfmf} (right). The large discrepancy at
$z=0$ is attributable to that already present at $z=1$, suggesting that
the assumption of constant $\eta(z,M)$ is not proper at $z>1$, unlike
the case of $\overline{\lambda} \simeq 1.0$ discussed earlier.

\begin{figure}
\epsscale{1.0}
\begin{center}
\includegraphics[angle=0,scale=0.5]{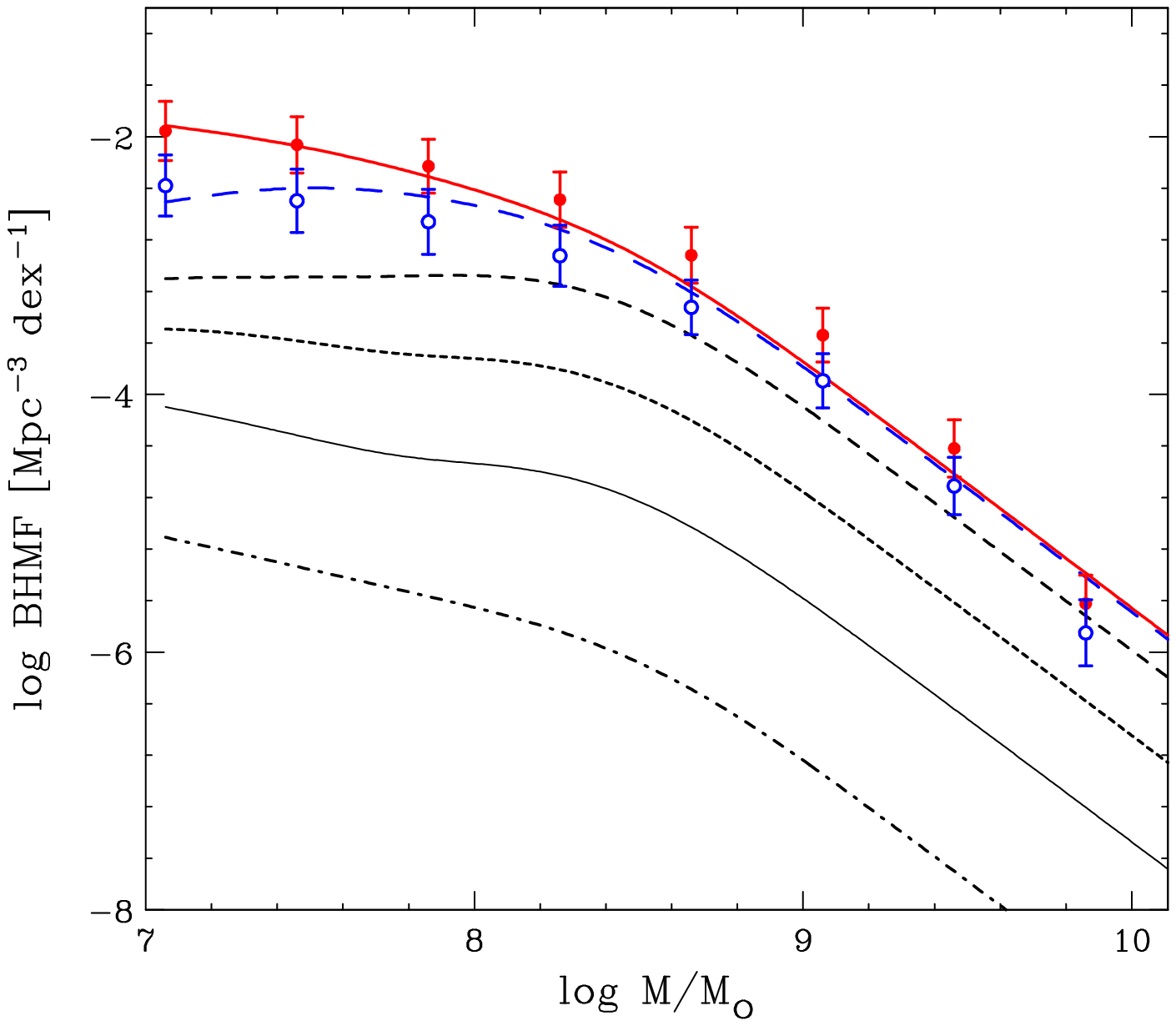}
\includegraphics[angle=0,scale=0.5]{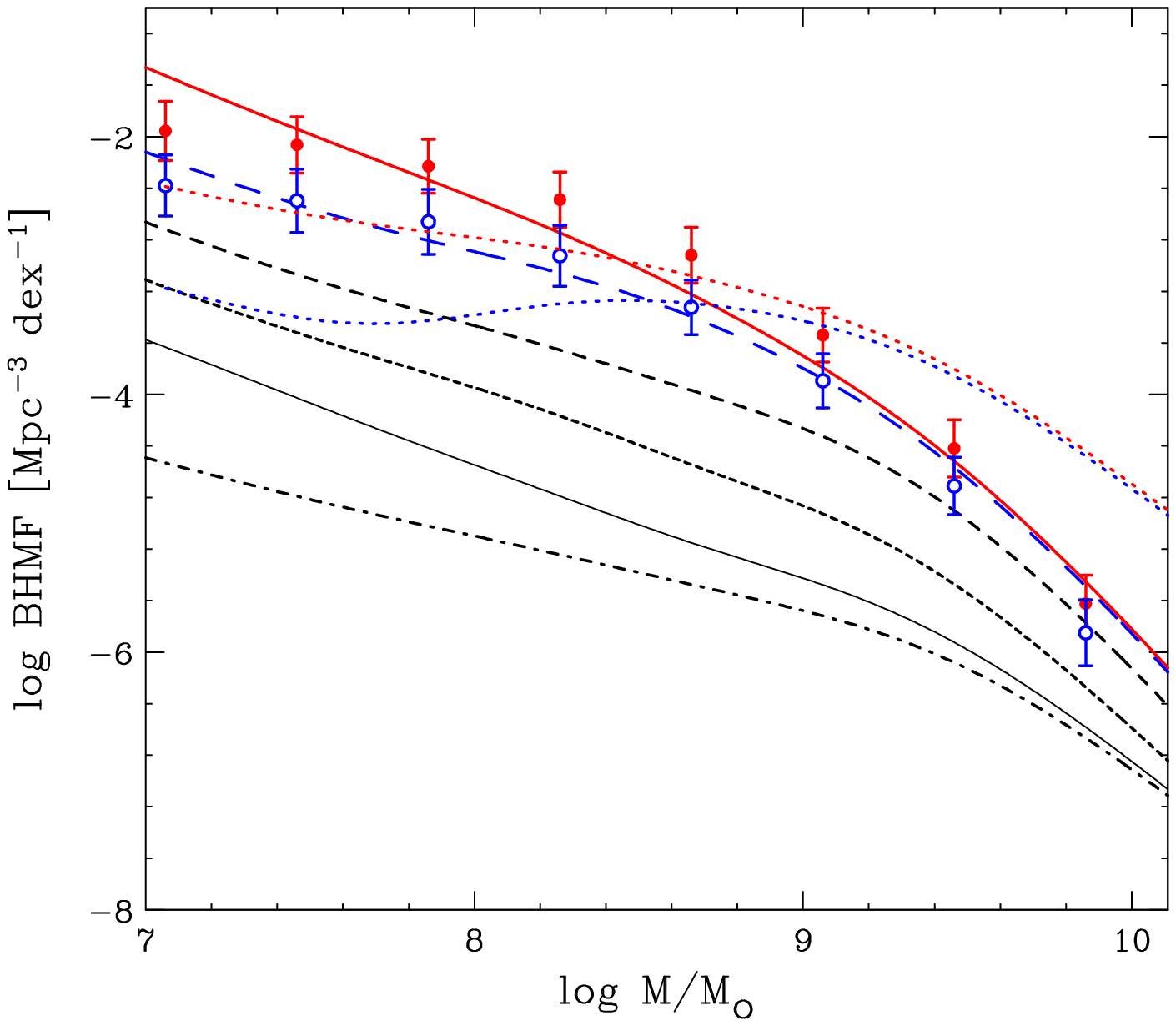}
\end{center}
\caption{
{\it Left}: the curves represent AGN relic mass functions of SMBHs calculated
from the continuity equation with a constant radiation efficiency of
$\eta = 0.053$ and an averaged Eddington ratio of log
$\overline{\lambda}=-0.014$ at $z=5$ (dot-dashed, black), $z=4$ (thin
solid, black), $z=3$ (short-dashed, black), $z=2$ (med-dashed, black),
$z=1$ (long-dashed, blue), and $z=0$ (thick solid, red).
The filled circles (red) and open circles (blue) represent the SMBH mass
functions at $z=0$ and $z=1$, respectively, revised from the
\citet{li11} data with the updated calibration between SMBH mass and
bulge mass by \citet{kor13}.
{\it Right}: the same but with a mass-dependent radiation efficiency in
the form of $\eta = 0.043 (M/10^8 \solarmass)^{0.54}$ and an averaged
Eddington ratio of log $\overline{\lambda}=-1.1$. The lower (blue) and
upper (red) dotted curves represent the best-fit results with a constant
radiation efficiency and log $\overline{\lambda}=-1.1$.
}
\label{fig-bhmfmf2}
\end{figure}

Recently, \citet{kor13} have updated the calibration between SMBH
mass and the luminosity, mass, or velocity dispersion of the bulge
component of the host galaxy in the local universe. 
This leads to an upward revision by a factor of $\sim$2--4
of SMBH masses that have been previously used. 
To examine the influences by this revision, we update the SMBH MFs
of \citet{li11}, by assuming the new SMBH-mass vs bulge-mass relation
given as equation (10) of \citet{kor13}, instead of that of \citet{har04}
adopted by \citet{li11}. The revision of the SMBH masses also affects
the deviation of the Eddington-ratio distribution of AGNs. We find that
the Eddington ratios in \citet{kol06} should be decreased by a factor of
$\approx$3 when we refer to equation (3) of \citet{kor13}, yielding an
updated mean value of log $\lambda \approx -1.1$.
We then repeat the same analysis as above by fitting AGN relic MFs to
these revised SMBH MFs. The results are shown in
Figure~\ref{fig-bhmfmf2}. When a
constant radiation efficiency is assumed, we obtain
$\eta=0.053^{+0.008}_{-0.006}$ and log $\overline{\lambda}=-0.14\pm0.07$
with $\chi^2/{\rm d.o.f.\ } = 11.2/14$ (Figure~\ref{fig-bhmfmf2} 
left). This Eddington
ratio is significantly larger than the observed value (log
$\lambda = -1.1$). When we fix $\overline{\lambda}=-1.1$ and
$\sigma_{{\rm log}\lambda} = 0.3$, the AGN relic MFs cannot reproduce
the observed MFs with $\chi^2/{\rm d.o.f.\ } = 140/15$. Introducing a 
power-law like mass dependence of the radiation efficiency again
significantly improves the fit, giving $\eta_8 =
0.043\pm0.006$ and $\delta=0.54\pm0.05$ with
$\chi^2$/{\rm d.o.f.\ } = 13.7/14 (Figure~\ref{fig-bhmfmf2} right). 

Thus, these arguments based on the new AGN BLF and updated SMBH
MFs are consistent with those by \citet{cao08}, \citet{li12}, and
\citet{sha13} that the radiation efficiency should increase with black
hole mass, at least at $z>1$. The possible contribution of mergers
neglected here only works to increase the predicted MF at the high mass
region, and hence does not essentially change this conclusion (see the
discussion of \citealt{sha13}). Importantly, we find that
the inferred radiation efficiency could be significantly reduced
compared with the previous estimates by the revision of the SMBH MFs.
Our results imply that, in relatively low mass (hence low luminosity)
AGNs with log $(M/\solarmass) \simlt 8.2$, where $\eta < 0.054$ 
on average, the standard accretion disk would be truncated
before reaching the radius corresponding to the innermost 
stable circular orbit of a non-rotating black hole and is
replaced by a RIAF. The inferred high radiation efficiencies of
higher mass AGNs suggest that their SMBHs are rotating, implying that
they grew predominantly by accretion.

We note, however, that the exact results depend on the assumption of the
Eddington ratio distribution function that is constant against
luminosity and redshift in the above analysis. Recent work using the
X-ray selected AGNs at $z\sim 1.4$ in the SXDS field by \citet{nob12}
suggests that the mean Eddington ratio is smaller at lower
luminosities. Furthermore, no consensus has been established on its
possible redshift evolution. To obtain robust conclusions, it is
important to determine the Eddington ratio distribution function and
the MF of AGNs over a wide range of luminosity and redshift based on 
accurate determination of their black hole masses.

\section{conclusions}
\label{sec-conclusions}

1. We have compiled so far the largest, highly complete sample of active
galactic nuclei (AGNs) detected in X-ray surveys performed with
\swift/BAT, \maxi, \asca, \xmm, \chandra, and \rosat, consisting of 4039
detections in the soft (0.5--2 keV) and/or hard ($>$ 2 keV) band. This
gives us the best opportunity to trace the cosmological evolution of
absorption properties and X-ray luminosity function (XLF) of all AGNs
with log \lx\ (2--10 keV) = 42--46, including both type-1 (unabsorbed)
and type-2 (absorbed) ones, from $z=0$ to $z=5$.

2. Using the latest compilation of spectral analysis of individual AGNs
detected in the \swift/BAT survey, we determine the shape of the
absorption (\nh) function in the local universe. We find that the
fraction of absorbed AGNs with log \nh\ = 22--24 among all Compton-thin
(CTN) AGNs with log \nh\ $\leq$ 24 is 0.43$\pm$0.03 at log \lx\ =
43.75. The distribution of photon index is peaked at
$\overline{\Gamma_1}=1.94\pm0.03$ for type-1 AGNs and
$\overline{\Gamma_2}=1.84\pm0.04$ for type-2 AGNs.

3. We confirm that the absorbed fraction of AGNs increases toward higher
redshifts by keeping the anti-correlation with the luminosity. At log
\lx\ = 43.75, the fraction of AGNs with log \nh\ = 22--24 among those
with log \nh\ $\leq$ 24 is proportional to $(1+z)^{0.48\pm0.05}$ up to
$z=2$.

4. To constrain the XLF of AGNs, we have developed a novel analysis
method where we perform a maximum likelihood fit directly to the list of
the observed count rate and redshift by taking into account selection
biases in each survey. Here we consider the evolution of the absorbed
fraction, the contribution of Compton-thick (CTK) AGNs with log \nh\
24--26, and AGN broad band X-ray spectra including reflection components
from tori based on the luminosity and redshift dependent unified scheme.

5. We find that the shape of the XLF at $z \sim 1-3$ 
is significantly
different from that in the local universe, showing flatter slopes in the
lower luminosity range below the break. Its cosmological evolution is
well described with the luminosity dependent density evolution (LDDE)
model, while the luminosity and density evolution (LADE) model fails
to fit the data. 

6. On the basis of the absorption function and XLF determined above, we
have newly constructed a population synthesis model of the X-Ray
Background (XRB), which can be regarded as a ``standard model'' that
well reproduces the source counts in both soft and hard bands, the
observed fractions of Compton-thick AGNs, and the spectrum of the hard
XRB.

7. To reproduce the hard XRB intensity in the 20--50 keV band within
current uncertainties, we constrain that the fraction of CTK AGNs with
log \nh\ = 24--26 to absorbed CTN AGNs with log \nh\ = 22--24 should be
0.5--1.6. This is also well consistent with the results of hard X-ray
surveys above $\simgt 10$ keV currently available.

8. We determine a bolometric luminosity function of AGNs by considering
the luminosity-dependent bolometric correction factor and its variation
from the XLF. The luminosity density of the whole AGNs has a peak around
$z\sim2$, where AGNs with bolometric luminosities of log $L = 46-47$
make the largest contribution. On the basis of Soltan's argument, 
the most recent estimate of
the local mass density of supermassive black holes (SMBHs) is reproduced by
adopting an averaged AGN radiation efficiency of $\approx$0.05, although
its mass dependence is suggested from the comparison of the AGN relic
mass function and observed ones at $z=0$ and $z=1$.

\acknowledgments

We would like to thank Murray Brightman for his help on the torus model,
Marco Ajello for providing the data of the XRB spectra in a machine
readable form, David Alexander for sending the area curve of the CDFN
survey, Bret Lehmer for his comments on the \logn s in the CDFS,
Yan-Rong Li for sending the data of the black hole mass function, and
John Kormendy, Francesco Shankar, Akylas Thanassis, 
Andy Strong, Richard Mushotzky, Chris
Done, David Ballantyne, Claudio Ricci, Agnese Del Moro, John Silverman,
Tohru Nagao, and Yuichi Terashima for discussions. We acknowledge the
efforts by the SXDS team that have helped the identification work of the
X-ray sample. This work was partly supported by the Grant-in-Aid for
Scientific Research 23540265 (YU). TM acknowledges support from by
UNAM-DGAPA Grant PAPIIT IN104113 and CONACyT Grant Cient\'ifica B\'asica
\#179662.


\ifnum1=1
\vspace{2cm}

\clearpage

\clearpage

\begin{deluxetable}{ccccccc}
\tabletypesize{\footnotesize}
\tablecaption{Surveys Used in the Analysis\label{table-survey}} 
\tablehead{
\colhead{Survey} &\colhead{Energy Band} &\colhead{Flux Limit ($\Gamma$)\tablenotemark{a}} 
&\colhead{Area} &\colhead{No. of AGNs\tablenotemark{b}} &\colhead{Completeness}&\colhead{Reference}\\ 
\colhead{} & \colhead{(keV)} &\colhead{(\ergs)} 
&\colhead{(deg$^2$)}&\colhead{}&\colhead{(\%)}&\colhead{}
}
\startdata
BAT9  &14--195& $8.4\times10^{-12}$ (1.6) & 30500 & 87 (85) & 100 & (1) \\
MAXI7 & 4--10 & $2.4\times10^{-11}$ (1.6) & 34000 & 38 (37) & 99 & (2) \\
AMSS  & 2--10 & $3.1\times10^{-13}$ (1.6) & 90.8  & 95 (95) & 97 & (3) \\
ALSS & 2--7 & $1.2\times10^{-13}$ (1.6) & 5.56 & 30 (30) & 100 & (4) \\
SXDS & 2--10 & $2.7\times10^{-15}$ (1.2) & 1.02 & 569 (567)& 99 & (5) \\
SXDS & 0.5--2 & $6.8\times10^{-16}$ (1.4) & 1.02 & 725 (703) & 99 & (5) \\
LH/XMM& 2--4.5 & $6.1\times10^{-15}$ (1.2) & 0.183 & 57 (57)& 91 & (6) \\
LH/XMM& 0.5--2 & $1.3\times10^{-15}$ (1.4) & 0.126 & 58 (57)& 88 & (7) \\
H2X   & 2--10 & $1.7\times10^{-14}$ (1.2) & 0.90 & 87 (87)& 92 & (8) \\
HBSS  & 4.5--7.5 & $2.2\times10^{-13}$ (1.6) & 25.17 & 62 (62)& 97 & (9)\\
CLASXS & 2--8 & $1.9\times10^{-14}$ (1.2) & 0.280 & 50 (50)& 96 & (10)\\
CLANS & 2--8 & $1.0\times10^{-14}$ (1.2) & 0.490 & 159 (158)& 95 & (10,11)\\
CLANS & 0.5--2 & $2.3\times10^{-15}$ (1.4) & 0.490 & 191 (183)& 95 & (10,11)\\
CDFN  & 2--8 & $2.1\times10^{-16}$ (1.0) & 0.124 & 307 (298)& 100 & (6,10)\\
CDFN  & 0.5--2 & $2.8\times10^{-17}$ (1.4) & 0.124 & 402 (234)& 100 & (7,10)\\
CDFS  & 2--8 & $1.1\times10^{-16}$ (1.0) & 0.129 & 358 (347)& 95 & (12) \\
CDFS  & 0.5--2 & $1.1\times10^{-17}$ (1.4) & 0.129 & 583 (299)& 93 & (12) \\
ROSAT\tablenotemark{c} & 0.5--2 & $3.5\times10^{-14}$ (1.8) & 0.3--21400
 & 722 (705) & 99 & (7) 
\enddata
\tablenotetext{a}{The smallest source flux in each sample, 
given in the 2--10 keV band for the
hard-band surveys above 2 keV (including BAT9) and in the 0.5--2 keV band
for the soft-band surveys. The photon index assumed to convert the
 count-rate into the flux is shown in the parenthesis.}
\tablenotetext{b}{The number of AGNs at $z=0.002-5.0$ after excluding
 sources with count rates smaller than $c_{\rm lim}(z)$ (see
 Section~\ref{sec-sample-luminosity}).}
\tablenotetext{c}{The completeness is calculated from the RBS, SA-N,
 NEPS, and RIXOS samples.}
\tablerefs{(1) \citet{tue08}; (2) \citet{ued11}; (3) \citet{aki03}; (4)
 \citet{aki00}; (5) \citet{aki14}; (6) \citet{has08} and references
 therein; (7) \citet{has05} and references therein; (9)
 \citet{del08}; (10) \citet{tro08}; (11) \citet{tro09}; (12) \citet{xue11}; }
 \label{tab-sample}
\end{deluxetable}

\begin{deluxetable}{cccccc}
\tablecaption{Parameters of Absorption Function \label{table-abs}}
\tablehead{
\colhead{$\psi_{43.75}^0$} &\colhead{$\beta$} 
&\colhead{$\psi_{\rm min}$}  &\colhead{$\epsilon$} &\colhead{$a1$}
&\colhead{$f_{\rm CTK}$}
}
\startdata
0.43$\pm$0.03\tablenotemark{a} & 0.24 (fixed) &0.2 (fixed)
& 1.7 (fixed) &0.48$\pm$0.05 &1.0 (fixed)
\enddata
\tablenotetext{a}{Errors are based on the \swift/BAT 9-month sample. It
is fixed at the best-fit value when performing ML fit to the whole sample (Section~\ref{sec-lf}).}
\tablecomments{Errors are 1$\sigma$ for a single parameter.}
\label{tab-abs}
\end{deluxetable}

\begin{deluxetable}{cccc}
\tablecaption{Parameters of Photon Index Function \label{table-photon}}
\tablehead{
\colhead{$\overline{\Gamma_1}$} &\colhead{$\Delta\Gamma_1$} &\colhead{$\overline{\Gamma_2}$} &\colhead{$\Delta\Gamma_2$} 
}
\startdata
1.94$\pm$0.03 & 0.09$\pm$0.05 & 1.84$\pm$0.04 & 0.15$\pm$0.06
\enddata
\tablecomments{Attached errors (1$\sigma$) are based on the \swift/BAT
 9-month sample. These are fixed at the best-fit values when performing ML fit to the whole sample (Section~\ref{sec-lf}).}
\label{tab-idx}
\end{deluxetable}

\begin{deluxetable}{cccccccccc}
\tabletypesize{\scriptsize}
\tablecaption{Best Fit Parameters of AGN Luminosity Functions\label{table-lf}} 
\tablehead{
\colhead{Band}
&\colhead{$A$\tablenotemark{a}} &\colhead{log $L_{*}$} &\colhead{$\gamma_1$} &\colhead{$\gamma_2$}
&\colhead{$p1^*$} &\colhead{$\beta_1$}
&\colhead{$z_{\rm c1}^*$} &\colhead{log $L_{\rm a1}$} &\colhead{$\alpha_1$}
}
\startdata
2--10 keV&
2.91$\pm$0.07 & 43.97$\pm$0.06  & 0.96$\pm$0.04 & 2.71$\pm$0.09
& 4.78$\pm$0.16 & 0.84$\pm$0.18
& 1.86$\pm$0.07 & 44.61$\pm$0.07 & 0.29$\pm$0.02\\
bolometric&
3.26$\pm$0.08 
 &45.48$\pm$0.09  &0.89$\pm$0.03 & 2.13$\pm$0.07
& 5.33$\pm$0.15 & 0.46$\pm$0.15
& 1.86\tablenotemark{b} &46.20$\pm$0.04 &0.29\tablenotemark{b}
\enddata
\tablenotetext{a}{In units of [$10^{-6}$ $h_{\rm 70}^3$ Mpc$^{-3}$].}
\tablenotetext{b}{Fixed.}
\tablecomments{Errors are 1$\sigma$ for a single parameter. Only free
 parameters are listed for the XLF; 
we fix $p2=-1.5$, $p3=-6.2$, $z_{\rm c2}^* =
3.0$, log $L_{\rm p}$ = log $L_{\rm a2} = 44 (45.67)$ for the XLF (BLF), 
and $\alpha2=-0.1$ (see Section~\ref{sec-lf}).}
\label{tab-lf}
\end{deluxetable}

\begin{deluxetable}{lccc}
\tabletypesize{\footnotesize}
\tablecaption{Comparison of Model Predictions\label{table-comparison}}
\tablehead{
\colhead{Changed Parameters} &\colhead{$a1$\tablenotemark{a}} &\colhead{\IXRB\tablenotemark{b}} &\colhead{$N(>S)$\tablenotemark{c}}\\
\colhead{} &\colhead{} &\colhead{($10^{-8}$ \ergss)} &\colhead{(deg$^{-2}$)}
}
\startdata
baseline model & 0.48$\pm$0.05& 6.39 &4300\\
$f_{\rm CTK}=2.0$ & 0.34$\pm$0.08& 6.87 &4500\\
$f_{\rm CTK}=0.5$ & 0.55$\pm$0.06& 5.72 &4120\\
$R_{\rm disk}=1.0$ & 0.39$\pm$0.07& 6.91 &4280 \\
$R_{\rm disk}=0.25$ & 0.54$\pm$0.06& 6.13 &4320\\
$\overline{\Gamma_1}=1.97$ and $\overline{\Gamma_2}=1.88$ & 0.51$\pm$0.06& 5.88 &4270\\
$\overline{\Gamma_1}=1.91$ and $\overline{\Gamma_2}=1.80$ & 0.44$\pm$0.07& 6.91 &4330\\
$\Delta\Gamma_1=0.14$ and $\Delta\Gamma_2=0.21$ & 0.48$\pm$0.05& 6.82&4340\\
$\Delta\Gamma_1=0.04$ and $\Delta\Gamma_2=0.10$ & 0.46$\pm$0.06& 6.09&4230
\enddata
\tablecomments{The other parameters are the same as in the baseline model.}
\tablenotetext{a}{The $a1$ parameter in the absorption function.}
\tablenotetext{b}{Predicted XRB Intensity in the 20--50 keV band.}
\tablenotetext{c}{Predicted AGN counts at $S=2.7\times10^{-16}$ \ergs\
 in the 2--8 keV band. The observed value in the CDFS is 4290$\pm$240 deg$^{-2}$\citep{leh12}.}
\label{tab-predictions}
\end{deluxetable}

\begin{deluxetable}{lcccccccc}
\tabletypesize{\scriptsize}
\tablecaption{Summary of Population Synthesis Models\label{table-modelsummary}}
\tablehead{
\colhead{References} &\colhead{XLF} &\colhead{$a1$\tablenotemark{a}} 
&\colhead{$\overline{\Gamma}$} &\colhead{$\Delta\Gamma$} 
&\colhead{$E_{\rm c}$} &\colhead{$f_{\rm CTK}$ (log \nh)\tablenotemark{b}} 
&\colhead{$R_{total}$\tablenotemark{c}} &\colhead{$E I_{\rm XRB}(E)$ at 30 keV}\\
\colhead{} &\colhead{} &\colhead{} 
&\colhead{} &\colhead{} 
&\colhead{(keV)} &\colhead{} 
&\colhead{} &\colhead{(keV cm$^{-2}$ s$^{-1}$ Str$^{-1}$)}
}
\startdata
(1) & U03 &0 &1.9 &0 &500 &0.63 (24--25) &1.0 & 50 \\
(2)\tablenotemark{d} & U03 &0.3 &1.9 &0 &375 &0.5 (24--25) &1.0 & 52 \\
(3) & H05 &0 &1.9 &0.2 &200 &1.05 (24--26) &$\approx$1.0& 40 \\
(4) & U03 &0.4 &1.9 &0 &300 &0.17 (24--26)&1.2 & 42 \\
(5)\tablenotemark{e} & U03 &0 &1.88 &0.15 &230 &0.1 (24--25)&$\approx$1.0 &44 \\
this work & this work &0.48 &1.94, 1.88\tablenotemark{f} &0.09, 0.15\tablenotemark{f} &300 &1.0 (24--26)&0.5+$R_{\rm torus}$ &45
\enddata
\tablenotetext{a}{The evolution index of the absorption fraction
 modelled as $\propto (1+z)^{a1}$.}
\tablenotetext{b}{The number ratio of CTK AGNs to obscured CTN AGNs with
 log \nh\ = 22--24. In the parenthesis the region of column densities (log \nh) considered for CTK AGNs is given.}
\tablenotetext{c}{The relative strength of the total reflection
 components as modelled by the pexrav code \citep{mag95}, $R_{\rm tot}\equiv\Omega/2\pi$ where $\Omega$ is the solid angle of the reflector.}
\tablenotetext{d}{Based on their figure~4 (i.e., the log \lx\ power-law
 parameterization for the absorption fraction, constant \nh\
 distribution, and local type-2 to type-1 AGN ratio of 4.0 are assumed).}
\tablenotetext{e}{Only a representative set of the parameters examined by them is shown here.}
\tablenotetext{f}{The first and second values correspond to those of type-1 and type-2 AGNs, respectively.}
\tablerefs{(1) \citet{ued03}; (2) \citet{bal06}; (3) \citet{gil07}; (4) \citet{tre09}; (5) \citet{aky12}}
\label{tab-models}
\end{deluxetable}
\fi

\end{document}